\documentclass[useAMS,usenatbib]{mn2e}
\usepackage{txfonts}
\usepackage{graphicx}
\usepackage{verbatim}
\usepackage{rotating}
\usepackage[flushleft]{threeparttable}
\usepackage{color}
\usepackage{calc}

\newcommand\finetilde{{\raise.17ex\hbox{$\scriptstyle\sim$}}} 

\begin{document}

\title[Wide-band Profile Domain Timing]{Wide-band Profile Domain Pulsar Timing Analysis}
\author[L. Lentati et al.]{\parbox{\textwidth}{L. Lentati$^{1}$\thanks{E-mail:
ltl21@cam.ac.uk}, M. Kerr$^{2}$, S. Dai$^{2}$, M. P. Hobson$^{1}$, R. M. Shannon$^{2,3}$, G. Hobbs$^{2}$, M. Bailes$^{4}$, N. D. Ramesh Bhat$^{3}$, S. Burke-Spolaor$^{5}$, W. Coles$^{6}$,
J. Dempsey$^{7}$ ,
P. D. Lasky$^{8}$,
Y. Levin$^{8}$,
R. N. Manchester$^{2}$,
S. Os{\l}owski$^{4,9,10}$,
V. Ravi$^{11}$,
D. J. Reardon$^{8}$,
P. A. Rosado$^{4}$,
R. Spiewak$^{4,12}$,
W. van Straten$^{4}$,
L. Toomey$^{2}$,
J. Wang$^{13}$
L. Wen$^{14}$
X. You$^{15}$
X. Zhu$^{14}$
 }\vspace{0.4cm}\\ %
$^{1}$ Astrophysics Group, Cavendish Laboratory, JJ Thomson Avenue,  Cambridge, CB3 0HE, UK\\
$^{2}$ Australia Telescope National Facility, CSIRO Astronomy \& Space Science, P.O. Box 76, Epping, NSW 1710, Australia\\
$^{3}$ International Centre for Radio Astronomy Research, Curtin University, Bentley, Western Australia 6102, Australia\\
$^{4}$ Centre for Astrophysics and Supercomputing, Swinburne University of Technology, P.O. Box 218, Hawthorn, Victoria 3122, Australia\\
$^{5}$ National Radio Astronomy Observatory Array Operations Center, P.O. Box O, Soccoro, NM 87801, USA\\
$^{6}$ Department of Electrical and Computer Engineering, University of California at San Diego, La Jolla, CA 92093, USA\\
$^{7}$ CSIRO Information Management \& Technology, PO Box 225, Dickson ACT 2602 \\
$^{8}$ Monash Centre for Astrophysics, School of Physics and Astronomy, Monash University, VIC 3800, Australia\\
$^{9}$ Fakult\"{a}t f\"{u}r Physik, Universit\"{a}t Bielefeld, Postfach 100131, 33501 Bielefeld, Germany\\
$^{10}$ Max Planck Institute for Radio Astronomy, Auf dem H\"{u}gel 69, D-53121 Bonn, Germany\\
$^{11}$ Cahill Center for Astronomy and Astrophysics, MC 249-17, California Institute of Technology, Pasadena, CA 91125, USA\\
$^{12}$ Department of Physics, University of Wisconsin-Milwaukee, P.O. Box 413, Milwaukee, WI 53201, USA\\
$^{13}$ Xinjiang Astronomical Observatory, Chinese Academy of Sciences, 150 Science 1-Street, Urumqi, Xinjiang 830011, China\\
$^{14}$ School of Physics, University of Western Australia, Crawley, WA 6009, Australia\\
$^{15}$ School of Physical Science and Technology, Southwest University, Chongqing, 400715, China}

\maketitle

\label{firstpage}

\begin{abstract}%
We extend profile domain pulsar timing to incorporate wide-band effects such as frequency-dependent profile evolution and broadband shape variation in the pulse profile.  We also incorporate models for temporal variations in both pulse width and in the separation in phase of the main pulse and interpulse.  We perform the analysis with both nested sampling and Hamiltonian Monte Carlo methods.  In the latter case we introduce a new parameterisation of the posterior that is extremely efficient in the low signal-to-noise regime and can be readily applied to a wide range of scientific problems.  We apply this methodology to a series of simulations, and to between seven and nine~yr of observations for  PSRs~J1713$+$0747, J1744$-$1134, and J1909$-$3744 with frequency coverage that spans 700-3600~MHz.  We use a smooth model for profile evolution across the full frequency range, and compare smooth and piecewise models for the temporal variations in DM.  We find the profile domain framework consistently results in improved timing precision compared to the standard analysis paradigm by as much as 40\% for timing parameters. Incorporating smoothness in the DM variations into the model further improves timing precision by as much as 30\%.  For PSR J1713+0747 we also detect pulse shape variation uncorrelated between epochs, which we attribute to variation intrinsic to the pulsar at a level consistent with previously published analyses.   Not accounting for this shape variation biases the measured arrival times at the level of $\sim$30~ns, the same order of magnitude as the expected shift due to gravitational-waves in the pulsar timing band.
\end{abstract}

\maketitle

\section{Introduction}

When the first pulsar was discovered in November 1967 \citep{1968Natur.217..709H}, the receivers used had a central frequency of 81.5~MHz, and a bandwidth of 1~MHz.  Today, observations with fractional bandwidths of $\ga$1/3 are common place  (e.g. \citealt{2009AAS...21460508R,2013sf2a.conf..327C,2013PASA...30...17M,2011A&A...530A..80S}).  The development of new instrumentation for pulsar timing such as the `Ultra-broadband receiver'\footnote{http://www.mpifr-bonn.mpg.de/research/fundamental/ubb} (Karuppusamy et. al. in prep.) and the `Parkes Ultra-Wideband Receiver' \citep{2015IAUGA..2256190M} will result in observations with instantaneous frequency coverage from 600~MHz up to 4~GHz.  This improvement in technology, however, has significantly increased the complexity of pulsar timing analysis.

In a typical observation of a pulsar an average pulse profile for that epoch is formed using the best available estimate of the timing model and using that to `fold' the individual pulses.  A model of the average profile, the `template' (see e.g.\ \cite{2013PASA...30...17M} for an example of forming templates)  is then used to estimate both the pulse time of arrival (ToA) and its uncertainty.  This is most commonly done via the `Fourier phase-gradient method' \citep{1992RSPTA.341..117T}  in which the phase offset between the folded profile data and the template is computed by taking the Fourier transform of both and performing a cross correlation between the two.

When considering large bandwidths, however, additional considerations are needed.  For example, as our line of sight to the pulsar varies with time, so too can the dispersion measure (DM) along that line of sight.  If an incorrect DM is assumed when determining the ToA this can cause a loss of precision by smearing the pulse and potentially bias the arrival times.  Profile evolution as a function of frequency has also been readily observed in many pulsars.  This can be a result of scattering in the ionized interstellar medium (IISM), instrumental effects, or intrinsic variation of the profile.  If the pulsar also scintillates as a result of scattering in the IISM (e.g. \citealt{1992RSPTA.341..151N}), then this will cause a change in the observed flux density of the pulse as a function of frequency.  When combined with profile evolution and DM variations, scintillation can further degrade the precision of the ToAs formed from the cross correlation process if the profile data is averaged over frequency, or if a template is used that does not incorporate profile evolution.

These factors will be complicated further by any shape variation in the observed pulse at each epoch that results from the averaging of a finite number of individual pulses.  At the single pulse level, profiles are known to exhibit significant variability (e.g. \citealt{1981ApJ...249..241H}).  As such,  as the sensitivity of observations improves and the instrumental noise decreases, this intrinsic stochasticity in the pulse shape will unavoidably become more significant.  In particular, if this process is a wide-band phenomenon then the significance of the shape variation will add coherently as the bandwidth increases.

Recently, several approaches have been proposed to mitigate some of the issues present in wide-band pulsar timing.  In both \cite{2014ApJ...790...93P} and \cite{2014MNRAS.443.3752L} two-dimensional extensions to the standard ToA forming process are introduced that fit for the DM at each observational epoch when estimating the ToA, and include profile evolution in the template.  Fitting for both the ToA and the DM in this way, however, introduces a significant number of potentially unneeded free parameters.  Ideally one would want to leverage any smoothness present in the DM variations over many years of observations in order to better constrain the model for DM at each epoch (e.g. \citealt{2014MNRAS.441.2831L}).  This, however, requires a joint analysis of all the profile data for a particular pulsar.

In this paper we extend the profile domain analysis framework described in \cite{2015MNRAS.447.2159L} and \cite{2015MNRAS.454.1058L} (henceforth L15a and L15b respectively, L15ab collectively) to include effects such as profile evolution and wide-band shape variation.  This framework allows for pulsar timing analysis to be carried out  directly on the profile data, rather than estimating ToAs  as in the standard pulsar timing paradigm.  This means that simultaneous estimation of, for example,  DM variations, pulse shape variation, profile evolution and the pulsar timing model can be carried out across the whole data set to ensure the greatest possible constraint.  This approach has already been used to model secular profile variation in the MSP J1643$-$1224 \cite{2016ApJ...828L...1S}, where the global analysis of both shape variation and timing instabilities was shown to improve the sensitivity of the data set to gravitational waves by an order of magnitude.

In L15ab all analysis was performed using the \textsc{MultiNest} \citep{2008MNRAS.384..449F, 2009MNRAS.398.1601F} and \textsc{PolyChord} \citep{2015MNRAS.453.4384H} sampling algorithms.  Both these samplers, however, are limited in the dimensionality of the problems that they can solve.  While \textsc{PolyChord} can efficiently sample from $\sim$200 dimensions, as our data sets become more complex we will need to be able to include many thousands of parameters.  As such in this work we also make use of Hamiltonian Monte Carlo (HMC) sampling methods using the Guided Hamiltonian Sampler (GHS, Balan, Hobson \& Ashdown in prep.).  In order to exploit this sampler over a wide range of problems, in Section~\ref{Section:NewLike} we first introduce a new parameterisation of the stochastic parameters used in L15ab which is significantly more efficient in the low signal-to-noise (S/N) regime.  While we will only use this parameterisation here in the context of profile domain pulsar timing analysis, its potential applications are extremely wide ranging.

In Section~\ref{Section:Like} we then describe our extensions to the profile domain formalism that incorporates wide-band effects, including optimizations that speed the likelihood evaluation time up by approximately two orders of magnitude relative to L15ab.  In Section~\ref{Section:Simulations} we apply this framework to a series of simulations of increasing complexity that include scintillation, profile evolution and DM variations.  We use these simulations to compare the timing precision obtained between the standard ToA based pulsar timing paradigm,  and our profile domain analyses.  In Sections~\ref{Section:Data} and~\ref{Section:Results} we then apply this approach to data sets for PSRs~J1713$+$0747, J1744$-$1134, and PSR J1909$-$3744.  These data sets span a range of observing frequencies from $\sim700$~MHz to 3600~MHz with observations taken over 7--9 years.  We both analyse the full data sets using the GHS and perform Bayesian model selection on the 2600--3600~MHz data using \textsc{PolyChord} to compare different descriptions of the evolution of the profile with frequency, and models that include stochastic, temporal pulse shape variation.  Finally in Section~\ref{Section:Conclusions} we offer some concluding remarks.

\section{A Low S/N model}
\label{Section:NewLike}

Given a data vector $\mathbf{d}$ of length $N_\mathrm{d}$ subject to Gaussian noise, one can write down the probability that the data is described by a model vector  $\mathbf{s}$, which can be considered a function of some parameters $\bmath{\theta}$ as:

\begin{equation}
\label{Eq:OldSignalLike}
\mathrm{Pr}(\mathbf{d} | \bmath{\theta}) = \frac{1}{\sqrt{(2\pi)^{N_\mathrm{d}} \mathrm{det}(\mathbfss{N})}}\exp\left(\frac{1}{2}(\mathbf{d} - \mathbf{s})^T\mathbfss{N}^{-1}(\mathbf{d} - \mathbf{s})\right),
\end{equation}
where without loss of generality we will consider $\mathbfss{N}$  to be a diagonal matrix with elements $N_{ij} = \sigma_i^2\delta_{ij}$, with $\sigma_i$ the standard deviation of the uncorrelated noise in data point $d_i$.

In particular, we will parameterise the model as the product of a vector of $N_\mathrm{m}$ amplitude parameters $\mathbf{a}$, and a  $N_\mathrm{d} \times N_\mathrm{m}$ matrix of basis vectors $\mathbfss{M}$.  These basis vectors could be Fourier modes as in \cite{2013PhRvD..87j4021L} that describe a stochastic gravitational wave background in pulsar timing data, or spherical harmonics as in \cite{2008MNRAS.389.1284T} where the model describes the fluctuations in the Cosmic Microwave Background.  In either case,  if we consider  a model vector, $\mathbf{s(\mathbf{a})} = \mathbfss{M}\mathbf{a}$, that describes a zero-mean stochastic process, one can then include a Gaussian prior on the amplitude parameters such that the probability that the model amplitudes, $\mathbf{a}$, are described by a set of hyper-parameters $\bmath{\varphi}$ is given by:

\begin{equation}
\label{Eq:OldPriorLike}
\mathrm{Pr}(\mathbf{a} | \bmath{\varphi}) = \frac{1}{\sqrt{(2\pi)^{N_\mathrm{m}} \mathrm{det}(\mathbf{\Psi})}}\exp\left(\frac{1}{2}\mathbf{a}^T\mathbf{\Psi}^{-1} \mathbf{a}\right),
\end{equation}
where $\mathbf{\Psi}$ is a diagonal matrix with elements $\Psi_{ij} = \varphi_i^2\delta_{ij}$, where $\varphi_i$ describes the standard deviation of the $i$th amplitude parameter  $a_i$.

One can then combine these terms to give the posterior probability of both the model amplitudes $\mathbf{a}$ and the hyper-parameters $\bmath{\varphi}$ given the data:

\begin{equation}
\label{Eq:OldLike}
\mathrm{Pr}(\mathbf{a}, \bmath{\varphi} | \mathbf{d}) = \mathrm{Pr}(\mathbf{d} | \bmath{a})\mathrm{Pr}(\mathbf{a} | \bmath{\varphi})\mathrm{Pr}(\bmath{\varphi}),
\end{equation}
where $\mathrm{Pr}(\bmath{\varphi})$ is the prior probability distribution for $\bmath{\varphi}$.

We now consider two toy problems using the posterior in Eq.~(\ref{Eq:OldLike}) which henceforth we refer to as `Model Parameterisation 1' (MP1).

\begin{figure*}
\begin{center}$
\begin{array}{cc}

\includegraphics[trim = 0 0 0 0, clip,width=90mm]{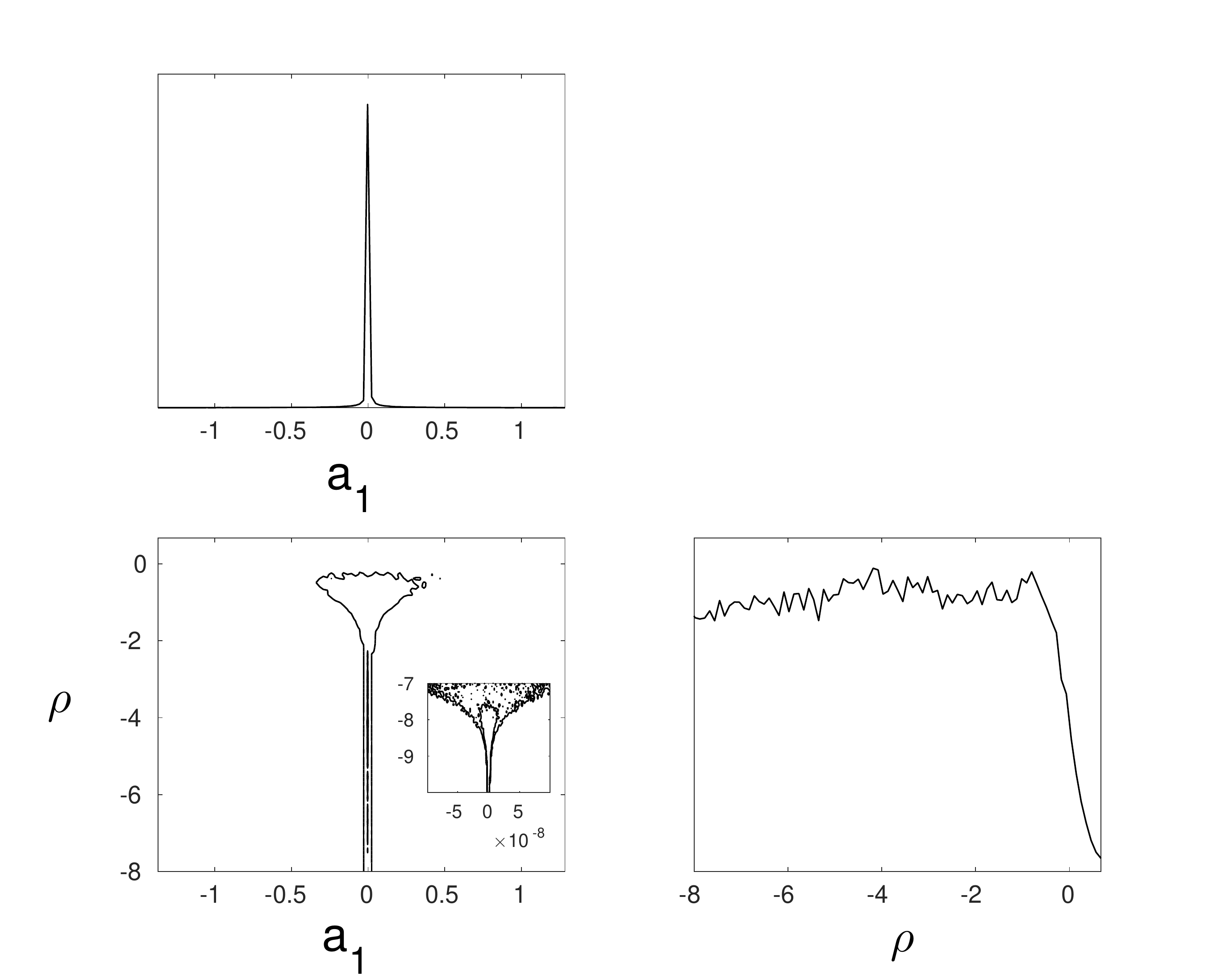} &
\includegraphics[trim = 0 0 0 0, clip,width=90mm]{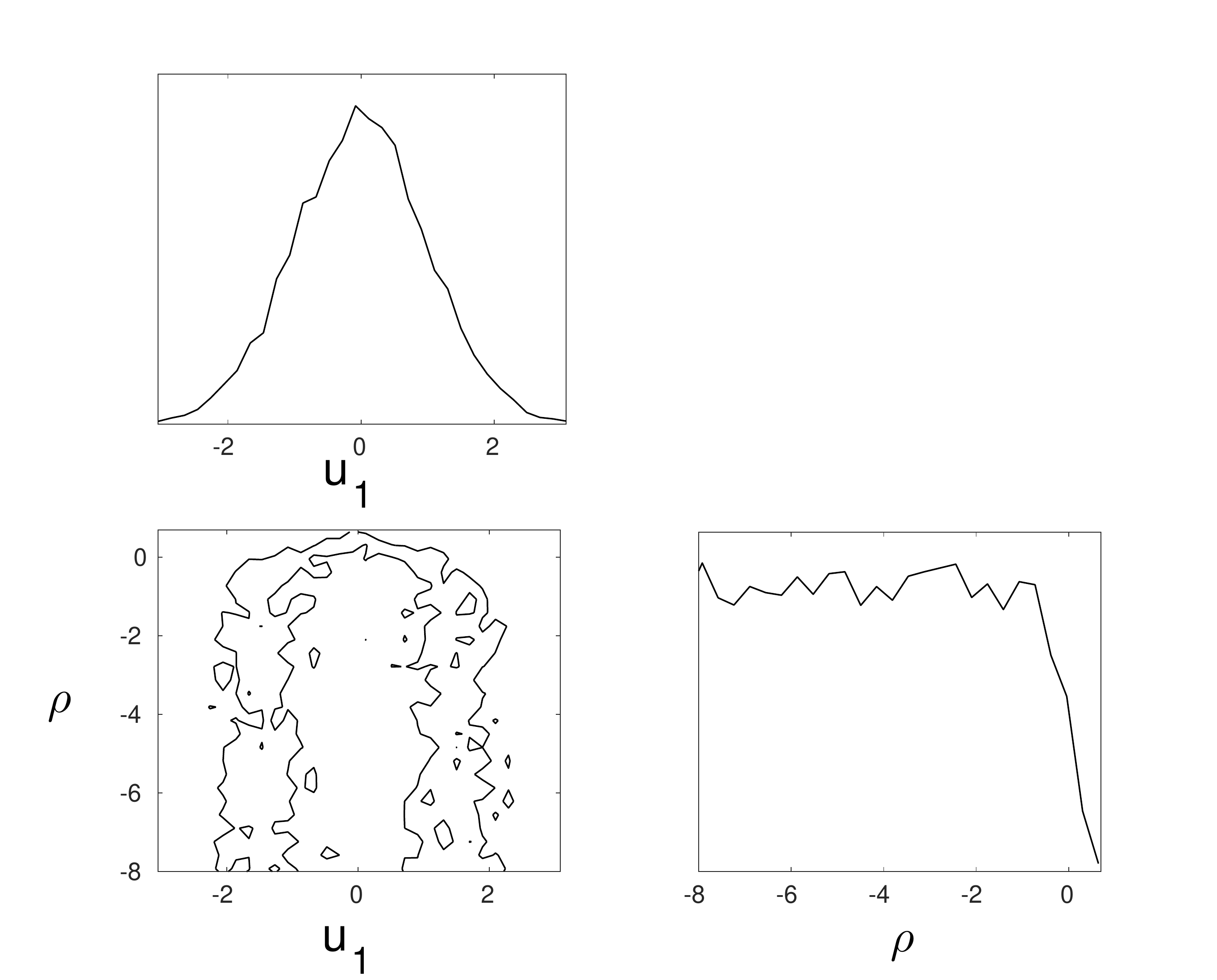} \\
\includegraphics[trim = 0 0 0 0, clip,width=90mm]{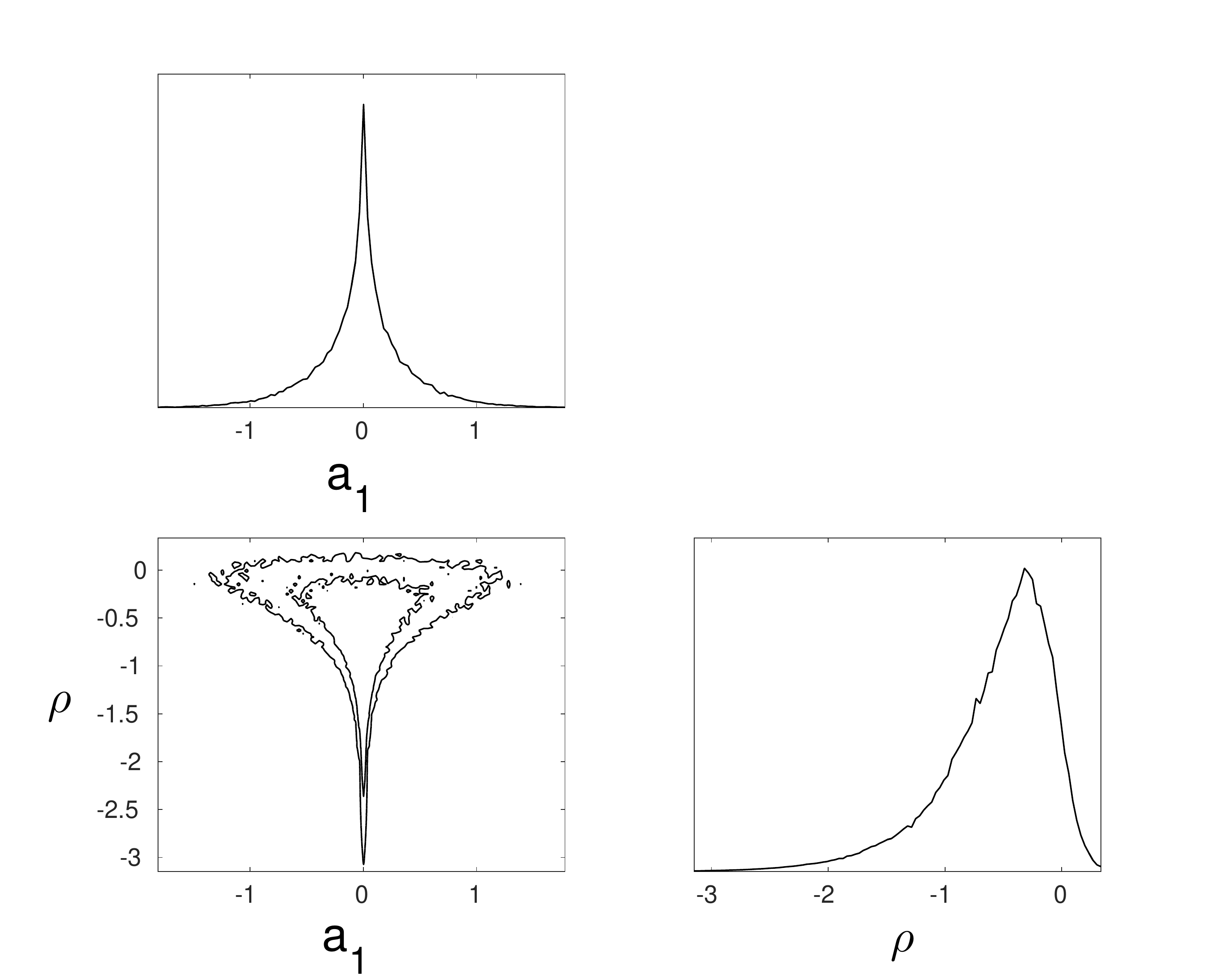} &
\includegraphics[trim = 0 0 0 0, clip,width=90mm]{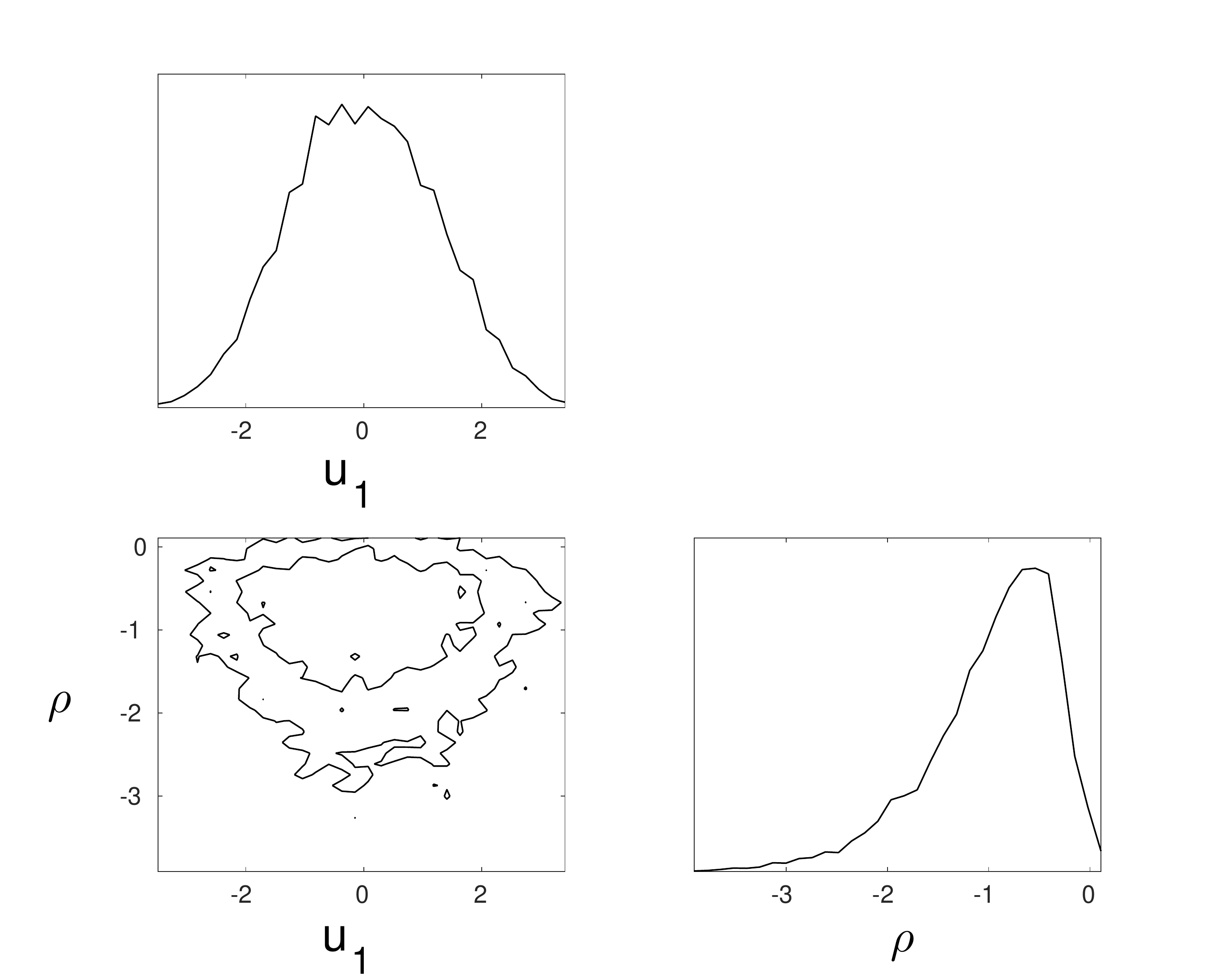} \\
\includegraphics[trim = 0 0 0 0, clip,width=90mm]{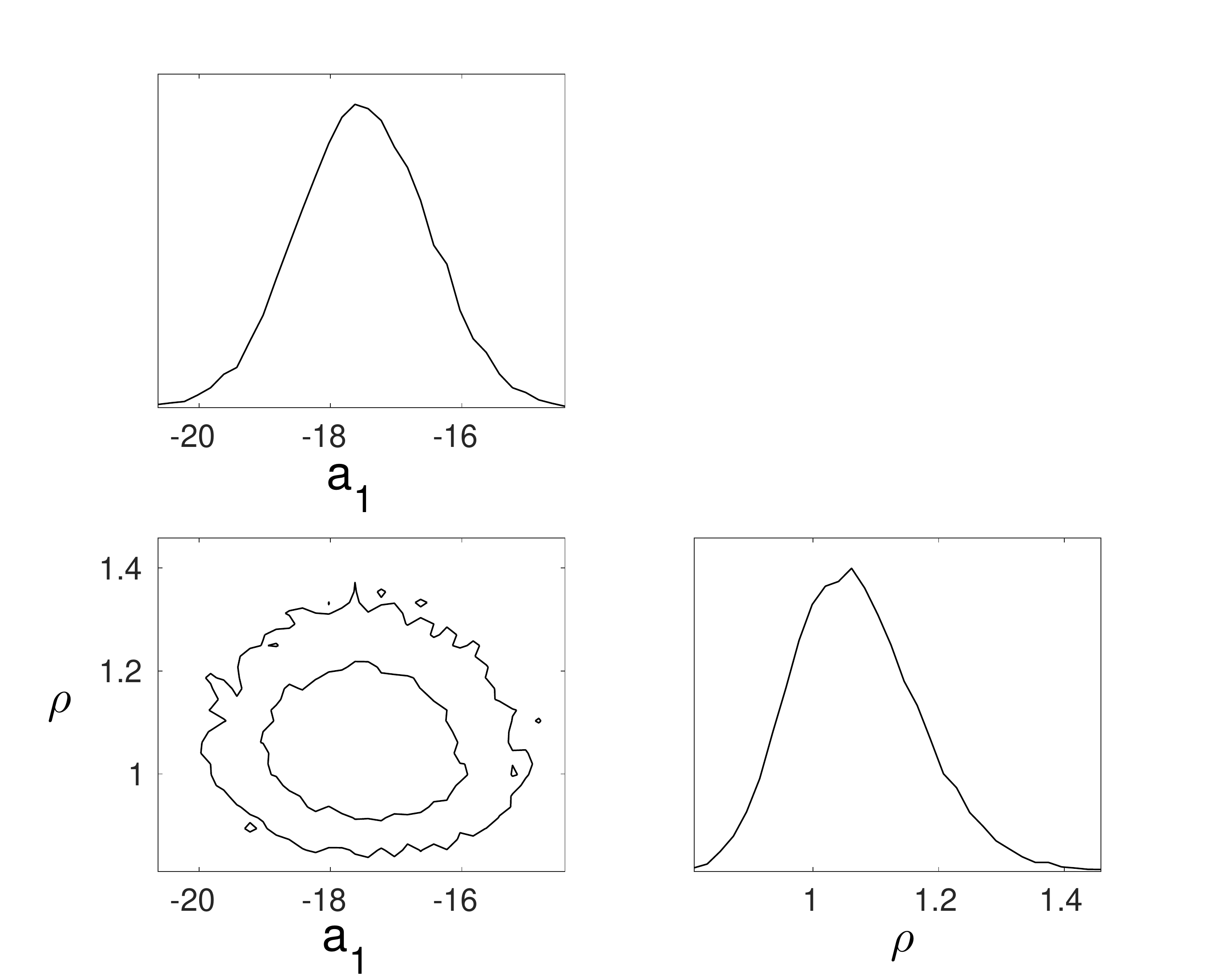} &
\includegraphics[trim = 0 0 0 0, clip,width=90mm]{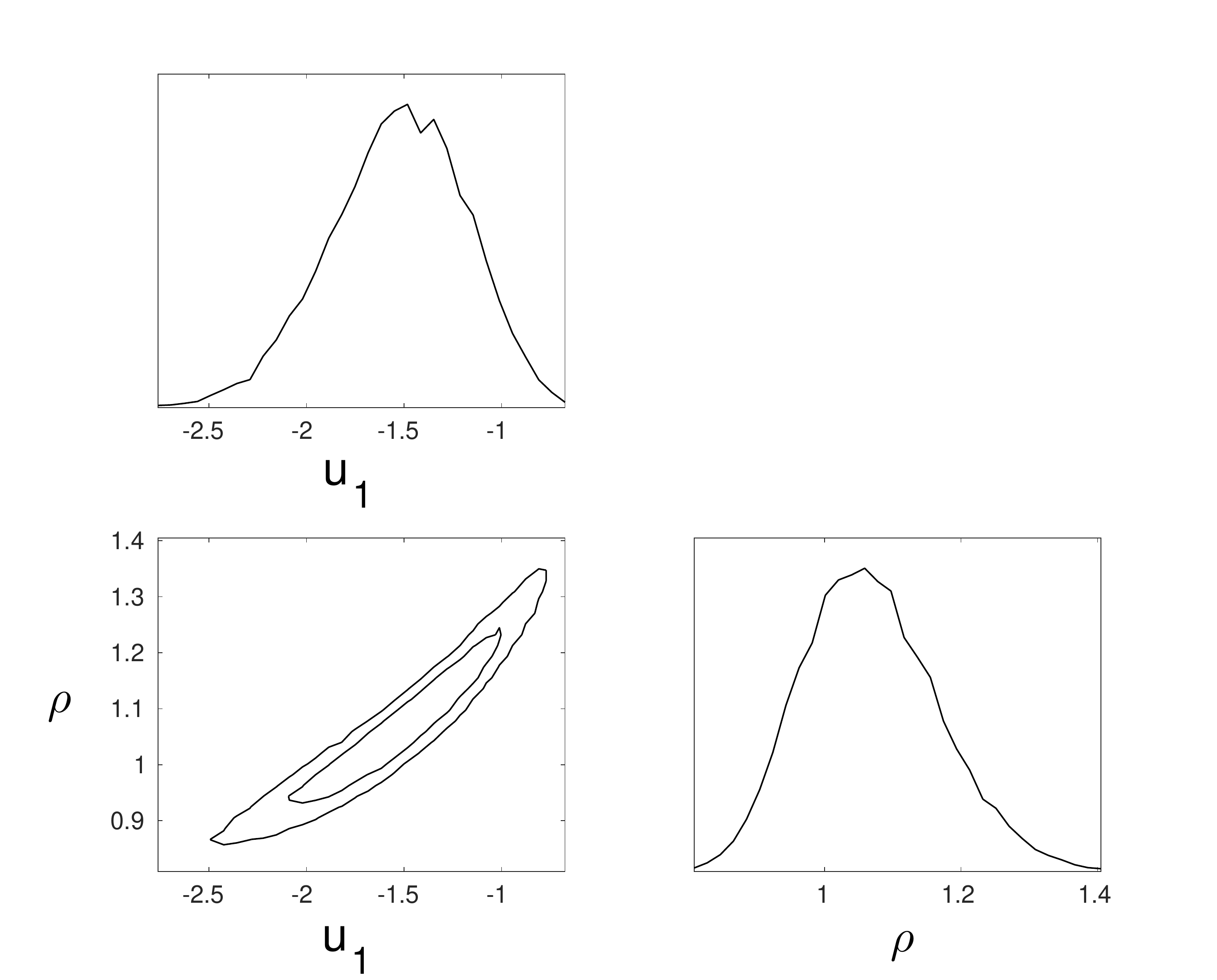} \\
\end{array}$
\end{center}
\caption{One- and two-dimensional posterior probability distributions for MP1  (left panels) and MP2 (right panels) when applied to two, 11-dimensional toy problems.  The top and middle panels are from a zero S/N example where we have used priors that are uniform in the log of $\varphi$ (top) and uniform in $\varphi$ (middle).  The bottom panels are from a high S/N problem where each amplitude parameter is $\sim$ 10$\sigma$.  We show only the first amplitude parameter $u_1$ or $a_1$ for MP2  and MP1 respectively.}
\label{figure:NLPlots}
\end{figure*}

In the first, which we refer to as problem T1, we consider a zero S/N scenario.  Here we take our data vector $\mathbf{d}$ to be of length ten, and include uncorrelated Gaussian noise with unit variance.  The matrix $\mathbfss{N}$ in Eq.~(\ref{Eq:OldSignalLike}) will thus be the identity matrix.  We also take our matrix of basis vectors $\mathbfss{M}$ to be the identity matrix, such that the model vector $\mathbf{s}$ describes an additional uncorrelated stochastic term in the data.  We include a single hyper-parameter $\varphi$ which describes the standard deviation of the model vector. In total we are therefore sampling from an 11-dimensional parameter space.  Note that the amplitude parameters are not intended to model the uncorrelated Gaussian noise, which is already factored into the matrix $\mathbfss{N}$.  Instead, the model amplitudes are describing an \textit{additional} uncorrelated  process in excess of the unit variance noise.  As such we would expect it to be consistent with zero in our analysis.

Rather than sample directly from the parameter $\varphi$ we instead sample from the parameter $\rho = \log_{10}\left(\varphi\right)$.  We perform this analysis using both a prior that is uniform in the amplitude of $\varphi$, and a prior that is uniform in the log of $\varphi$. The posterior probability distributions obtained using the \textsc{MultiNest} algorithm for the first model amplitude, $a_1$, and $\rho$ using the log-uniform prior and uniform prior are shown in the top-left and middle-left panels of Fig.~\ref{figure:NLPlots} respectively.  One can immediately see that these are highly non-trivial parameter spaces to search over in the zero S/N case.  When using a prior that is uniform in the log of $\varphi$, all scales are \textit{a priori} equally probable.  Thus, when no signal is present, any amplitude less than the noise level will have equal weight in the posterior.  The sampler must therefore explore progressively smaller values of $a_1$ up to the lower bound of the prior, which here we have set to be $-8$.  This covers approximately eight orders of magnitude in the model amplitude $a_1$.  When using a prior that is uniform in the amplitude of $\varphi$ this issue is somewhat alleviated, as the prior distribution favours larger values of $\rho$.  However, it is still challenging to sample from, especially when using sampling methods that use a single step size for each parameter, which is not appropriate in this case, in order to fully explore the parameter space.

In order to solve these issues we now describe a reparameterisation of Eqns.~ \ref{Eq:OldSignalLike} and~\ref{Eq:OldPriorLike}.  Here, rather than sample from the parameters $\mathbf{a}$, we instead sample from the related parameters $\mathbf{u}$ where for the $i$th model amplitude we will have:

\begin{equation}
a_i = u_i\varphi,
\end{equation}
where as before $\varphi$ is the model parameter describing the standard deviation amplitude parameters $\mathbf{a}$.  In order to still sample uniformly in the original parameters, $\mathbf{a}$, we then include an additional term, the determinant of the Jacobian describing the transformation from $a_i$ to $u_i$.  The Jacobian in this case has elements:

\begin{equation}
J_{i,j} = \varphi\delta_{i,j,},
\end{equation}
with $\delta_{i,j,}$ the kroneckar delta.  The determinant is therefore:

\begin{equation}
\mathrm{det}\left(\mathbfss{J}\right) = \prod_{i=0}^{m}\varphi,
\end{equation}
which acts to cancel exactly with the determinant of the matrix $\mathbf{\Psi}$ in Eq.~(\ref{Eq:OldPriorLike}).  Henceforth we refer to this as `Model parameterisation 2' (MP2).

The top-right and middle-right panels of Fig.\ref{figure:NLPlots} show the one- and two-dimensional posterior probability distributions for the first amplitude parameter, $u_1$, and $\rho$ for the same toy problem T1, using log-uniform and uniform priors in $\varphi$ for the two panels respectively.  The difference between the old and new likelihoods is clear, with a single step size in $\mathbf{u}$ now being appropriate across the range of $\rho$ sampled in both cases.  When performing the sampling using \textsc{MultiNest} we find this parameterisation results in approximately a factor 5-10 decrease in the number of samples required compared to the original parameterisation.

In the bottom-left and bottom-right panels of Fig.\ref{figure:NLPlots} we then show the same parameters obtained from a second, high S/N, toy problem in which our data vector has an additional uncorrelated stochastic component with a standard deviation of ten.  As in the first toy problem we still include the unit-variance noise component which is accounted for in the noise matrix $\mathbfss{N}$.  In this problem each amplitude parameter will therefore be of the order 10~$\sigma$, and will take values in the approximate range of $-30$ to 30.  The hyperparameter $\varphi$ will be approximately 1.  In this case the new parameterisation is less effective, as the model amplitudes and power spectrum coefficients become correlated, whereas in the original parameterisation they are completely uncorrelated.

In the following work when using the GHS we will use  both parameterisations dependent upon whether the model in question is in the high or low signal-to-noise regime.

\section{A profile domain model}
\label{Section:Like}

The methods used in this work are drawn from those presented in L15ab.  The key difference here is that, while in these previous works frequency-averaged profiles were used for each observational epoch,  here we will be using multi-channel profile data.   As a result, in the following section we will extend the existing formalism to incorporate band-wide shape variation and profile evolution.  In addition, the use of multi-channel data significantly increases the number of profiles that must be dealt with in the analysis.  In the data sets used in Section~\ref{Section:Results} there are between $12\,000$ and $21\,000$ profiles, compared to 300 used in the analysis of PSR J1909$-$0747 in L15b.  We therefore incorporate a shapelet interpolation scheme into our analysis that speeds up the likelihood evaluation by approximately two orders of magnitude for these larger data sets, with no detectable loss of timing precision.  We describe this process in Section~\ref{Section:Interpolate}.

For a full description of the general profile domain framework we refer the reader to  L15ab.  Below we will give details of how the methodology has been changed to accommodate wide-band observations, and any differences that arise when performing the analysis with the GHS.

\subsection{A Profile Model}
\label{section:Shapelets}

As in L15ab we construct our profile model using the shapelet basis \citep{2003MNRAS.338...35R}.  A shapelet profile is described by a position $t$, a scale factor $\Lambda$, and a set of $n_\mathrm{max}$ amplitude parameters, with which we can construct the set of  basis functions:

\begin{equation}
B_n(t;\Lambda) \equiv \Lambda^{-1/2} \Phi_n(\Lambda^{-1}t),
\end{equation}
with $\Phi_n(t)$ given by:

\begin{equation}
\Phi_n(t) \equiv \left[2^nn!\sqrt{\pi}\;\right]^{-1/2} H_n\left(t\right)\;\exp\left(-\frac{t^2}{2}\right),
\end{equation}
where $H_n$ is the $n$th Hermite polynomial.   We can then write our profile model, $s(t, \mathbf{\zeta}, \Lambda)$,  as the sum:

\begin{equation}
\label{Eq:oldshapefunction}
s(t, \mathbf{\zeta}, \Lambda) = \sum_{n\mathrm{=0}}^{n_{\mathrm{max}}} \zeta_n(\nu)B_n(t;\Lambda),
\end{equation}
where $\zeta_n$ are the shapelet amplitudes, which we have explicitly written as a function of the observing frequency $\nu$, and where $n_{\mathrm{max}}$ is the number of shapelet basis vectors included in the model.

In our analysis, as in L15ab, we use the shapelet basis to describe the overall profile shape, and then include an amplitude parameter for each profile in the data set that scales this average model.  One component of the shapelet model therefore acts as the reference for the rest.  In the work that follows we use the zeroth-order term,  leaving only $n_{\mathrm{max}}-1$  free parameters $\zeta_n$ which are the amplitudes for the shapelet components with $n>0$ and we take $\zeta_0 = 1$.  Each profile will also have an an arbitrary baseline offset, which we denote $\gamma$.  Written in this way Eq. \ref{Eq:oldshapefunction} becomes:

\begin{equation}
\label{Eq:shapefunction}
s(t, A, \mathbf{\zeta}, \Lambda, \gamma) = A\sum_{n\mathrm{=0}}^{n_{\mathrm{max}}} \zeta_n(\nu)B_n(t;\Lambda) + \gamma,
\end{equation}
with $A$ the overall scaling factor for a particular profile.

Note that one is free to use multiple shapelets to model a single profile.  We do this in Section~\ref{Section:Results} in our analysis of PSR J1744$-$1134, for which there are two well separated profile components corresponding to the main pulse and the interpulse.    In this case we parameterise the amplitudes of the interpulse components relative to the zeroth order Gaussian term from the main pulse, and include a parameter $\delta \phi$ corresponding to their separation in units of time.  For simplicity, we henceforth write the set of parameters that describe the profile as $\theta \equiv (A, \mathbf{\zeta}, \Lambda, \delta \phi, \gamma)$.  Our model in this case will then be:

\begin{equation}
\label{Eq:twoshapefunction}
s(t, \theta) = A\left(\sum_{n\mathrm{=0}}^{n_{\mathrm{max}}} \zeta_n(\nu)B_n(t;\Lambda) + \sum_{m\mathrm{=0}}^{m_{\mathrm{max}}} \zeta_m(\nu)B_m(t+\delta \phi;\Lambda)\right) + \gamma,
\end{equation}
where the index $n$ refers, as before, to the main pulse, and the index $m$ refers to the interpulse, which includes $m_\mathrm{max}$ amplitude parameters in the model.  While PSR J1909$-$3744 is also known to have an interpulse (eg., \citealt{2015MNRAS.449.3223D}),  as it is extremely weak we do not consider it  in the profile model.

\subsection{Shapelet interpolation}
\label{Section:Interpolate}

Evaluating the shapelet model for a large number of profiles rapidly becomes extremely computationally intensive for even a modest numbers of shapelet components ($n_{\mathrm{max}} \sim 10$).  In L15b the profile model for PSR J1909$-$3744 required approximately 30 components, and evaluating this model dominated the likelihood evaluation time.  For large numbers of profiles, this approach is not computationally tractable.  We therefore adopt an interpolation scheme, where the shapelet basis vectors are precomputed on a grid from $t=0$ up to the duration of a single phase bin for the maximum likelihood value of the scale factor $\Lambda$, determined using the fully folded profile data.  When performing the sampling we then linearise the width parameter and include this in our model simultaneously with the rest of the profile parameters, a process we describe in Section~\ref{Section:Evolution}. We need only compute the grid up to the duration of a single phase bin as for shifts of greater than one bin the interpolated model can be rotated by the relevant integer number of bins.   The interpolation interval of this grid can be chosen based on the precision of the data set being analysed.  When performing the sampling, rather than evaluate the shapelet basis vectors directly, we instead use the set of gridded basis functions that are closest to the ToA predicted by our model.

\begin{figure}
\begin{center}$
\begin{array}{c}
\includegraphics[trim = 50 50 0 50, clip,width=80mm]{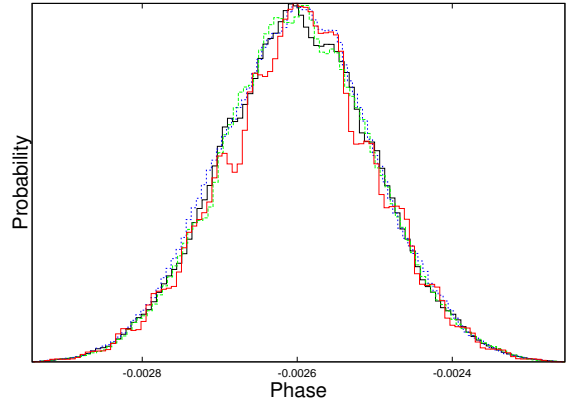} \\
\end{array}$
\end{center}
\caption{One-dimensional posterior probability distribution for the ToA in phase for a single epoch for PSR J1909$-$3744 using different interpolation intervals.  Colours represent: No interpolation (black), 1~ns (green), 32~ns (blue), 128~ns (red).  The effect of interpolation only begins to impact parameter estimates beyond the largest interval.}
\label{figure:InterpPhase}
\end{figure}

In Fig.~\ref{figure:InterpPhase} we show the estimated ToA in phase for a single profile from our PSR J1909$-$3744 data set when using no interpolation, and when choosing the closest possible interpolation interval to 1~ns (2879 interpolation bins), 32~ns (90 interpolation bins), and 128~ns (23 interpolation bins).  For the 1~ns and 32~ns interpolation interval, no significant departure from the uninterpolated case can be seen, and the parameter estimates and uncertainties derived for the ToA are almost identical.  For an interpolation interval of 128~ns the effects of quantisation become visible, and the parameter estimates and uncertainties begin to diverge.

For all the analysis that follows we use an interpolation interval of 1~ns, chosen to be sufficiently small that no bias enters our analysis as a result of the interpolation process.

\subsection{Evaluating the timing model}

In order to extend the timing framework described in L15ab  to incorporate multi-channel profile data we first consider the data in terms of a set of $N_\mathrm{e}$ epochs. Each epoch $i$ then has $N_{\mathrm{c},i}$ channels, such that the total number of profiles  $N_\mathrm{p} = \sum_{i=1}^{N_\mathrm{e}} N_{\mathrm{c},i}$.  The profile in the $j$th channel of the $i$th epoch then consists of a set of $N_{i,j}$ values representing the flux density of the profile as measured at a set of times $\mathbf{t_{i,j}}$.  As in L15ab we write our profile model, $\mathbf{s_{i,j}}$,  as a function of:

\begin{enumerate}
\item  the timing model parameters for the pulsar in question $\mathbf{\epsilon}$,
\item the overall phase offset $\phi$,
\item the profile parameters $\theta$.
\end{enumerate}
We will therefore have:

\begin{equation}
\label{Eq:newshapefunction}
s_{i,j}(t, \mathbf{\theta}, \mathbf{\epsilon}, \phi) = s(t-\tau(\mathbf{\epsilon})_{i,j}-\phi, \mathbf{\theta}),
\end{equation}
where $\tau(\mathbf{\epsilon})_{i,j}$ is the ToA predicted by the set of timing parameters $\mathbf{\epsilon}$.  As in L15b we compute the correction to the  solar system barycenter for each profile after including the phase offset and any instrumental offsets.

We can then write the likelihood that the data is described only by the timing model parameters, the phase offset, the shapelet parameters and baseline offsets as:

\begin{eqnarray}
\label{Eq:TimeLike}
\mathrm{Pr}(\mathbf{d} | \mathbf{\epsilon}, \phi, \mathbf{\theta}) &\propto& \prod_{i=1}^{N_\mathrm{e}}\prod_{j=1}^{N_{\mathrm{c},i}} \frac{1}{\sqrt{\mathrm{det}\mathbf{N_{i,j}}}}\\
&\times& \exp{\left[-\frac{1}{2}(\mathbf{d_{i,j}} - \mathbf{s_{i,j}})^T\mathbfss{N}_\mathbf{{i,j}}^{-1}(\mathbf{d_{i,j}} - \mathbf{s_{i,j}})\right]} \nonumber,
\end{eqnarray}
where $\mathbfss{N}_\mathbf{{i,j}}$  is the white noise covariance matrix for the profile corresponding to the $j$th channel in the $i$th epoch, with elements $(N_{i,j})_{mn} = \sigma_{i,j}\delta_{mn}$, with $\sigma_{i,j}$ the root-mean-square (RMS) deviation of the uncorrelated radiometer noise in the profile.

When using \textsc{PolyChord}, in order to decrease the dimensionality of the problem, we obtain an estimate of $\sigma_{i,j}$ from the off-model region of the profile data, where the fractional amplitude of the model profile is less than $0.1\%$.  As in L15b, this value will then be modified by a global scaling factor, referred to as PFAC, that will be an additional free parameter in the analysis.  We note that in traditional pulsar timing, a parameter called EFAC is used to scale the ToA uncertainties.  We refer to our scaling parameter as PFAC in the profile domain case because, in principle, they should have an equivalent effect.  When performing the analysis with the GHS, however, we do not include a PFAC, as $\sigma_{i,j}$ is a free parameter in our analysis for every profile.

\subsection{Models for Profile Evolution}
\label{Section:Evolution}

\textit{A priori}, little is known about the pulsar emission mechanism or in what way the pulse shape will vary with frequency.   In the `radius-to-frequency mapping' model,  emission is considered to be narrowband at a given altitude in the pulsars magnetosphere.  In this framework, the emission frequency decreases with altitude, which can lead to a change in both  the separation in phase,  and in the width, of different components in the average pulse profile (e.g., \citealt{1978ApJ...222.1006C}).

Given such a theoretical framework, one could potentially build a frequency-dependent profile model by searching over a basis that consists of a number of independent  components where the evolution of the width and separation in phase for each component are determined by free parameters in the analysis.  Such a model, however, will be highly non-linear and contain significant degeneracies making the parameter space difficult to explore.

In our analysis we therefore consider two models for the evolution of the profile with frequency.  The first is a general polynomial expansion of the shapelet amplitudes with frequency.  While this model is not physically motivated, it is capable of describing any potential smooth profile evolution, and as with the model for the average profile, is linear in the amplitude parameters thus making it simple to sample from.  The model is defined such that for the $p$ terms in the polynomial we can write the $i$th shapelet amplitude $\zeta_i(\nu)$ as:

\begin{equation}
\zeta_i(\nu) = \sum_{k=0}^p(\nu - \nu_c)^k\zeta_{i,k},
\end{equation}
where $\nu_c$ is an arbitrary reference frequency, and $\zeta_{i,k}$ is the amplitude parameter for the $k$th polynomial of the $i$th term in the shapelet model.  The second model is a simple linear change in the width of the profile with frequency.  We can incorporate this model into our analysis by making the scale factor, $\Lambda$, a function of frequency, such that  Eq.~(\ref{Eq:shapefunction}) becomes:

\begin{equation}
\label{Eq:fitwidthshapefunction}
s(t, \nu, A, \mathbf{\zeta}, \Lambda_c, \lambda) = A\sum_{n\mathrm{=0}}^{n_{\mathrm{max}}} \zeta_n(\nu)B_n(t;\Lambda(\nu, \lambda, \Lambda_c)),
\end{equation}
where the profile width as function of frequency is given by:

\begin{equation}
\Lambda(\nu, \lambda, \Lambda_c) = \Lambda_c(1+\lambda(\nu-\nu_c)),
\end{equation}
\c{with $\Lambda_c$ the width at the reference frequency $\nu_c$.}
In order to avoid recomputing the interpolated shapelet basis vectors for different widths, we can expand our shapelet model for small changes in the width $\lambda$.  We can define a `width profile' as:

\begin{eqnarray}
\label{Eq:WProfile}
w_{i,j}(t, \mathbf{\theta}) &=& \frac{1}{\Lambda_c} \sum_{n\mathrm{=0}}^{n_{\mathrm{max}}}\zeta_n(\nu)\left[\left(t_\beta^2 - \frac{1}{2}\right)B_n(t;\Lambda_c)\right. \nonumber \\
&-& \left.\sqrt{2n}\;t_\beta B_{n-1}(t;\Lambda_c)\right],
\end{eqnarray}
with
\begin{equation}
t_\beta = \frac{t}{\Lambda_c},
\end{equation}
such that our altered shapelet model will be given by:

\begin{equation}
\label{Eq:WWidthShiftedShapelet}
s'_{i,j}(t, \nu, \mathbf{\theta}, \lambda) = s_{i,j}(t,  \mathbf{\theta}) + \lambda (\nu-\nu_c) w_{i,j}(t, \mathbf{\theta}).
\end{equation}

\subsection{Pulse Jitter}
\label{Section:Jitter}

As in L15b, we define pulse jitter as a shift in the arrival time of the mean profile relative to that predicted by the pulsar's timing model where the shifts are uncorrelated between epochs.   Note that the term `jitter' is often used to denote any stochastic shape variation, such as `stochastic wide-band impulse modulated self-noise' (SWIMS) \citep{2011MNRAS.418.1258O}. In subsection~\ref{Section:Stochasticity} we will discuss a more general model for epoch-to-epoch variation in the profile shape to model such processes.  While stochastic shape variation that is uncorrelated between observational epochs would result in a white noise process in the ToAs, uncorrelated shifts in the arrival times could also result from the high-frequency tail of flat spectrum timing noise, or small systematic offsets between observations.  This approach makes the assumption that the jitter amplitudes at each epoch are Gaussian distributed, where the standard deviation of that distribution is a free parameter in the analysis.

When sampling with \textsc{PolyChord} we take the same approach as in L15b where this shift is linearised.  In this case for the $j$th channel in the $i$th epoch, with profile model $s_{i,j}(t, \mathbf{\theta})$, we can write down a `jitter profile' $j_{i,j}(t, \mathbf{\theta})$ given by:

\begin{eqnarray}
\label{Eq:JitterProfile}
j_{i,j}(t, \mathbf{\theta}) &=&  \frac{1}{\sqrt{2}\Lambda}\sum_{n\mathrm{=0}}^{n_{\mathrm{max}}}\zeta_n(\nu_j)\left(\sqrt{n}B_{n-1}(t;\Lambda)\right.\\ \nonumber
&-& \left.\sqrt{n+1}B_{n+1}(t;\Lambda)\right),
\end{eqnarray}
such that our shifted shapelet model will be given by:

\begin{equation}
\label{Eq:ShiftedShapelet}
s'_{i,j}(t, \mathbf{\theta}, \delta t_i) = s_{i,j}(t,  \mathbf{\theta}) +\delta t_i j_{i,j}(t, \mathbf{\theta}).
\end{equation}
Note that the only change that was required to make this model for pulse jitter band wide was to define one shift, $\delta t_i$,  for each epoch $i$.  For a narrow band description of pulse jitter, where the shifts are uncorrelated between different frequency channels, one would simply define a separate shift per channel $j$ as $\delta t_{i,j}$.

When performing the analysis with the GHS, however, we use the non-linear model, and so include the shift parameter $\delta t$ directly in our expression for the profile model:

\begin{equation}
\label{Eq:jittershapefunction}
s'_{i,j}(t, \mathbf{\theta}, \Lambda, \delta t_i) = A_i\sum_{n\mathrm{=0}}^{n_{\mathrm{max}}} \zeta_nB_n(t-\delta t_i;\Lambda).
\end{equation}

Regardless of which sampler is used, the standard deviation of the $N_\mathrm{e}$ shift parameters $\mathbf{\delta t}$  is then incorporated into the analysis by including a Gaussian prior on the amplitude parameters.  As in L15b this is achieved by defining the covariance matrix $\mathbfss{J}$ of the jitter amplitudes as:

\begin{equation}
\label{Eq:jitterPrior}
J_{ij} = \left< \delta t_i\delta t_j\right> = \mathcal{J}_i\delta_{ij},
\end{equation}
where the hyper-parameter $\mathcal{J}_i$ is the standard deviation of the arrival time shifts due to the jitter model at epoch $i$.  When using the linearised model we then normalise the amplitudes as in L15b such that the shift amplitudes can be described by a single standard deviation $\mathcal{J}_i$ for all observations in units of seconds.

We note that one of the advantages of sampling with the GHS is that it is not necessary to  marginalise analytically over any of the model parameters.  As such it is much simpler to include a non-Gaussian prior for the amplitude parameters describing the pulse jitter by following the same process as in \cite{2014MNRAS.444.3863L}.  This will be investigated further in subsequent work.

In Section~\ref{section:Shapelets} when considering a pulse profile that consisted of a main pulse and an interpulse, we explicitly wrote the shapelet model as the sum of these two separate components.  In this case the total jitter profile will similarly be given by the sum of the two separate jitter profiles.  In Section~\ref{Section:Results} in our analysis of PSR J1744$-$1134 this allows us to compare two models for pulse jitter.  First, where both the main pulse and the interpulse are shifted by the same amount at each epoch, and second, where the main pulse is able to shift independently of the interpulse.  Henceforth we refer to this second case as 'Interpulse Jitter',  which is therefore equivalent to modelling epoch-to-epoch variation in the parameter $\delta \phi$ in Eq.~(\ref{Eq:twoshapefunction}), which described the separation of the main and interpulse in the profile model.

\subsection{Width Jitter}

In addition to pulse jitter, we can use the expression for small changes in the profile width given in Eq.~(\ref{Eq:WProfile}) to model uncorrelated epoch-to-epoch variation in the profile width.  As before we can define the $i$th altered epoch given a change in width $\lambda_i$ as:

\begin{equation}
\label{Eq:WidthShiftedShapelet}
s'_{i,j}(t, \mathbf{\theta}, \lambda_i) = s_{i,j}(t,  \mathbf{\theta}) + \lambda_iw_{i,j}(t, \mathbf{\theta}).
\end{equation}

As for the pulse jitter, we then include a constraint on the amplitudes of the width jitter parameters $\mathbf{\lambda}$  by fitting for the standard deviation of the distribution.  The covariance matrix $\mathbfss{W}$ of the width jitter amplitudes is written:

\begin{equation}
\label{Eq:WjitterPrior}
W_{ij} = \left< \lambda_i\lambda_j\right> = \mathcal{W}_i\delta_{ij},
\end{equation}
where the hyper-parameters $\mathbf{\mathcal{W}}$ represent the standard deviation of the changes in the profile width at epoch $i$.

\subsection{Profile Stochasticity}
\label{Section:Stochasticity}

In addition to a model for pulse jitter, L15b introduced models for variation in the shape of the profile that were uncorrelated between epochs.    Here we will refer to this profile stochasticity in terms of uncorrelated (UC) and phase-correlated (PC) variations in the pulse shape, where in both cases the correlations we refer to are in pulse phase.

\subsubsection{Uncorrelated Stochasticity}

In L15b a `stochastic envelope' was defined which represented  an increase in the uncorrelated noise in the on pulse region.  This modelled `self-noise' in the profile data (e.g. \citealt{2014MNRAS.444.3863L}), representing shape variation on scales smaller than the width of a phase bin.  In L15b the stochastic envelope was defined to have the same shape as the mean profile, and was proportional to the amplitude of the profile.

In the analysis presented here we will consider a two component model for the stochastic envelope:

\begin{enumerate}
\item As in L15b, a term that is proportional to the profile, and
\item a second term that is constant across the on-pulse region, and is proportional to the amplitude of the pulse.
\end{enumerate}

We therefore define two scaling parameters, $\alpha$ and  $\beta$, which represent this increase in the variance, which add in quadrature to our instrument noise $\sigma_{i,j}$ such that the new variance in a bin $k$ for epoch $i$ and channel $j$ is given by:

\begin{equation}
\label{Eq:HFStoc}
\hat{\sigma}_{ijk}^2 = \sigma_{ij}^2 + \beta^2s_{i,j}(t, \mathbf{\theta})_k^2 + \alpha^2A_{i,j}^2\delta_{p},
\end{equation}
where the delta function $\delta_{p}$ is 1 for the on pulse region, and 0 for the off pulse region, and $A_{i,j}$ is the amplitude of the model profile for epoch $i$ and channel $j$.

When sampling with \textsc{PolyChord} we marginalise analytically over the individual profile amplitudes.  As such, as in L15b we use a maximum likelihood estimate of the amplitudes,  $A_{i,j}$,  in order to scale both terms in the stochastic envelope as in Eq.~\ref{Eq:HFStoc}.  This maximum likelihood estimate is given by:

\begin{equation}
\hat{A}_{i,j} = \frac{F_\mathrm{d}}{F_\mathrm{m}},
\end{equation}
with $F_\mathrm{d}$ the flux in the on-pulse region of the profile data, and $F_\mathrm{m}$ the flux in the profile model setting $A_{i,j} = 1$ . With the GHS, however, we sample from the profile amplitude numerically, and thus simply include the amplitude parameters directly in Eq.~(\ref{Eq:HFStoc}).

\subsubsection{Phase-Correlated Stochasticity}
\label{Section:LowFreqStoc}

To model low-frequency stochastic shape changes in the profile we use the same shapelet basis as for the template.  Our goal is then to robustly determine the power spectrum of the shape changes as a function of the scale (or order) of the component in the model.  We note that this is a very general model for epoch-to-epoch shape variation in the profile data.  As such the shape variation induced by the presence of  either width jitter or  pulse jitter could be modeled using this approach. However, when performing Bayesian analysis, if the data really is well described by a simple model for pulse jitter then this will be reflected in the posterior probability distributions for the stochastic hyper-parameters.  Qualitatively, this is simply because it is more probable that a single hyper-parameter describing the standard deviation of the pulse jitter is the appropriate amplitude, than it is that the many parameters which describe the low-frequency stochasticity  all take appropriate values.

We define the set of shape variation power spectrum coefficients $\mathbf{\mathcal{S}}$, such that for a given scale $i$ the covariance matrix will be given by:

\begin{equation}
\label{Eq:SjitterPrior}
S_{ij} = \left< (\zeta_i - \bar{\zeta}_i)(\zeta_j - \bar{\zeta}_j)\right> = \frac{1}{P_{d,k}}\mathcal{S}_i\delta_{ij},
\end{equation}
where $\bar{\zeta}_i$ is the amplitude of the $i$th shapelet coefficient from the average profile. The normalising factor $P_{d,k}$ is the power in the on pulse region at epoch $k$.  When sampling with \textsc{PolyChord} we estimate this scaling parameter for each profile from the data, taking $P_{d,k} = F^2_{d,k}$, whereas when sampling with the GHS we use the model amplitudes being sampled.  In either case, this gives us the power spectrum coefficients $\mathcal{S}_i$ in units of the fraction of the total power in the profile.

\subsection{Marginalising analytically over the linear parameters}
\label{section:Margins}

As in L15ab when performing our analysis with \textsc{PolyChord} we will marginalise analytically over the linear amplitude parameters in order to decrease the parameter space.  We will now describe this process for our wide-band models, for which the implementation is sufficiently different to the narrow band models described in L15b to warrant detailing below.

In total we will be marginalising over:

\begin{enumerate}
\item the arbitrary offset for each profile $\gamma_{i,j}$,
\item the overall amplitude for each profile $A_{i,j}$,
\item the band-wide pulse jitter amplitudes $\delta t_i$,
\item the band-wide width jitter amplitudes $w_i$,
\item the low-frequency profile stochasticity amplitudes $\zeta_i$.
\end{enumerate}

Given $N_\mathrm{c}$ channels in a particular epoch $i$, where the profiles in each channel $j$ have $N_{\mathrm{b}}$ phase bins, we will have $ N_\mathrm{c}N_{\mathrm{b}}$ phase bins in total for the entire epoch.

As the number of linear parameters we wish to marginalise over is $N_m$, we therefore define the $(2N_\mathrm{c}+N_\mathrm{m}) \times N_\mathrm{c}N_\mathrm{b}$ matrix $\mathbfss{P}_\mathbf{i}$, for which the first $2N_\mathrm{c}$ rows will have a block diagonal format given by:

\begin{eqnarray}
(P_i)_{2(j-1)+1,N_\mathrm{b}(j-1)+k} &=& 1 \nonumber \\
(P_i)_{2(j-1)+2,N_\mathrm{b}(j-1)+k} &=& (s_{i,j})_k, \nonumber \\
\end{eqnarray}
for each channel $j$, and phase bin $k$, and will be zero otherwise.

The remaining $N_\mathrm{m}$ rows are filled with the basis vectors describing pulse jitter, width jitter and stochastic profile variation.  For example we will have:

\begin{eqnarray}
(P_i)_{2N_\mathrm{c}+1,N_\mathrm{b}(j-1)+k} &=& j_{i,j}(t, \mathbf{\theta})_k  \nonumber \\
\end{eqnarray}
and similarily for the other stochastic parameters.

We then define the diagonal, $(2N_\mathrm{c}+N_\mathrm{m})  \times (2N_\mathrm{c}+N_\mathrm{m}) $ matrix $\mathbf{\Psi_i}$, which is zero for the elements corresponding to the overall amplitude and baseline offsets, and is equal to $\mathcal{J}_i$, $\mathcal{W}_i$ and $\mathcal{S}_i$ for the pulse jitter, width jitter and profile stochasticity parameters.

From here the marginalisation process occurs as in L15b.  Adopting the notation of that paper, we define the matrix $\mathbf{\Sigma}_\mathbf{i}$  for each epoch as $\mathbfss{P}_\mathbf{i}^T\mathbfss{N}_\mathbf{i}^{-1}\mathbfss{P}_\mathbf{i} + \mathbf{\Psi_i}^{-1}$  and define $\mathbf{P}_\mathbf{i}^T\mathbf{N}_\mathbf{i}^{-1}\mathbf{d}_\mathbf{i}$ as $\mathbf{\bar{d}}_\mathbf{i}$.

Our final probability distribution for the non-linear parameters is then given by:

\begin{eqnarray}
\label{Eq:MarginAmp}
\mathrm{Pr}(\mathbf{\theta}, \mathbf{\epsilon}, \mathbf{\mathcal{S}}, \mathbf{\mathcal{J}} | \mathbf{d}) &\propto& \prod_{i=1}^{N_e} \frac{\mathrm{det} \left(\mathbf{\Sigma}_\mathbf{i}\right)^{-\frac{1}{2}}}{\sqrt{\mathrm{det}\left(\mathbfss{N}_\mathbf{i}\right)}} \nonumber \\
&\times& \exp\left[-\frac{1}{2}\left(\mathbf{d}_\mathbf{i}^T\mathbfss{N}_\mathbf{i}^{-1} \mathbf{d}_\mathbf{i} - \mathbf{\bar{d}}_\mathbf{i}^T\mathbf{\Sigma}_\mathbf{i}^{-1}\mathbf{\bar{d}}_\mathbf{i}\right)\right].
\end{eqnarray}

\subsection{Sampling with the GHS}
\label{section:GHS}

HMC sampling methods have been successfully applied to a wide range of scientific problems with extremely high dimensionality  (e.g., over $10^6$ in \citealt{2008MNRAS.389.1284T}).   HMC uses local gradient information about the likelihood surface to draw on the mathematical framework used to describe the motion of particles in potential wells.   In doing so, the random walk behaviour exhibited by conventional MCMC methods is suppressed and the sampler remains efficient even in high-dimensional problems.

One of the main drawbacks with the HMC method that has limited its acceptance within the scientific community is the large number of tuning parameters required in order to effectively explore the parameter space.  In particular, the step size for each parameter and the number of steps $n_s$ in the trajectory must be selected, typically requiring tuning runs.  If the step size is too small, computational time is wasted taking many small steps, while if it is too large the acceptance rate will decrease.  Similarly, if the number of steps is too small, successive samples will be too close to one another in parameter space leading to high correlation within the chain, while taking too many steps will lead the HMC to follow trajectories that loop around the parameter space.

The GHS is designed to eliminate much of this tuning aspect.   It makes use of the Hessian of the problem probability distribution calculated at its peak to set the step size for each parameter.  The number of steps is then drawn from a uniform distribution, U(1, $n_\mathrm{max}$), with $n_\mathrm{max}$ of ten found to be suitable for all tested problems.  A single global scaling parameter for the step size is then the only tunable parameter, chosen such that the acceptance rate for the GHS is $\sim$ 68\%. In order to perform the sampling with the GHS we therefore need the following:

\begin{enumerate}
\item the gradient of negative log likelihood for each parameter,
\item the peak of the joint probability distribution,
\item the Hessian calculated at that peak.
\end{enumerate}

While in principle one might be tempted to include as much information in the Hessian as possible, in practice we do not follow this approach for two reasons.

Firstly many of the parameters are almost completely uncorrelated.  Storing the full Hessian in such cases therefore wastes both system memory and computing time.  The latter effect is a result of the matrix-vector multiplications required by the GHS in each likelihood calculation that are of the size of the Hessian.   We therefore divide our Hessian into  $N_p+1$ blocks.  We have one  $3\times3$ block for each profile, which includes the elements of the Hessian for the overall amplitude, baseline offset and system noise for that profile. The Hessian for all remaining parameters, which includes the timing model, profile model and any stochastic parameters, is then stored as a final block.

Secondly, the Hessian can become numerically unstable when including the cross terms for certain parameters.  For example, we have found that when using the MP2  likelihood, including the full cross correlation between the amplitude parameters that define the model vector $\mathbf{s}$, and the timing model parameters can significantly reduce the sampling efficiency as a result of this instability.

\subsubsection{Obtaining the maximum likelihood solution}

In order to obtain a suitable point at which to calculate the Hessian we use a two-stage process.  We first use a Nelder-Mead optimisation algorithm \citep{NelderMead65} to find the maximum-likelihood parameter estimates using the likelihood in Eq.~(\ref{Eq:MarginAmp}).  Using these parameters we then analytically solve for the remaining linear parameters that were marginalised over in Section~\ref{section:Margins}.

\subsubsection{Calculating gradients and the Hessian}

While we will not list the gradients and elements of the Hessian for all parameters, we will provide details for some of the key parameters below.  In particular, the gradients and elements of the Hessian for the non-linear timing model and all parameters that result in shifts of the profile warrant some consideration.

We find the basis vectors used by \textsc{Tempo2}\footnote{https://bitbucket.org/psrsoft/tempo2} to evaluate the linear timing model to be sufficient to compute the gradient and Hessian of the non-linear timing model parameters.

In order to calculate the gradient for shifts in the arrival time of the profile we can use the jitter profile, $\mathbf{j}$,  given in Eq.~(\ref{Eq:JitterProfile}).  For example, denoting the profile residuals after subtracting the model at a particular point in parameter space as $\mathbf{r}$, and the basis vector for the $i$th timing model parameter, $\epsilon_i$,  as $\mathbf{M_i}$, we can approximate the log-likelihood for small changes in this parameter as:

\begin{equation}
\label{Eq:ApproxLike}
L(d | \epsilon_i) = \sum_{j=0}^{N_\mathrm{p}}\frac{1}{2}\left(\mathbf{r_j} - A_jM_{ij}\epsilon_i\mathbf{j_j}\right)^T\mathbfss{N}_j^{-1}\left(\mathbf{r_j} - A_jM_{ij}\epsilon_i\mathbf{j_j}\right),
\end{equation}
with $A_j$ the amplitude, and $M_{ij}$ the value of the basis vector $\mathbf{M_i}$ for the $j$th profile.  We can then compute the gradient of Eq.~(\ref{Eq:ApproxLike}) with respect to $\epsilon_i$:

\begin{equation}
\label{Eq:ApproxLikeGrad}
\frac{\mathrm{d}L(d | \epsilon_i)}{\mathrm{d}\epsilon_i} = \sum_{j=0}^{N_\mathrm{p}} -A_jM_{ij}\mathbf{j_j}^T\mathbfss{N}_j^{-1}\left(\mathbf{r_j} - A_jM_{ij}\epsilon_i\mathbf{j_j}\right),
\end{equation}
and the second derivative:
\begin{equation}
\label{Eq:ApproxLikeGrad}
\frac{\mathrm{d^2}L(d | \epsilon_i)}{\mathrm{d}\epsilon_i^2} = \sum_{j=0}^{N_\mathrm{p}} (A_jM_{ij})^2\mathbf{j_j}^T\mathbfss{N}_i^{-1}\mathbf{j_j} .
\end{equation}

\section{Simulations}
\label{Section:Simulations}

\begin{table}
\caption{Simulated Timing Model Parameters}
\begin{tabular}{ll}
\hline
Parameter & Value \\
\hline
Right ascension, $\alpha$ (hh:mm:ss)\dotfill &  17:13:49.5325456 \\
Declination, $\delta$ (dd:mm:ss)\dotfill & +07:47:37.49994 \\
Pulse frequency, $\nu$ (s$^{-1}$)\dotfill & 218.81184044143712135 \\
First derivative of pulse frequency, $\dot{\nu}$ (s$^{-2}$)\dotfill & $-$4.08412$\times 10^{-16}$  \\
Dispersion measure, DM (cm$^{-3}$pc)\dotfill & 16.0 \\
\hline
\end{tabular}\label{Table:SimParams}
\end{table}

In order to test the efficacy of the analysis method described in the preceding sections, we apply it to three simulations of increasing complexity.  Details of these simulations are given below:

\begin{figure*}
\begin{center}$
\begin{array}{cc}
\includegraphics[trim = 50 50 0 50, clip,width=80mm]{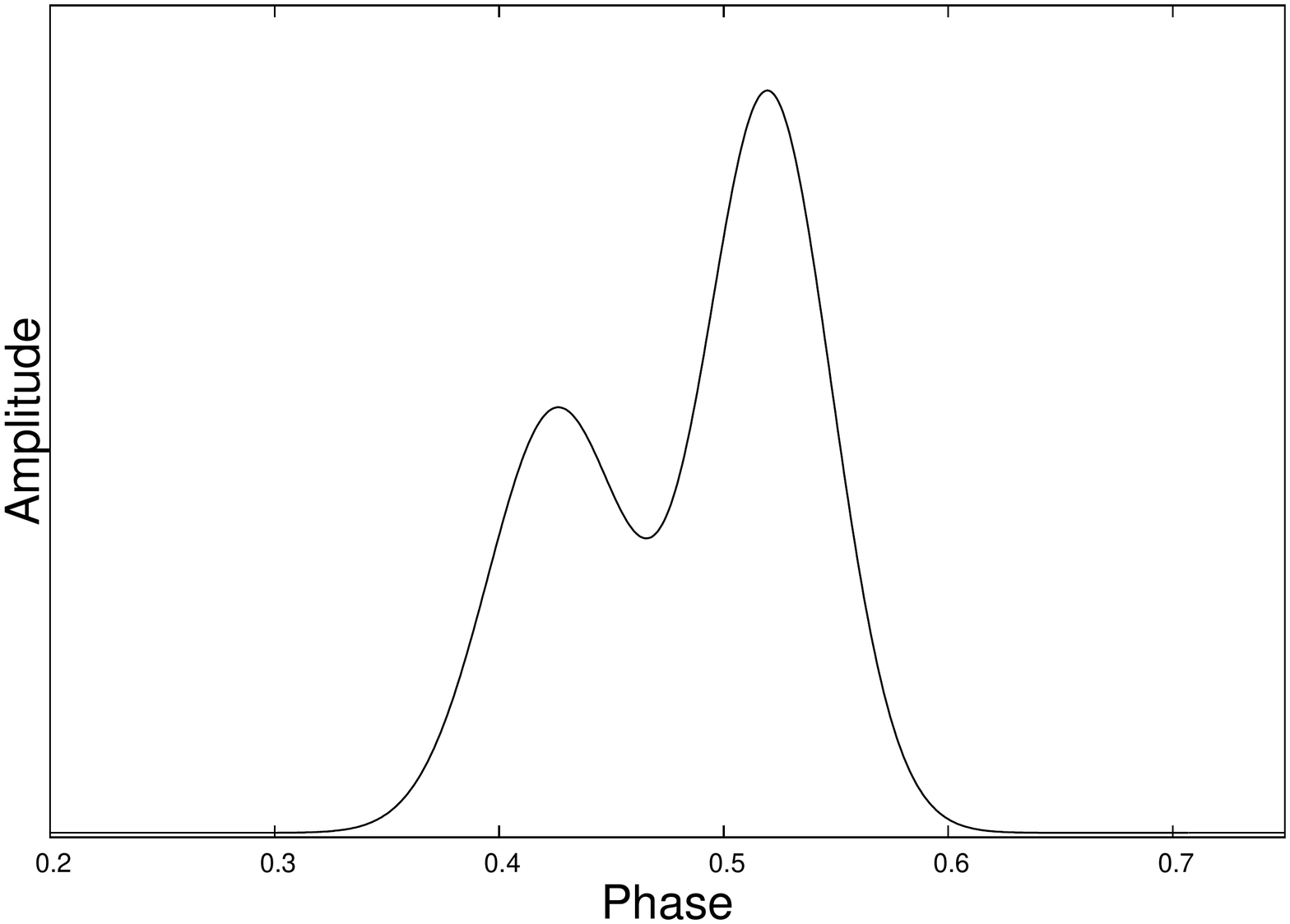} &
\includegraphics[trim = 50 50 0 50, clip,width=80mm]{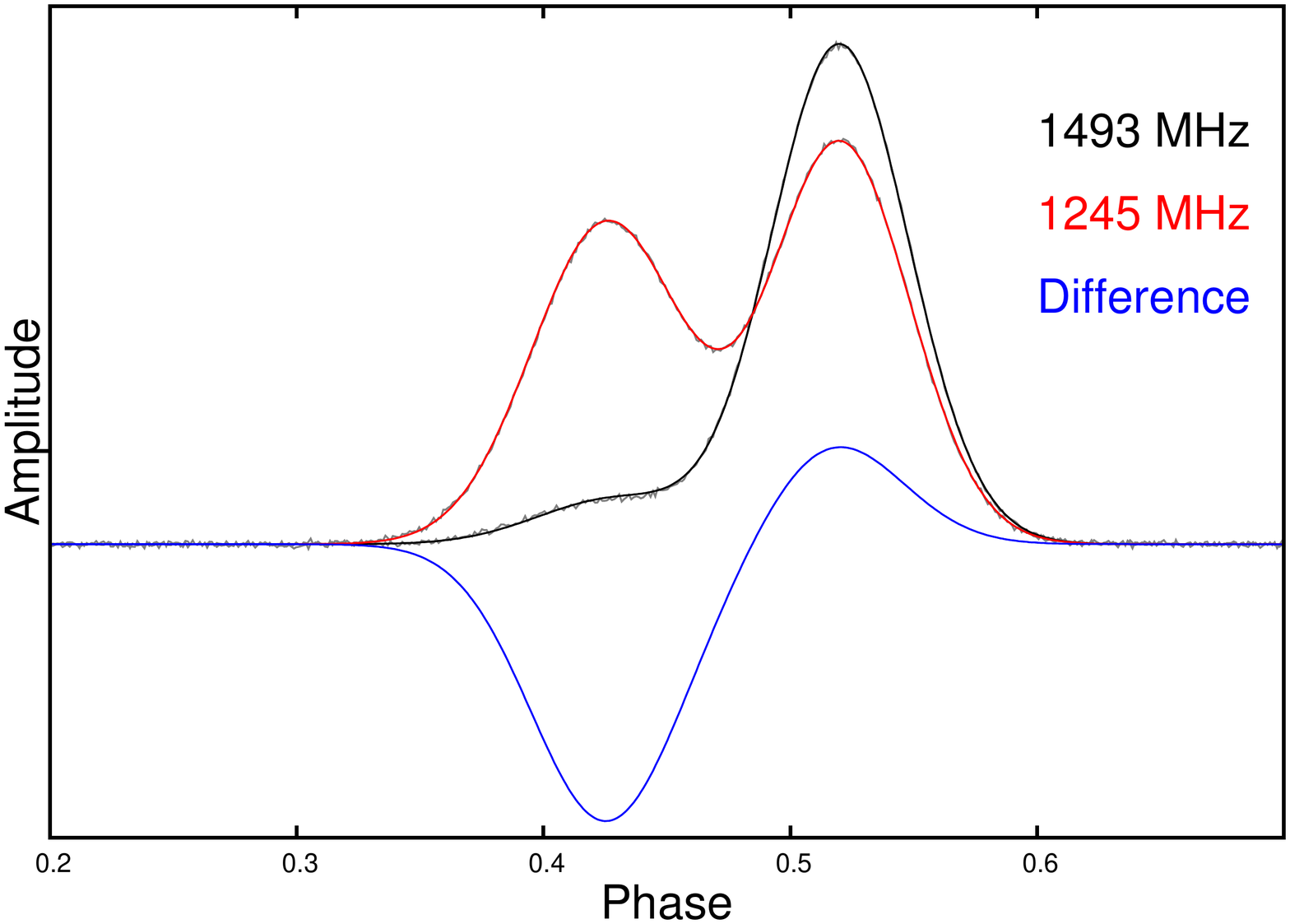} \\
\includegraphics[trim = 50 50 0 50, clip,width=80mm]{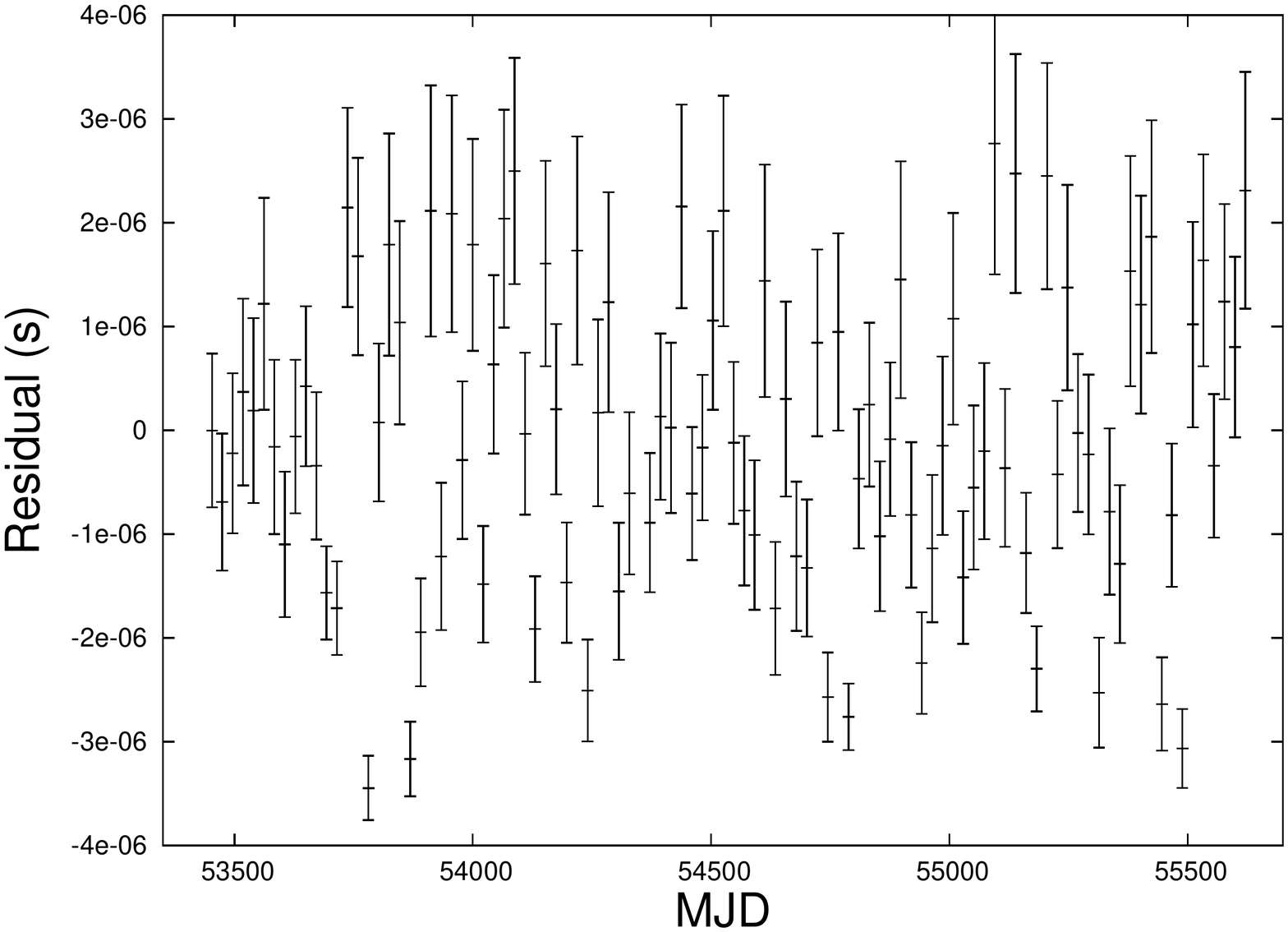} &
\includegraphics[trim = 50 50 0 50, clip,width=80mm]{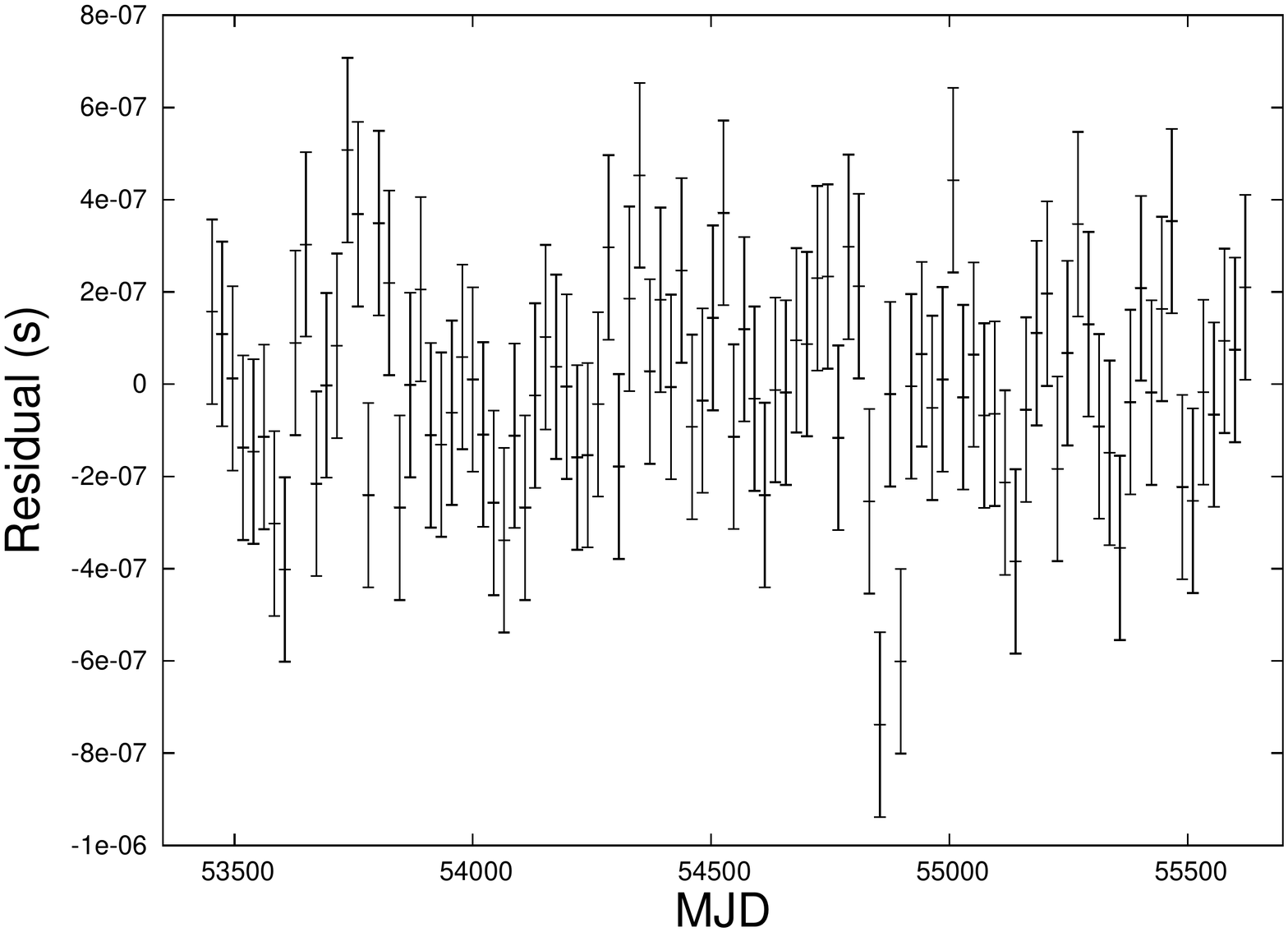} \\
\includegraphics[trim = 50 50 50 50, clip,width=90mm]{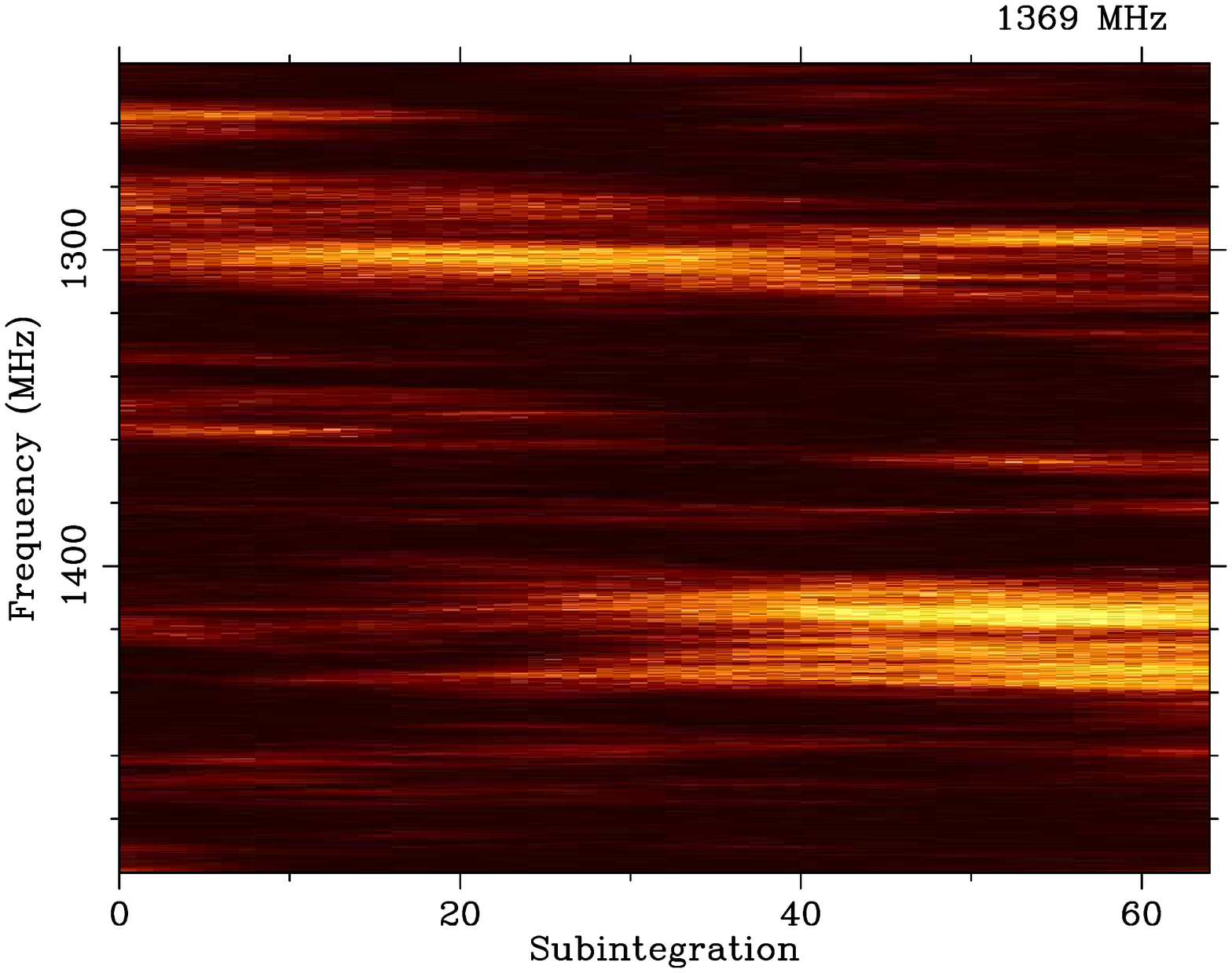} &
\includegraphics[trim = 0 0 0 0, clip,width=80mm]{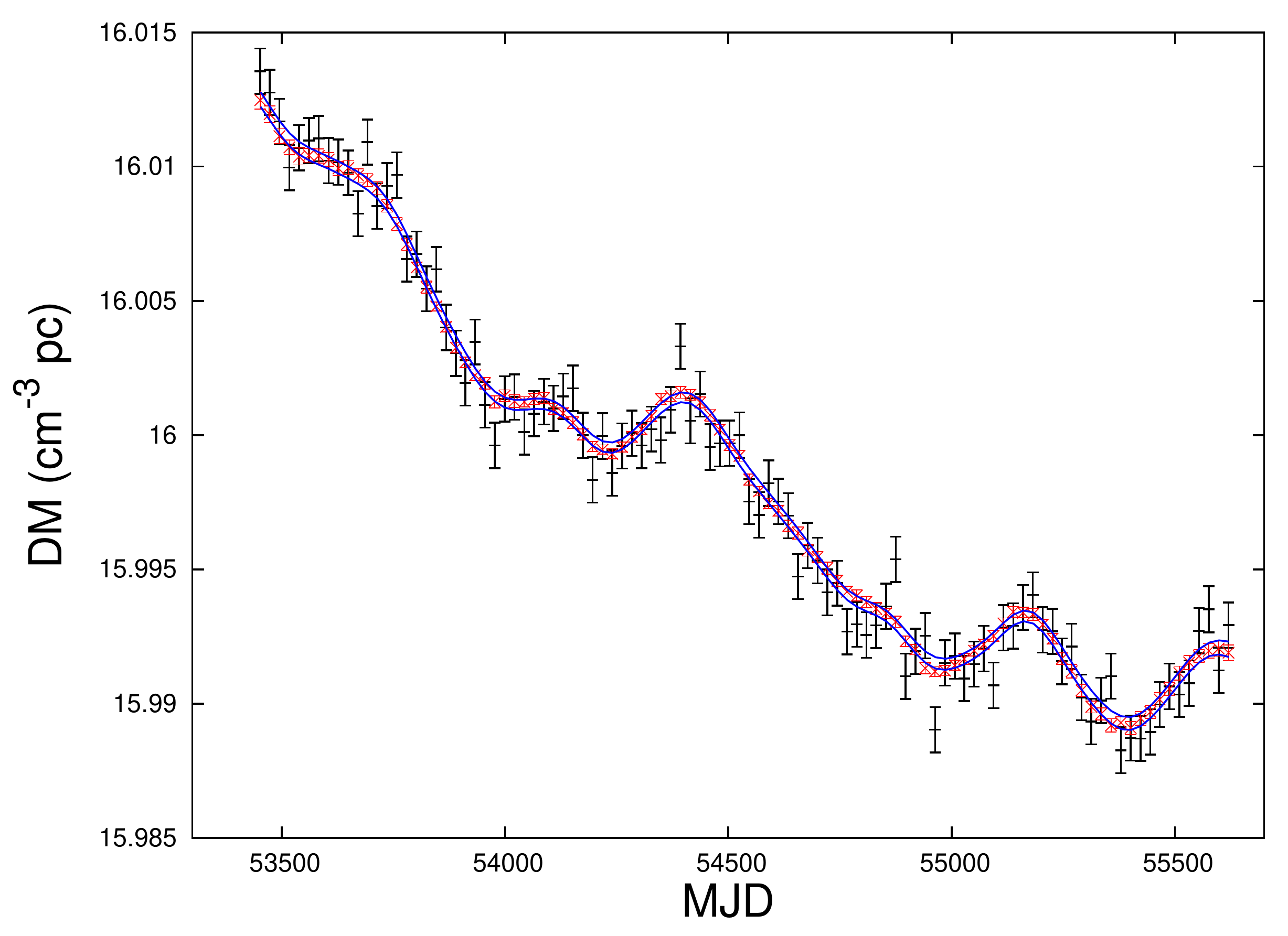} \\
\end{array}$
\end{center}
\caption{(Top left) Model profile used in Simulation~1.  (Top right) Evolving profile model used in Simulations~2 and ~3.  The profile evolves linearly across the band from the red to the black curve.  (Middle) Timing residuals from Simulation~2 when fully frequency averaging the profile data before forming the ToAs (left panel), and when forming ToAs from 8~MHz channels, and then averaging the residuals for each epoch after fitting for the timing model (right panel). In the latter case we include `FD' parameters in the timing model to act as a proxy to profile evolution in the ToA analysis.  (Bottom left) Example of the scintillation for one epoch in Simulation~2. (Bottom right) DM variations in Simulation~3 measured independently at each epoch (black points with error bars), measured using the DMX parameterisation in a profile domain analysis (red points with error bars), and using a smooth model for the DM variations in the profile domain (blue line representing the 1~$\sigma$ confidence interval). }
\label{figure:SimPlotss}
\end{figure*}

\begin{itemize}
\item{Simulation 1:} \\
\\Simulation~1 consists of 100 observational epochs covering a total time span of six years. Each epoch assumes 256~MHz of bandwidth with a central frequency of 1369~MHz, which is split into 32, 8-MHz channels.  The model profile used is shown in Fig.~\ref{figure:SimPlotss} (top left), and does not evolve over this frequency range.  Each epoch includes 64 one-minute subintegrations, with a total integrated S/N $\sim$1000 for each epoch.  We  simulate a simple timing model including only position, period, period derivative and DM with values for all parameters chosen to be consistent with PSR~J1713+0747.  We list the simulated timing model parameters in Table~\ref{Table:SimParams}. \\
\item{Simulation 2:} \\
\\As Simulation~1, however we include both profile evolution across the observed band and scintillation.  The simulated profile at the top and bottom of the band is shown in Fig.~\ref{figure:SimPlotss} (top right), while an example of the simulated scintillation for one epoch is shown in Fig.~\ref{figure:SimPlotss} (bottom left).  We simulate the dynamic spectrum of interstellar scintillation according to \cite{2016MNRAS.462.3115D}. Our simulations are in the regime of strong scintillation, and are valid for a thin scattering layer of homogeneous isotropic turbulence with a Kolmogorov spectrum.\\
\item{Simulation 3:} \\
\\As Simulation~2 however we also include significant DM variations, consistent with those observed in PSR~J1045$-$4509 \citep{2016MNRAS.455.1751R}. The measured DM at each epoch is shown in  Fig.~\ref{figure:SimPlotss} (bottom right, black points with error bars).\\
\end{itemize}

For each simulation we obtain fully frequency-averaged ToAs using the timing model in Table~\ref{Table:SimParams} and also ToAs for the 32 channels separately.  We then perform a standard timing analysis using the ToAs for these two data sets and a profile domain analysis on the profiles used to form the 8-MHz ToAs.

We list the relative timing precision, given as the mean ratio of the uncertainties in the timing model parameters, obtained from these different analyses for each of the three simulations in Table~\ref{Table:SimPrecision}.  In Simulation~1 we find no evidence for any stochastic parameters in either the ToA or profile domain, and obtain completely consistent results from each of the analysis methods.  This is to be expected, as the assumptions made when forming the ToAs (no profile evolution, stationary DM) were correct.

This is not the case in either of the subsequent simulations.  In Simulation~2 the combination of profile evolution and scintillation results in significant loss of precision in the fully frequency-averaged ToAs.   This can be seen in Fig.~\ref{figure:SimPlotss} (middle-left panel), where we show the large amount of scatter present in the timing residuals.  When modelling this scatter in the frequency-averaged ToA analysis we use an EQUAD parameter (an additional white noise term that adds in quadrature with the ToA uncertainty) with a value of $1.34\times10^{-6}$~s. Given the typical uncertainty on the frequency-averaged ToAs is $0.8\times10^{-6}$~s this is a significant decrease in the precision of the data.   The impact of the profile evolution and scintillation is much less significant in the 8-MHz ToAs.  In Fig.~\ref{figure:SimPlotss} (middle-right panel) we show the epoch-averaged residuals when modelling the profile evolution using the `FD' parameterisation  \citep{2015ApJ...813...65T}.  The FD parameters model profile evolution as a shift in the arrival time given by:

\begin{equation}
\Delta_\mathrm{FD}= \sum_{i=1}^n c_i\log\left(\frac{\nu}{1\mathrm{GHz}}\right)^i,
\end{equation}
with the $c_i$ free parameters to be fit for.  In our analysis we find only the first term is required to model the profile evolution in our simulations in the ToA domain with a value of $(33.4 \pm 1.0) \times 10^{-5}$.  Higher order terms are highly correlated, possibly due to the narrow overall bandwidth of the simulation, and we find including them does not improve the fit.

\begin{figure}
\begin{center}$
\begin{array}{c}
\includegraphics[trim = 50 50 0 50, clip,width=80mm]{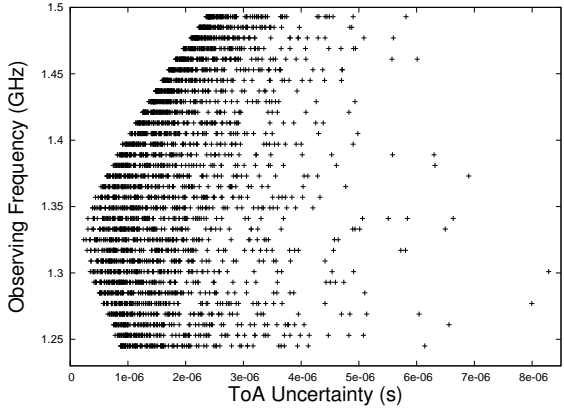} \\
\end{array}$
\end{center}
\caption{Formal ToA uncertainties for the 8MHz ToAs from Simulation~2.  A clear trend towards larger error bars can be seen towards the highest and lowest frequencies as a result of a mismatch between the average template, and the evolving profile.}
\label{figure:Sim2Errors}
\end{figure}

We find no evidence for an EQUAD using the ToAs formed from the 8~MHz channels when including the first FD parameter, and obtain timing precision that is a factor eight better than the fully frequency-averaged ToAs.  In our profile domain analysis we model the profile evolution directly as a linear change in the profile parameters (cf. Section~\ref{Section:Evolution}), as opposed to using a proxy as with the FD parameterisation.  In this case we find an additional 25$\%$ improvement in the timing precision compared to the 8-MHz ToAs.  This improvement comes because, in the ToA domain, we have formed ToAs using a template that does not incorporate profile evolution, and are modeling that evolution purely as a shift in the arrival times. In Fig.~\ref{figure:Sim2Errors} we  show the ToA uncertainties obtained from using a single template across the observing band.  There is a clear trend towards larger uncertainties towards the edge of the band, as the mismatch between the template and the profile increases.  In our ToA domain analysis, while we include EFAC and EQUAD parameters that scale, and add in quadrature to the error bars, no models currently include smooth, frequency-dependent scaling of the ToA uncertainties.  This leads to an overall decrease in sensitivity compared to the profile domain analysis, where we are correctly modeling the evolution as a change in shape.

\begin{table}
\caption{Measured timing precision in simulations relative to the 32 channel ToA analysis. For Simulation~3 we do not include the parameters describing the DM variations.}
\begin{tabular}{ll}
\hline
\multicolumn{2}{c}{Simulation 1} \\
\hline
Analysis & Relative Timing Precision \\
\hline
Frequency Averaged ToAs & 1.0 \\
32 Channel ToAs & 1.0 \\
Profile Domain & 1.0 \\
\hline
\multicolumn{2}{c}{Simulation 2} \\
\hline
Analysis & Relative Timing Precision \\
\hline
Frequency Averaged ToAs (EQUAD) & 8.0 \\
32 Channel ToAs (FD) & 1.0 \\
Profile Domain (PE) & 0.75 \\
\hline
\multicolumn{2}{c}{Simulation 3} \\
\hline
Analysis & Relative Timing Precision \\
\hline
Frequency Averaged ToAs (EQUAD) & 9.0 \\
32 Channel ToAs  (DMX, FD)  & 1.0 \\
Profile Domain  (DMX, PE) & 0.73 \\
Profile Domain - (Smooth DM, PE) & 0.43 \\
\hline
\end{tabular}\label{Table:SimPrecision}
\end{table}

Finally in Simulation~3, we see a further decrease in the timing precision obtained with the fully frequency-averaged ToAs relative to both  the 8-MHz ToAs and the profile domain analysis.  In this simulation we have included significant DM variations, and thus the effect of profile evolution and scintillation is compounded by the fact that we are no longer dedispersing at the correct DM for each epoch.  This leads to further perturbation of the frequency averaged profile, and thus increases the scatter in the residuals.  We first model the DM variations using the DMX parameterisation \citep{2013ApJ...762...94D}, in which a piecewise-constant DM(t) model is included in the analysis, with a separate value for each epoch in the data set.  In this case we find a similar improvement in timing precision when going from the 8-MHz ToAs to the profile domain as in Simulation~2.  We find a further 60\% improvement in precision when using a smooth model for the DM variations in our profile domain analysis.  We implement a model for DM variations using  the same approach as described in L15a, and find both MP1 and MP2  give consistent results.

\begin{figure}
\begin{center}$
\begin{array}{c}
\includegraphics[trim = 50 50 0 50, clip,width=80mm]{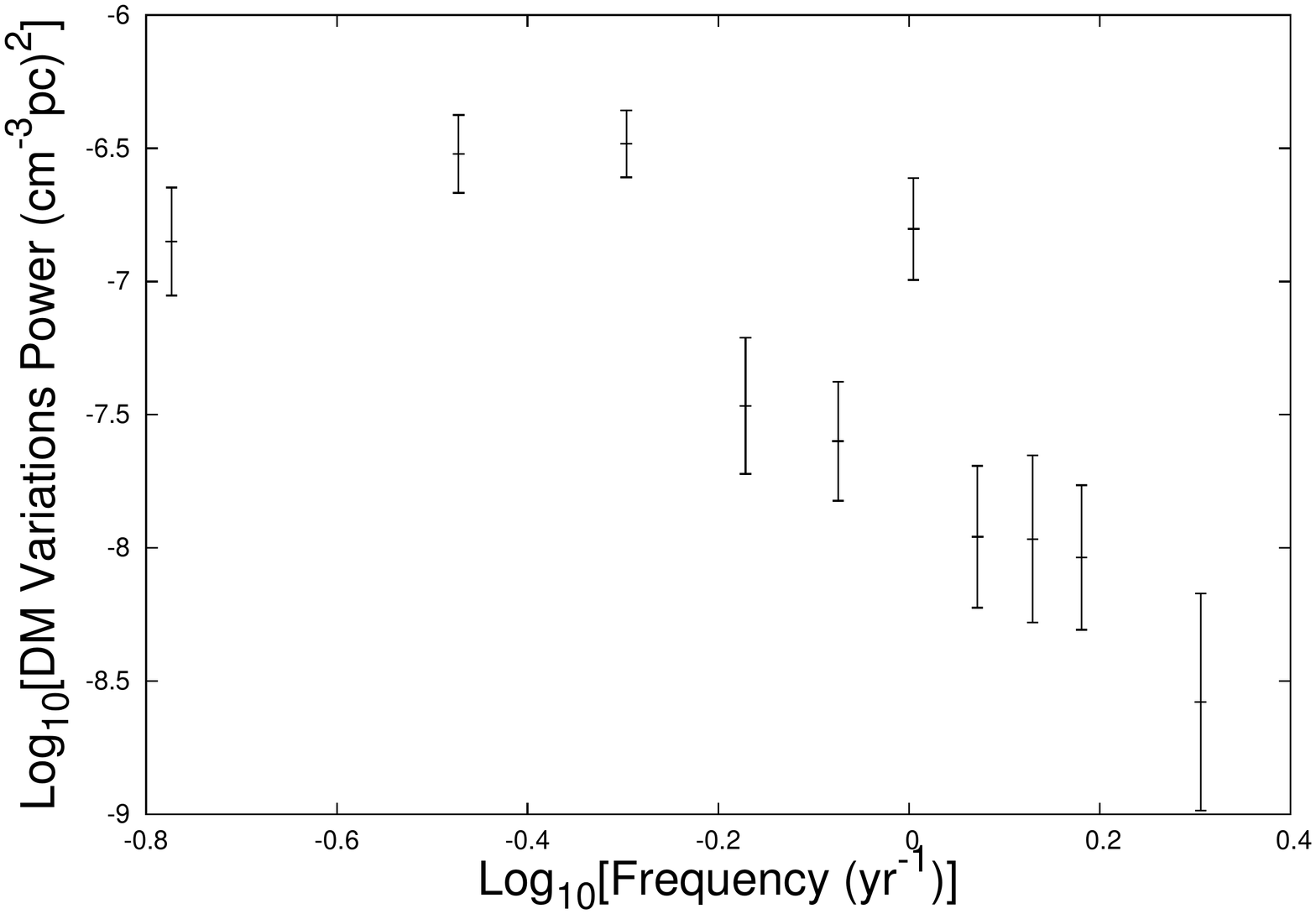} \\
\includegraphics[trim = 50 50 0 50, clip,width=80mm]{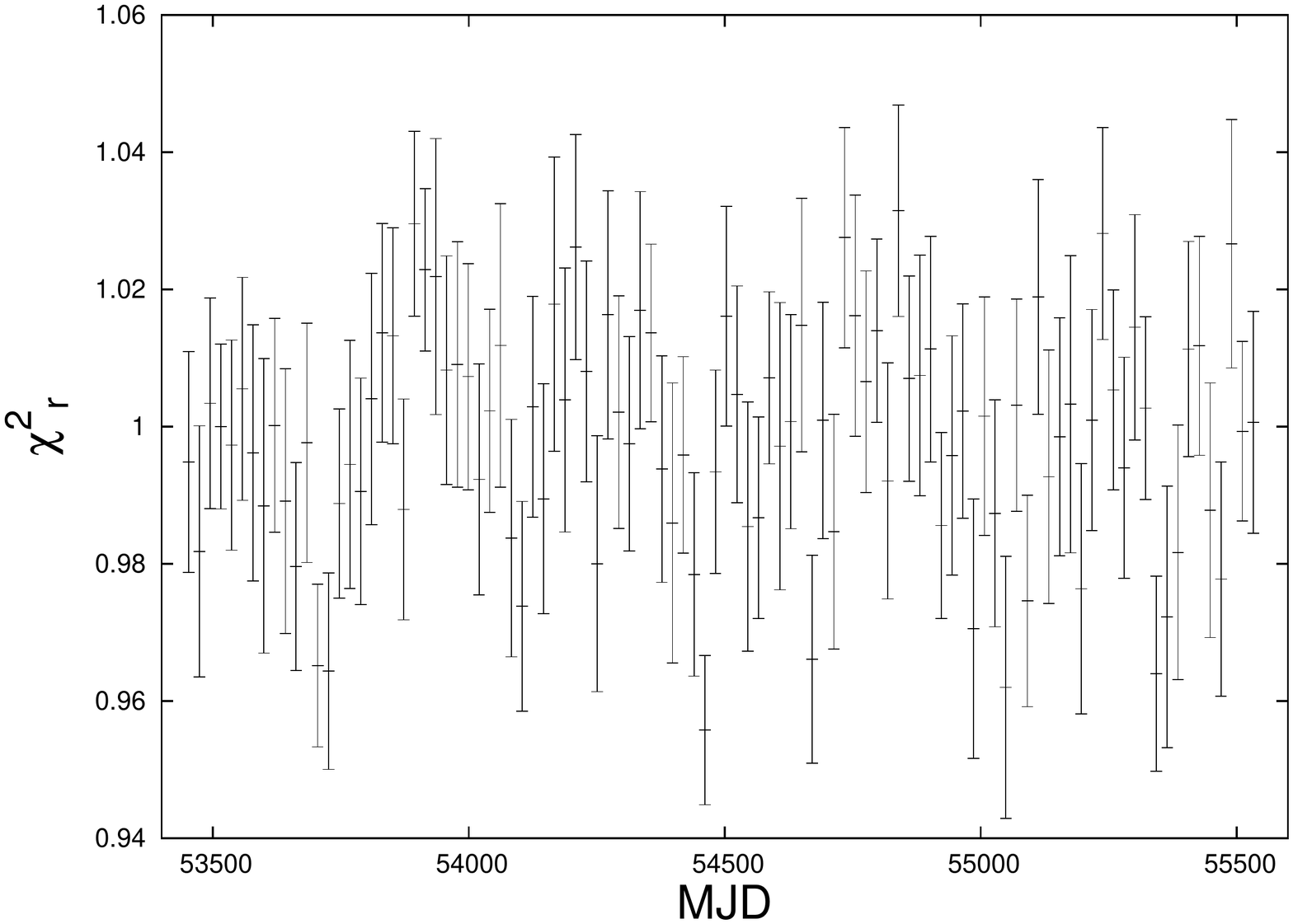} \\
\end{array}$
\end{center}
\caption{(Top) Mean parameter estimates and 1~$\sigma$ uncertainties on the DM power spectrum coefficients for those frequencies in the model that are detected with high significance.  In total we find only twelve frequencies from $1/T$ to $12/T$ are necessary in the model to describe the higher order DM variations, with $T$ the length of the data set. (Bottom) Mean value and standard deviation of the reduced $\chi^2$ for the on-pulse region of the profile data at each epoch in Simulation~3.}
\label{figure:DMPowers}
\end{figure}

In Fig.~\ref{figure:SimPlotss} (bottom right) we show the parameter estimates from the DMX model (red points with errors) and the one sigma confidence interval on the signal from the smooth DM model (blue lines) obtained in the profile domain. We also compare these with the result of performing a two dimensional analysis on each epoch independently where we fit for both a phase offset and the DM at that epoch (black points with errors).  In Fig.~\ref{figure:DMPowers} (top panel) we show the mean parameter estimates and 1~$\sigma$ uncertainties on the DM power spectrum for those frequencies in the model that are detected with high significance.  In total we find only twelve frequencies from $1/T$ to $12/T$ are necessary in the model to describe the higher-order DM variations, with $T$ the length of the data set.  The bottom panel of Fig.~\ref{figure:DMPowers}  shows the mean reduced $\chi^2$, $\chi^2_r = \chi^2/\sqrt{N_\mathrm{b}}$ for the on-pulse region of the profile data from the 32~channels at each epoch using this smooth model for the DM.   We find the results are consistent with a value of 1.0 within the uncertainties.  It is this significantly reduced parameter space compared to the 100~DMX parameters that results in the improvement in timing precision between the two models.  We note that this improvement will depend upon the complexity of the DM variations present in the data set.  If there are significant non-stationary fluctuations in the DM (e.g. \citealt{2015ApJ...808..113C, 2013MNRAS.429.2161K}) that require more complex models (e.g. \citealt{2016MNRAS.458.2161L}) then the difference between the two models will decrease.


Both the DMX and the smooth DM model provide significantly better constraints than modelling the DM independently at each epoch.  In \cite{2014ApJ...790...93P} and \cite{2014MNRAS.443.3752L} the DM is estimated in this way in order to obtain ToA estimates that are not biased by dedispersing with an incorrect DM.  The degree of improvement in modeling the DM  coherently across the entire data set will naturally depend upon the bandwidth and the S/N of that data set.  However, as the bandwidth and quality of observations improve so too will the number of parameters required to model the data.  For example, lower frequency observations will require phase and both time variable DM and scattering terms to be modeled simultaneously for each epoch.  If a pulsar shows detectable, band-wide shape variation, a statistical description of that variation should be included in the model in order to robustly estimate the remaining parameters of interest.  Other factors, such as frequency-dependent DM across a wide bandwidth \citep{2016ApJ...817...16C} will also eventually need to be considered. However,  by estimating these parameters simultaneously over the whole data set as in our profile domain framework, one can always ensure that the optimal result is achieved.

\section{Data Sets}
\label{Section:Data}

\begin{table}
\caption{Details of the individual pulsar data sets. $N_E$ is the total number of observing epochs, $N_P$ is the total number of profiles in the data set, and $\sigma_\mathrm{w}$ is the weighted rms of the ToAs formed from the profile data.}
\begin{tabular}{ccccc}
\hline\hline
Data Set             &    Timespan      &	$N_E$		&    $N_P$  &  $\sigma_\mathrm{w}$ \\
PSR                 & yr                &                           &        & $\mu$s    \\
\hline
J1713+0747	         &	6.74	&	482	& 13643	  & 0.55\\
J1744$-$1134	     &	6.74	&	496	& 13610	  &   1.4 \\
J1909$-$3744         &	9.02	&	 695   & 19194	  &  0.43\\
\hline
\end{tabular}
\label{Table:evidenceDefs}
\end{table}


We perform our analysis using observations of PSRs~J1713$+$0747, J1744$-$1134, and J1909$-$3744 made with the 64-m Parkes radio telescope.  Data were collected using two receiver packages, a co-axial system at 10~cm and 40~cm, and the centre pixel of a multi-beam 20\,cm system.  Data at 3100\,MHz (co-axial, hereafter `10~cm') and 1369\,MHz (multi-beam, hereafter `20~cm') were recorded using a digital polyphase filterbank (PDFB4) with a typical resolution of 1024 channels and 1024 bins and respective bandwidths of 1024\,MHz and 256\,MHz.  Data from the lower co-axial band, centred at 732\,MHz (hereafter `40~cm'), were recorded over a 64\,MHz bandwidth with a similar polyphase system, DFB3, until its demise in April of 2014.  Subsequent data employed CASPSR, a coherently dedispersing system producing 512-channel, 1024-bin output.  Data from the DFBs were averaged over 1-minute sub-integrations, while CASPSR data maintain an 8-second resolution.   Specific details of the three data sets are given in Table~\ref{Table:evidenceDefs} and more details about the observing system and data reduction are given in \cite{2013PASA...30...17M}.  For completeness, we outline the data reduction below.

We carry out data reduction using the \textsc{Psrchive} package (Hotan et. al 2004).  We remove channels within 5\% of the band edge as they have low gain and may contain aliased signals.  Narrow-band radio-frequency-interference (RFI) is identified and removed with a median bandpass filter, and impulsive broadband RFI is identified by eye and removed by deleting affected sub-integrations.  We note that the 10~cm band is largely free of both varieties of RFI.  The 20~cm band is affected primarily by the passage of global positioning system satellites through the telescope sidelobes and by occasional strong, impulsive RFI from aircraft.  The 40\,cm band was largely free from RFI until April of 2015 when a new mobile phone base station went online, necessitating a shift of observing band frequency and leaving strong contamination.

Observations are carried out with a cadence of approximately $14\pm10$ days for the co-axial and multi-beam systems separately.  While the 10~cm and 40~cm observations occur simultaneously for the majority of epochs, the 20~cm observations are interleaved between them, so that the overall cadence is higher, at $6 \pm 7$ days, however we note that the true distribution is non-Gaussian.  Before each pulsar observation, we record modulated noise injected at the receiver frontend, allowing estimation and correction of differential gain and phase between the voltage probes through the amplification and downconversion system.  At each observing epoch we observe the radio galaxy Hydra A, allowing bandpass and flux density calibration.  We find that all quantities vary only modestly with time.  We apply these corrections to the raw data to produce flux- and polarization-calibrated pulse profiles.  For the 20\,cm data, there is evidence for elliptical polarization/cross-coupling of the receiver feed, causing the observed pulse profile to depend slightly on the parallactic angle of the source.  We correct these variations using a model of the parallactic-angle dependence obtained from long observations of PSR~J0437$-$4715; see \citet{2013PASA...30...17M} for more details.  After calibration the sub-integrations were time averaged for each epoch.

There are several discrete time offsets (known as jumps) in the data sets.  In particular, a firmware upgrade for the DFB systems at MJD 55319 resulted in an offset for each spectrometer mode which we include as free parameters in our analysis.

\section{Results}
\label{Section:Results}

In the following three subsections we describe the results obtained when using our wide-band profile domain timing technique on the three data sets described in Section~\ref{Section:Data}.  In our analysis we have considered both the 10~cm data on its own, for which we used \textsc{PolyChord} to perform evidence comparisons with the analytic likelihood, and also the full data sets covering 3~GHz of bandwidth where the sampling was performed with the GHS using the numerical likelihood.  In the latter case we perform only parameter estimation and do not obtain evidence values for model comparison.

In Section~\ref{Section:PEvo} we compare the two models for profile evolution described in Section~\ref{Section:Evolution} which we denote as models:

\begin{description}
\item[(E1)] A polynomial expansion of the shapelet amplitudes that describe the mean profile with frequency, where the change in each amplitude is independent, and
\item[(E2)] a one-dimensional change in the width of the profile as a function of frequency.
\end{description}
In Section~\ref{Section:PStoc} we discuss the results for our models for profile stochasticity, for which we consider:

\begin{description}
\item[(UC)] Profile stochasticity uncorrelated in phase, and
\item[(PC)] Phase-correlated profile stochasticity.
\end{description}

In Section~\ref{Section:PJitter} we describe the results for the different jitter models we consider in our analysis.  We denote these as:

\begin{description}
\item[(PJ)] Pulse jitter as described in Section~\ref{Section:Jitter},
\item[(WJ)] Width jitter, and
\item[(IPJ)] Interpulse jitter.
\end{description}
In Section~\ref{Section:DMVar} we compare the piecewise DMX and smooth power-law models for DM variations in both our profile and ToA domain analysis, and finally in Section~\ref{Section:TimingPrecision} we compare the timing precision from the different approaches.

For ease of notation when discussing evidence comparisons, we denote our most basic model, which includes only the pulsar timing model, a model for the mean profile, and a PFAC parameter, as model~(M0).

\subsection{Profile Evolution}
\label{Section:PEvo}

\subsubsection{10~cm data}

We find all three data sets show extremely significant support for profile evolution across the 10~cm band.  We find an increase in the log evidence, which we denote $\Delta \mathcal{Z}$,  of over 100 relative to any model that does not incorporate profile evolution constructed from the elements listed in Section~\ref{Section:Results}.

\begin{figure}
\begin{center}$
\begin{array}{c}
\includegraphics[width=80mm]{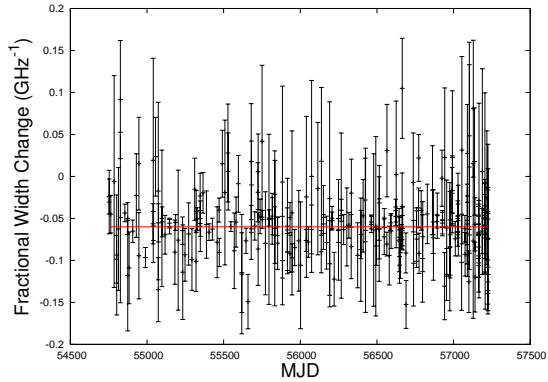}
\end{array}$
\end{center}
\caption{The mean parameter estimate and 1~$\sigma$ uncertainties for the evolution of the profile width as a function of frequency for PSR J1909$-$3744.  Parameter estimates are obtained separately for each epoch from a three-dimensional analysis where the arrival time, DM, and width change is evaluated simultaneously, with other timing parameters fixed to the maximum-likelihood estimates obtained from the analysis of the full data set.  The horizontal line is set to the mean value of the evolution determined from the full analysis of $-0.0592$. }
\label{figure:1909WidthCheck}
\end{figure}

\begin{figure*}
\begin{center}$
\begin{array}{c}
\includegraphics[trim = 0 0 0 0, clip,width=130mm]{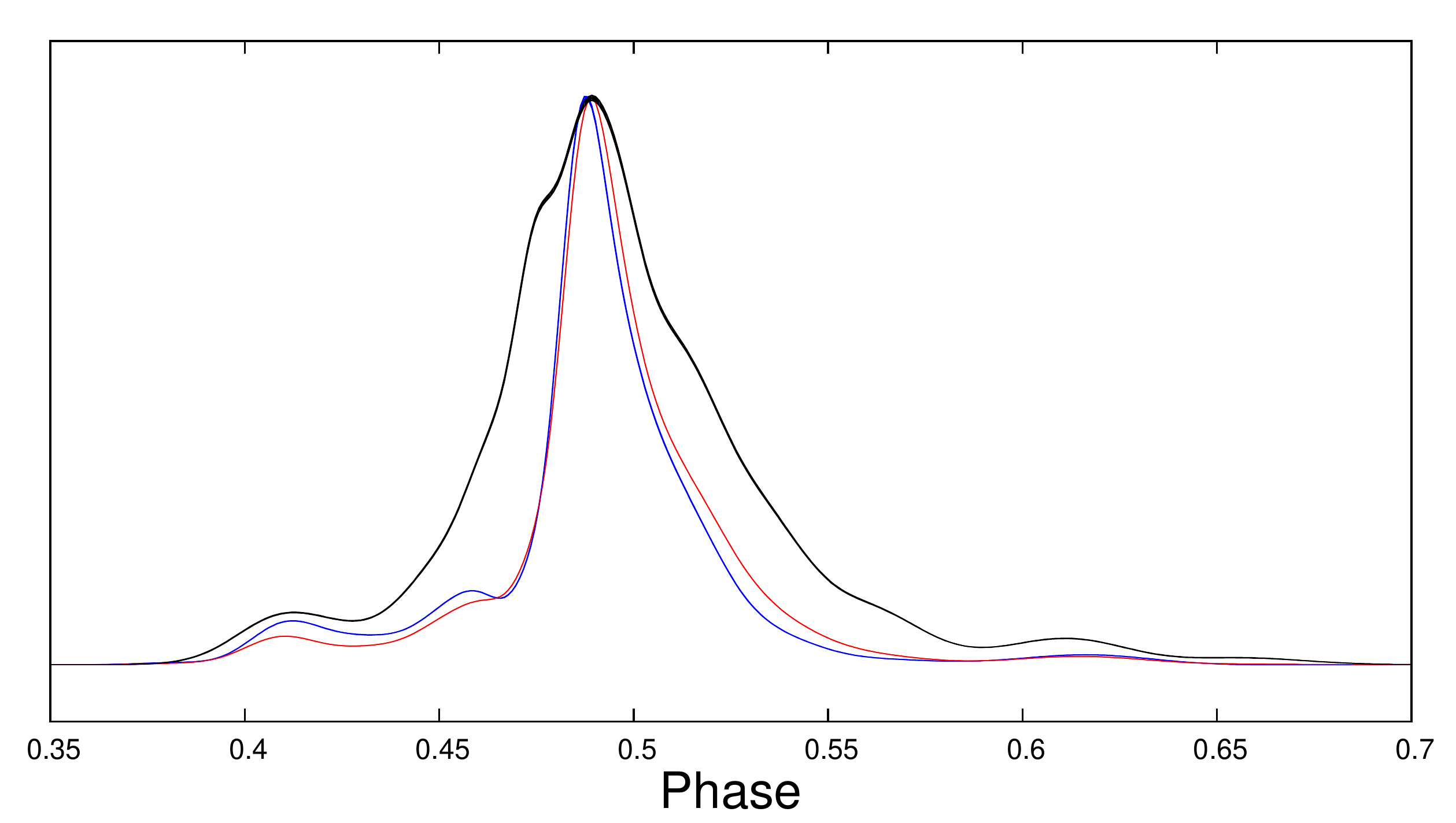} \\
\includegraphics[trim = 0 0 0 0, clip,width=130mm]{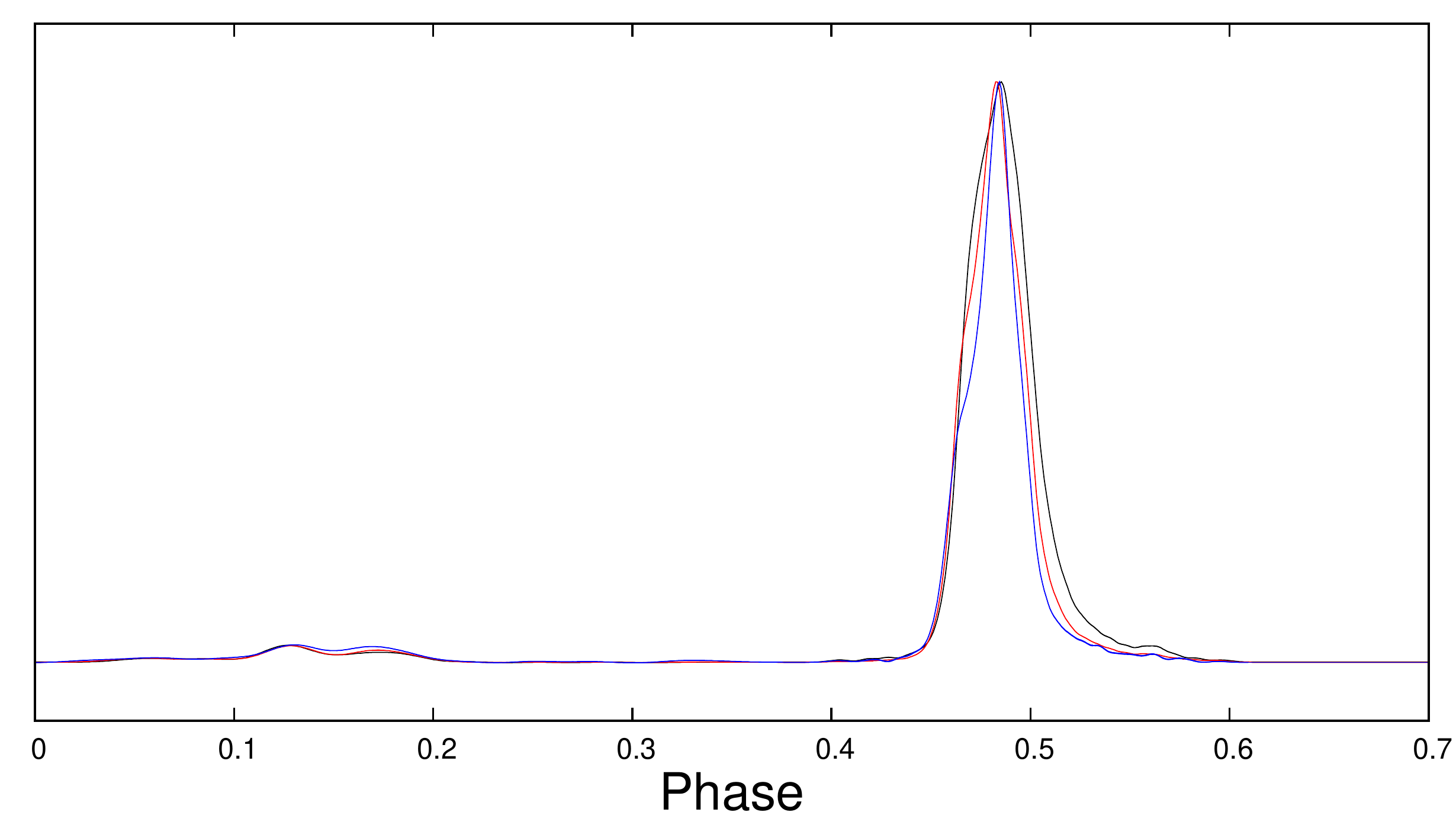} \\
\includegraphics[trim = 0 0 0 0, clip,width=130mm]{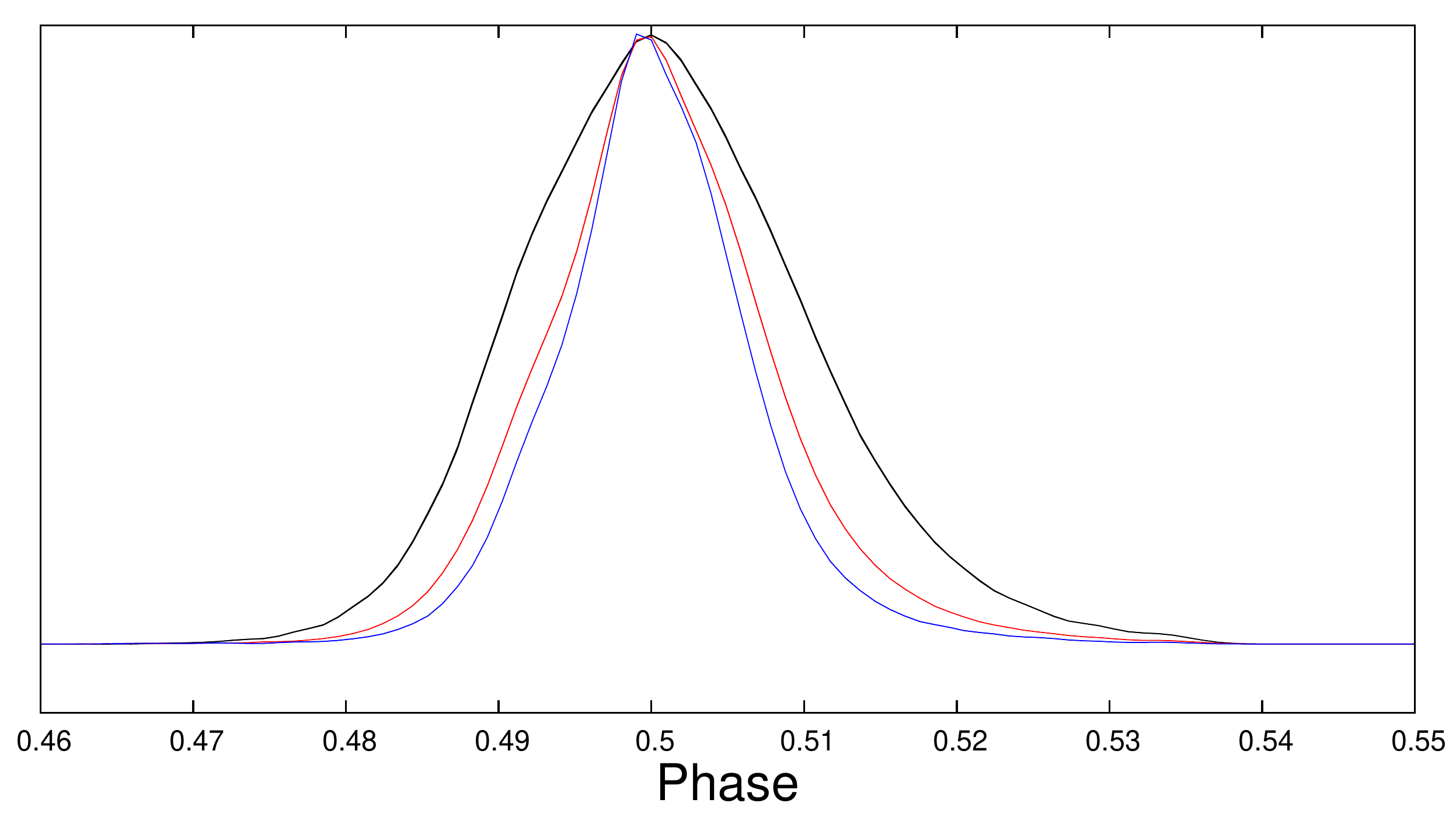} \\
\end{array}$
\end{center}
\caption{Model pulse profiles for PSR J1713+0747 (top), J1744$-$1134 (middle) and J1909-3744 (bottom), evaluated at frequencies of 700~MHz (black lines), 1400~MHz (red lines) and 2800~MHz (blue lines).  These frequencies are approximately the centers of the 40~cm and 20~cm bands, and the bottom of the 10~cm band respectively. In each case the overall phase parameter is set to $\phi = 0.5$ (cf. Eq.~\ref{Eq:newshapefunction}). }
\label{figure:J1909ProfEvo}
\end{figure*}

We find that for model~(E2), the fractional change in the profile width is $0.01396 \pm 0.0007$, $-0.039 \pm 0.003$, $-0.0592 \pm 0.0013$~GHz$^{-1}$ for PSR J1713+0747,  J1744$-$1134, and J1909$-$3744 respectively.  Here a negative value reflects the fact that the profile narrows as the frequency increases.  In all three data sets, however, we find the evidence supports the more complex model for profile evolution, with $\Delta \mathcal{Z}$ = 387, 260, and 4 in favour of model~(E1) for the three data sets respectively.  Apart from PSR J1909$-$3744, the profile evolution is thus dominated by very general shape change with frequency, as opposed to width changes, implying that little physical interpretation can be drawn from the exact value of the simple 1-parameter model.  Even in PSR J1909$-$3744, however,  deviations from the simple model are still significant.

One might expect a change in the width as a function of frequency if the DM has not been modelled appropriately, and the profile has been smeared across the band.  We therefore perform a consistency check on the PSR J1909$-$3744 data set in which we perform a three-dimensional analysis on each epoch separately.  In each case we fit for the ToA, the DM, and the change in width as a function of frequency.  In Fig.~\ref{figure:1909WidthCheck} we plot the mean parameter estimate  and 1-$\sigma$ uncertainties for the evolution of the profile width that results from this analysis.  We find that the evolution measured at each epoch is consistent with the estimate determined from our global analysis, and obtain a mean and standard deviation from our per epoch analysis of $-0.06 \pm 0.02$~GHz$^{-1}$.  That the uncertainty in this case is an order of magnitude larger than the global analysis is simply the result of performing an incoherent analysis across multiple epochs, as opposed to a fully coherent analysis with the timing model.

\begin{figure}
\begin{center}$
\begin{array}{c}
\includegraphics[trim = 65 50 20 50, clip,width=80mm]{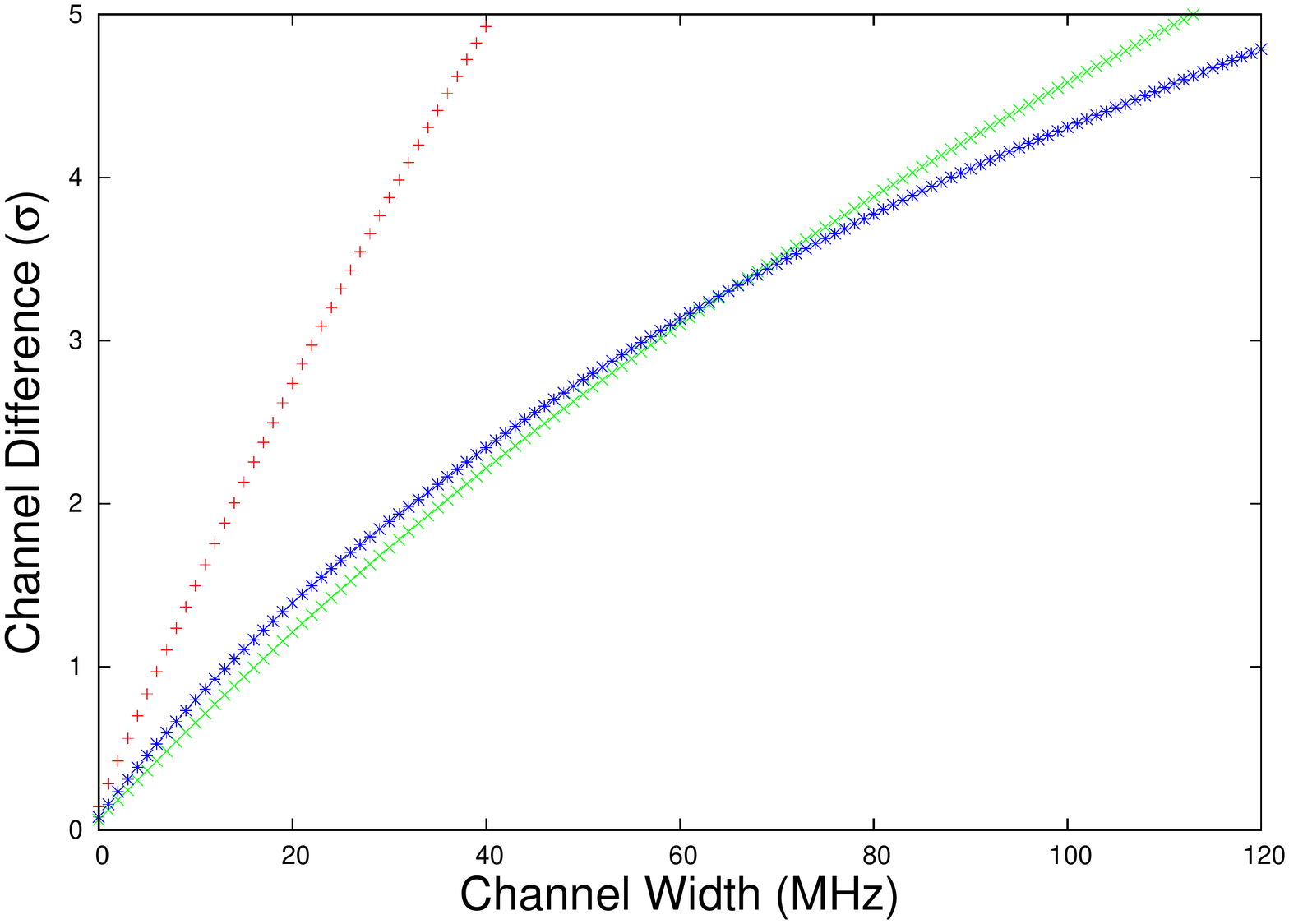} \\
\includegraphics[trim = 65 50 20 50, clip,width=80mm]{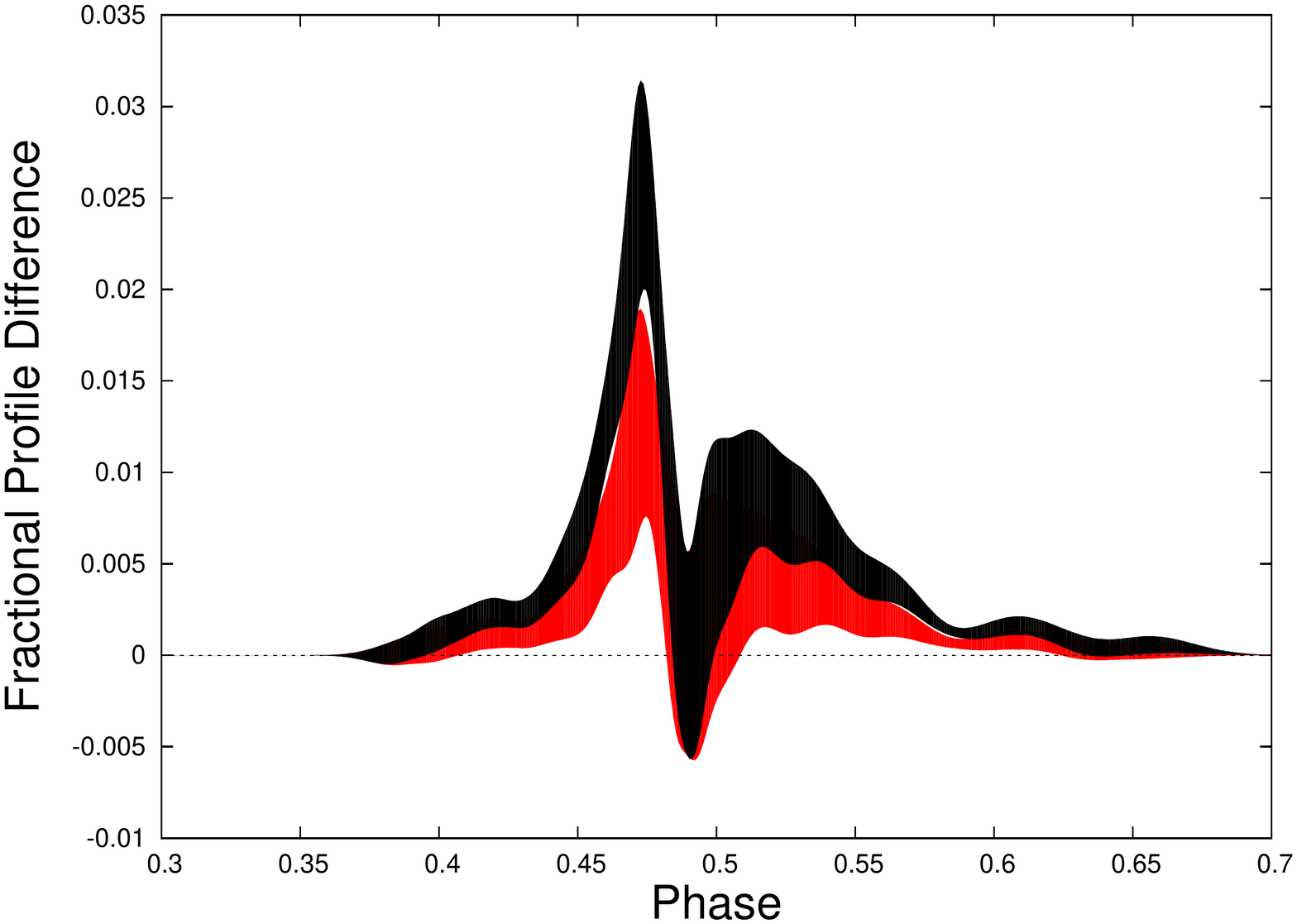} \\
\end{array}$
\end{center}
\caption{(Top panel) The maximum change in the profile model due to frequency evolution as a function of channel bandwidth for PSRs J1713+0747 (red + symbols), J1744$-$1134 (green x symbols), and J1909$-$3744  (blue $\ast$ symbols) for the 10~cm band.  The difference is measured in terms of the confidence intervals on the profile model returned by our profile domain analysis.  The curvature is a result of the uncertainty in the model for profile evolution, which increases as a function of frequency, gradually dominating over the uncertainty in the mean profile, which is constant as a function of frequency. (Bottom panel) Difference between the model for profile evolution for PSR J1713+0747 between frequencies of 724 and 728~MHz (red interval), and between frequencies of 724 and 732~MHz (black interval). }
\label{figure:FreqSIg}
\end{figure}

\subsubsection{700--3600~MHz data}

In Fig.~\ref{figure:J1909ProfEvo} we show the mean model pulse profiles for PSRs J1713+0747 (top panel), J1744$-$1134 (middle panel), and J1909$-$3744 (bottom panel) at a frequency of 700~MHz (black lines), 1400~MHz (red lines) and 2800~MHz (blue lines).  We find that a cubic expansion of the shapelet coefficients with frequency is sufficient to model the profile evolution for both  PSRs J1744$-$1134  and J1909-3744, however PSR J1713+0747  required an additional quartic term.   This is commensurate with the number of FD parameters required for each of the pulsars when performing a ToA domain analysis, with PSR J1713+0747 warranting the first three terms in the FD model, compared to two for PSR J1744$-$1134 and one for PSR J1909$-$3744.

We can use the confidence intervals for our models of profile evolution in each data set to determine the channel width such that neighbouring channels are still consistent to some statistical level.  We illustrate this in two ways in Fig.~\ref{figure:FreqSIg}.
First (top panel), using the 10~cm band we show the maximum change in the profile as a function of observing frequency in terms of the uncertainties in the profile model.  We find that for PSRs J1744$-$1134 and J1909$-$3744 a channel width of approximately 40~MHz is required such that the maximum difference in the profile model  between neighbouring channels at any point in phase is less than 2~$\sigma$.  That these values are similar despite the difference in timing precision between the two data sets is simply a result of the magnitude of the profile evolution being much greater in the PSR J1744$-$1134 data set, while the S/N of the data set is much larger for PSR J1909$-$3744.  For PSR J1713+0747, however, we find that a narrower channel width of 20~MHz is required to maintain a similar level of statistical consistency.
Secondly (bottom panel), we show the difference in the profile model for PSR J1713+0747 between frequencies of 724 and 728~MHz (red interval), and between frequencies of 724 and 732~MHz (black interval).  One can clearly see that even over small fractional changes in observing frequency ($\sim 5.5\times10^{-3}$) there is still significant detectable profile evolution.  We note that, as we were sampling simultaneously for our profile model and our model for DM, these uncertainties account for the covariances that exist between our models for these two processes.  Thus the observed evolution in the profile is not attributable to incorrect estimation of the DM as a function of time.


\subsection{Profile Stochasticity}
\label{Section:PStoc}

\begin{figure}
\begin{center}$
\begin{array}{c}
\includegraphics[trim = 0 50 0 60, clip, width=80mm]{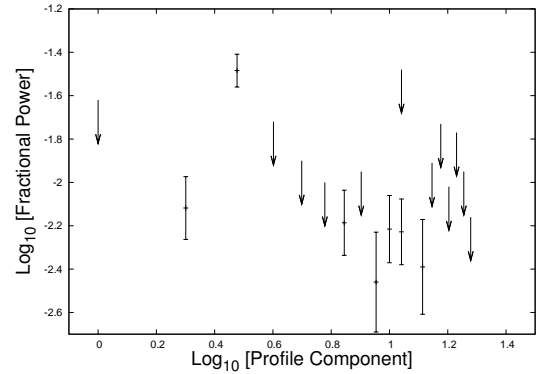}\\
\end{array}$
\end{center}
\caption{Power spectrum of the shape variations in PSR J1713+0747 from the 10~cm data obtained using \textsc{PolyChord} .  Arrows indicate 95\% upper limits, points with error bars indicate significant detections. Increasing profile component reflects smaller scales in phase. }
\label{figure:J1713Shape}
\end{figure}

\subsubsection{10~cm data}

\begin{figure*}
\begin{center}$
\begin{array}{cc}
\includegraphics[trim = 0 0 0 0, clip,width=80mm]{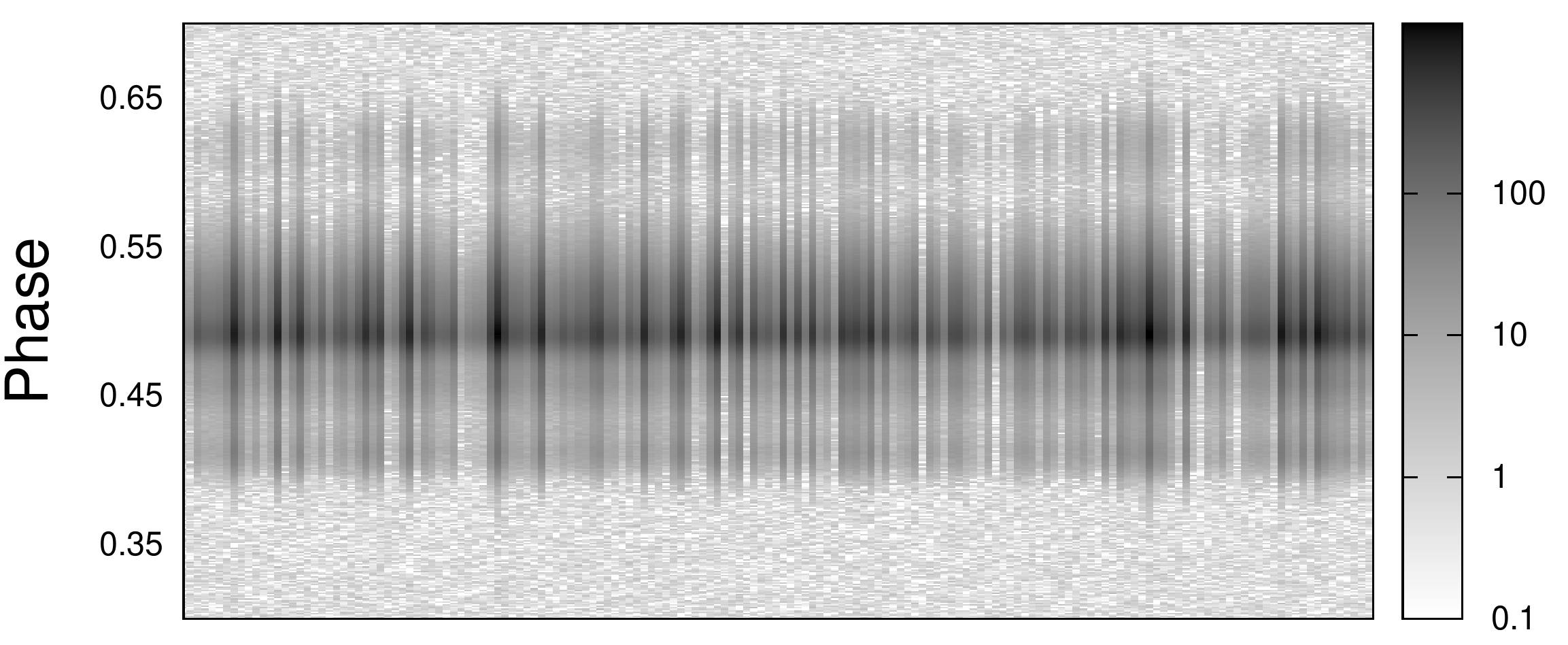} &
\includegraphics[trim = 0 0 0 0, clip,width=80mm]{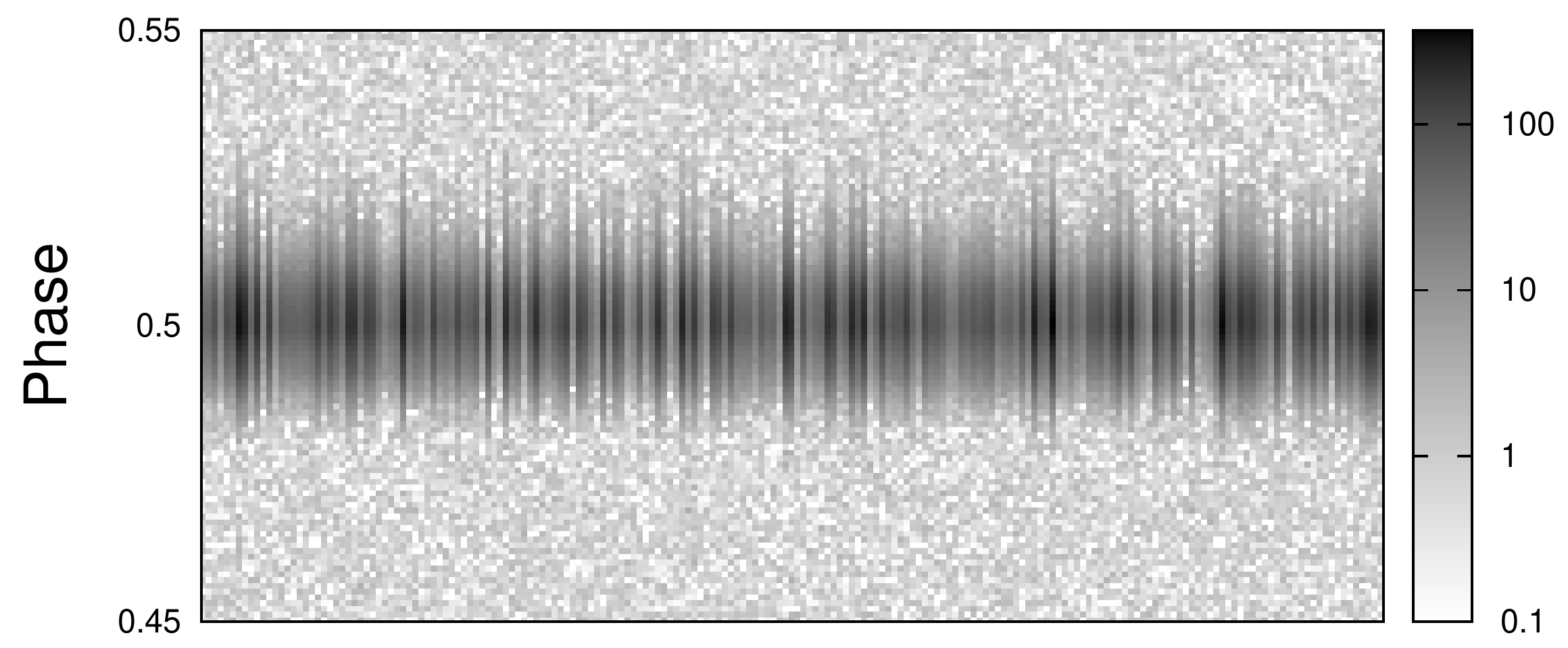} \\
\includegraphics[trim = 0 0 0 0, clip,width=80mm]{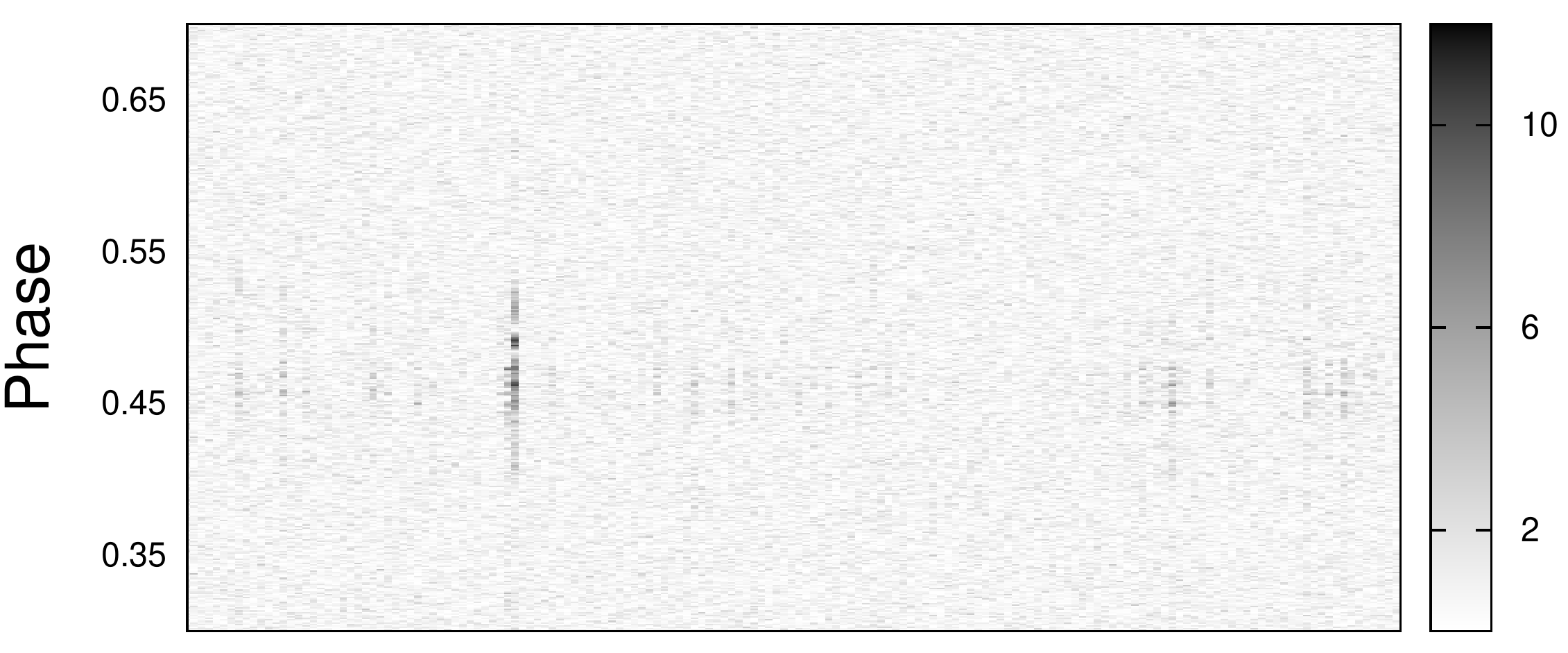} &
\includegraphics[trim = 0 0 0 0, clip,width=80mm]{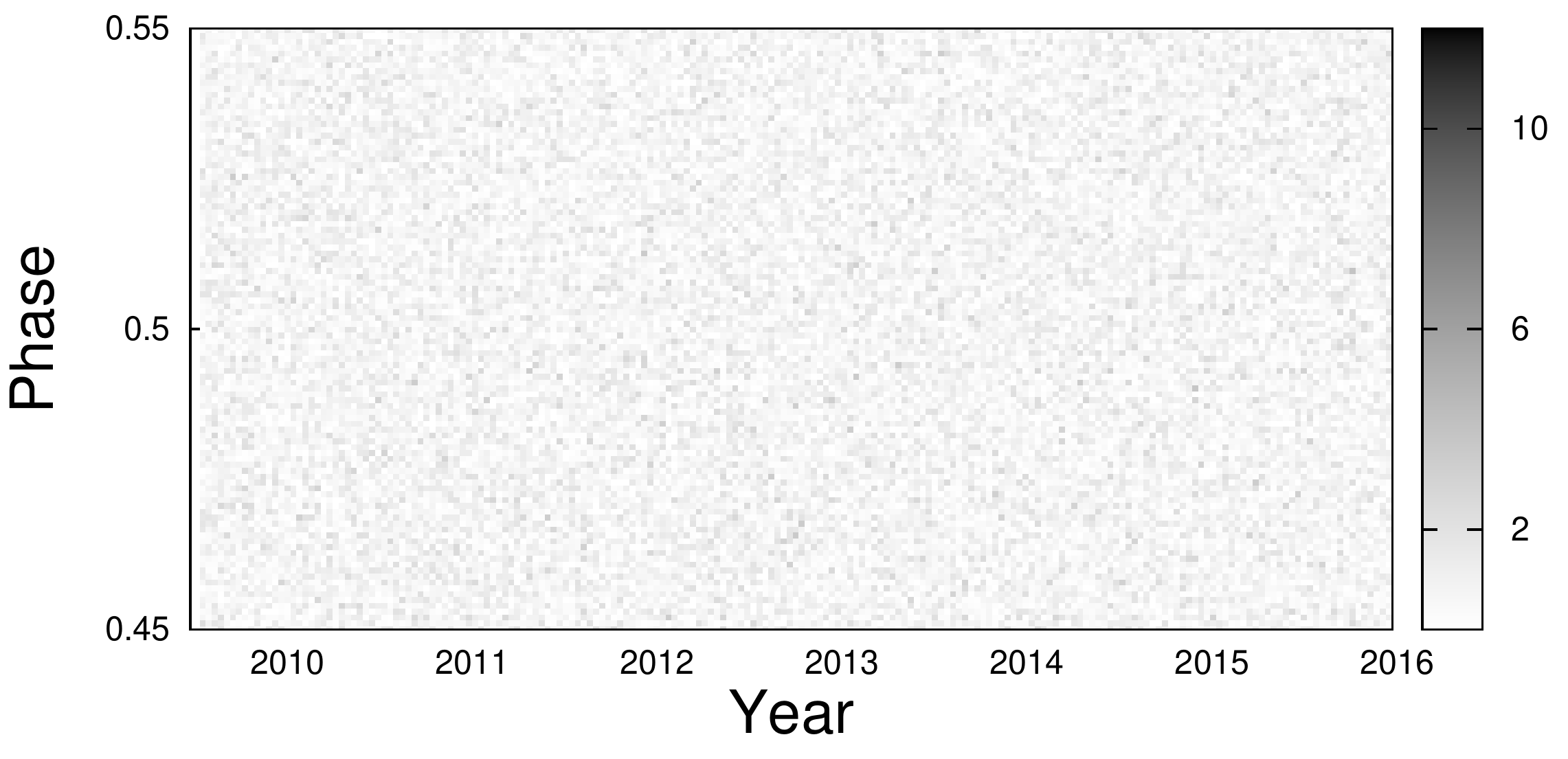} \\
\includegraphics[trim = 0 0 0 0, clip,width=80mm]{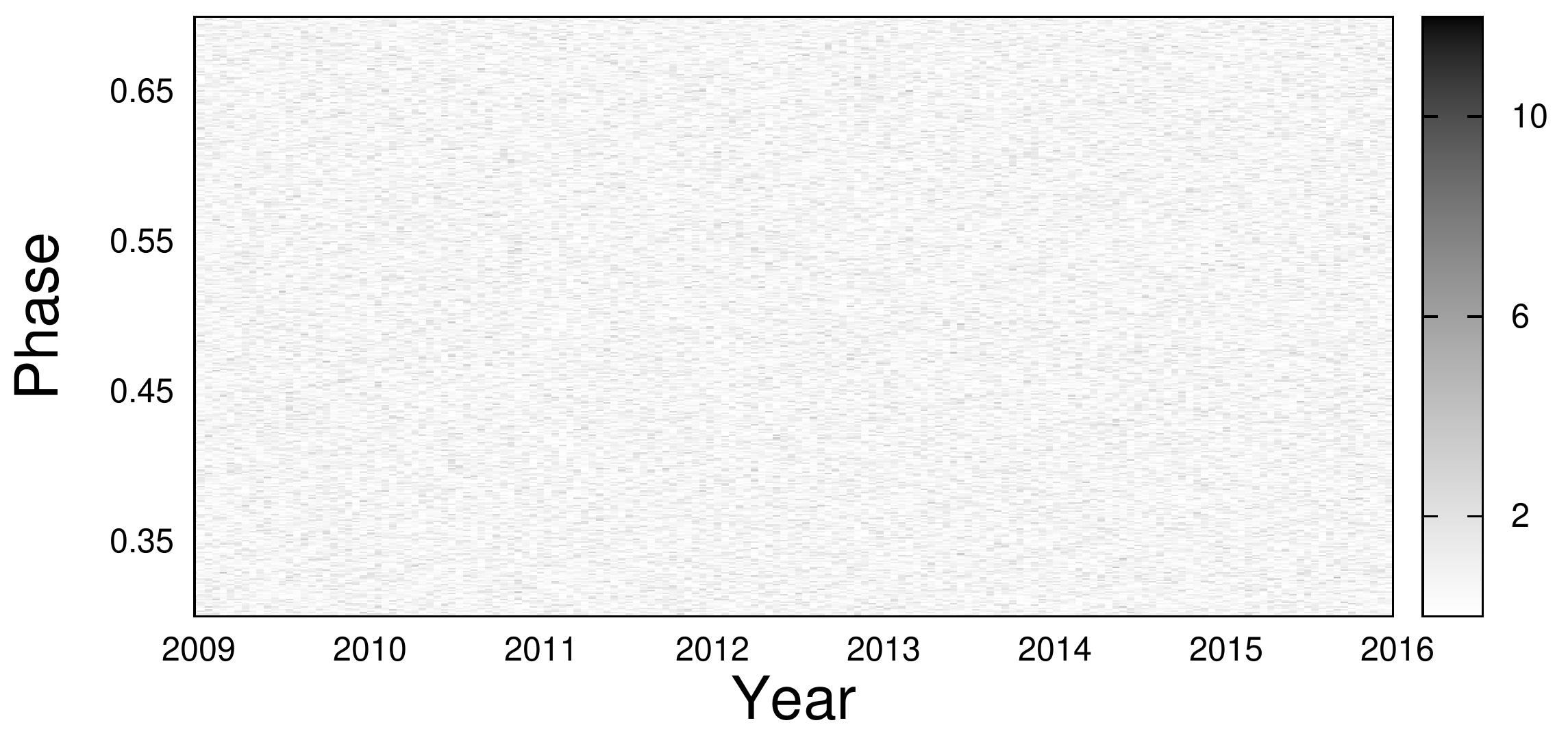} &
\hspace{-0.5cm}
\includegraphics[trim = 0 0 0 0, clip,width=70mm]{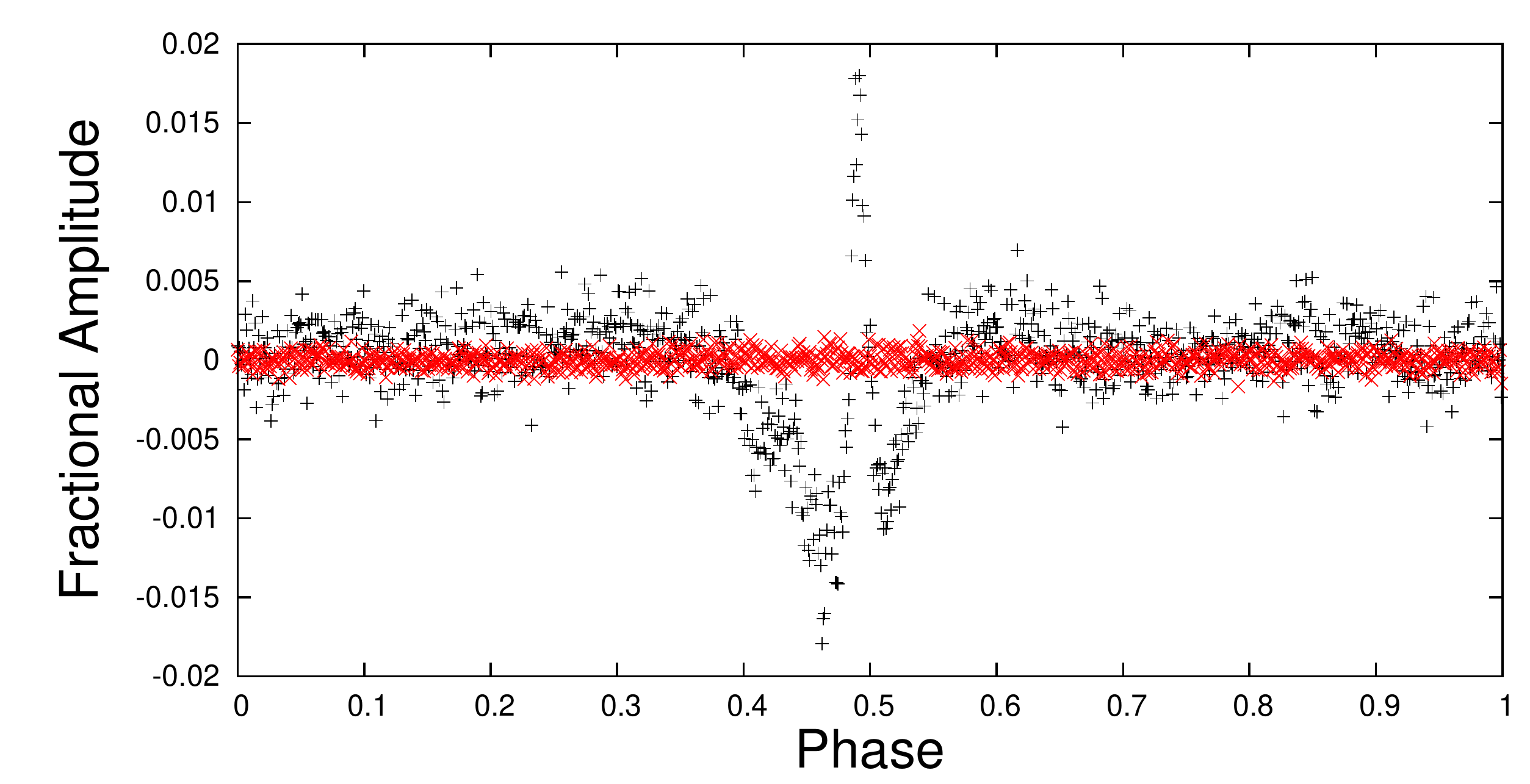}\\
\end{array}$
\end{center}
\caption{Log of absolute intensity (top panels), and absolute value of the profile residuals  after subtracting out the mean profile model (middle panels) for the 20~cm profile data for PSR J1713+0747 (left) and for the 10~cm data for PSR J1909$-$3744 (right) in units of S/N.  For the 20~cm PSR J1713+0747 profile data  we then additionally show the absolute value of the residuals after subtracting out our model for stochasticity in the pulse profile  (bottom-left panel), and the residuals from the fully time-averaged 20~cm data (bottom-right panel, red points) with the residuals from the epoch for which the most significant profile stochasticity was detected (black points).  }
\label{figure:HighSNProf}
\end{figure*}

When analysing the 10~cm data with \textsc{PolyChord} we detect significant profile stochasticity only in the PSR J1713+0747 data set for which we obtain an increase in the log evidence for an (M0, E1, PC) model compared to an (M0, E1) model of approximately 200.  For PSRs J1744$-$1134	and J1909$-$3744 we find no increase in the log evidence for either the (PC) or (UC) model parameters.  In Fig.~\ref{figure:J1713Shape} we show the power spectrum for the shape variation as a function of the component in the shapelet model from this  \textsc{PolyChord} analysis.  From left to right the increasing order of the shapelet coefficient can be thought of as representing progressively smaller scales in phase.  We find the profile stochasticity to be at the level of approximately 1\% in the individual components,  and find the shape variation is detectable in individual epochs with over 3$\sigma$ significance in approximately 15\% of observations.

If the origin of this shape variation were not intrinsic to the pulsar, we might expect that the signal would not be coherent across the full frequency band.  As such, we use the 10~cm data to compare models for which the profile stochasticity is a coherent process across the band, with a model where it is incoherent (i.e, where the realisation of the change in shape is allowed to vary from channel to channel).  We find the evidence is significantly in favour of the coherent model, with an increase in the log evidence of approximately 60 compared to the incoherent model.

Whether this variation is intrinsic to the pulsar, or is related to systematic effects in the data is impossible to determine with only a single telescope.  However, as a consistency check we calculate the impact the maximum likelihood model for the variation has on the time of arrival of the pulse.  We find this to be at the 10-30~ns level for the most significant detections, consistent with previously published estimates for this pulsar (e.g., \citealt{2014MNRAS.443.1463S, 2015ApJ...813...65T}.

While significant, we do not yet find that this profile stochasticity has a significant impact on the timing analysis, with consistent parameter estimates and uncertainties with or without the (PC) model.  This consistency can be explained by estimating the ToA for the epoch with the most significant shape variation either assuming that the profile noise is white, or including the maximum-likelihood parameter estimates for the PC power spectrum in the model.  We find that the log evidence increases by 9, and the mean arrival time shifts by 35~ns when including the (PC) model, however the ToA uncertainty in both cases is 58~ns.  We therefore see that even for the most significant example, the data set is not yet sensitive enough for the bias in arrival time when ignoring profile stochasticity to impact the timing results.

The most sensitive 95\% upper limit for an isotropic stochastic gravitational-wave background is currently $A < 10^{-15}$ at a reference frequency of 1~yr$^{-1}$  \citep{2015Sci...349.1522S}.  A background of this amplitude would induce a shift in the time of arrival of a pulse at the level of $\sim$100~ns.  The change in the arrival time when accounting, or not, for this shape variation in the PSR J1713+0747 data set is already a significant fraction of this level.  Observations from the largest radio telescopes in the world (e.g. \citealt{2016MNRAS.458.3341D, 2016MNRAS.455.1751R, 2015ApJ...813...65T}) have therefore already reached the stage where correctly modeling such shape variation has become of significant importance in order  to reach the greatest levels of sensitivity to gravitational waves.

\subsubsection{700--3600~MHz data}

\begin{figure*}
\begin{center}$
\begin{array}{c}
\includegraphics[trim = 0 0 0 0, clip, width=150mm]{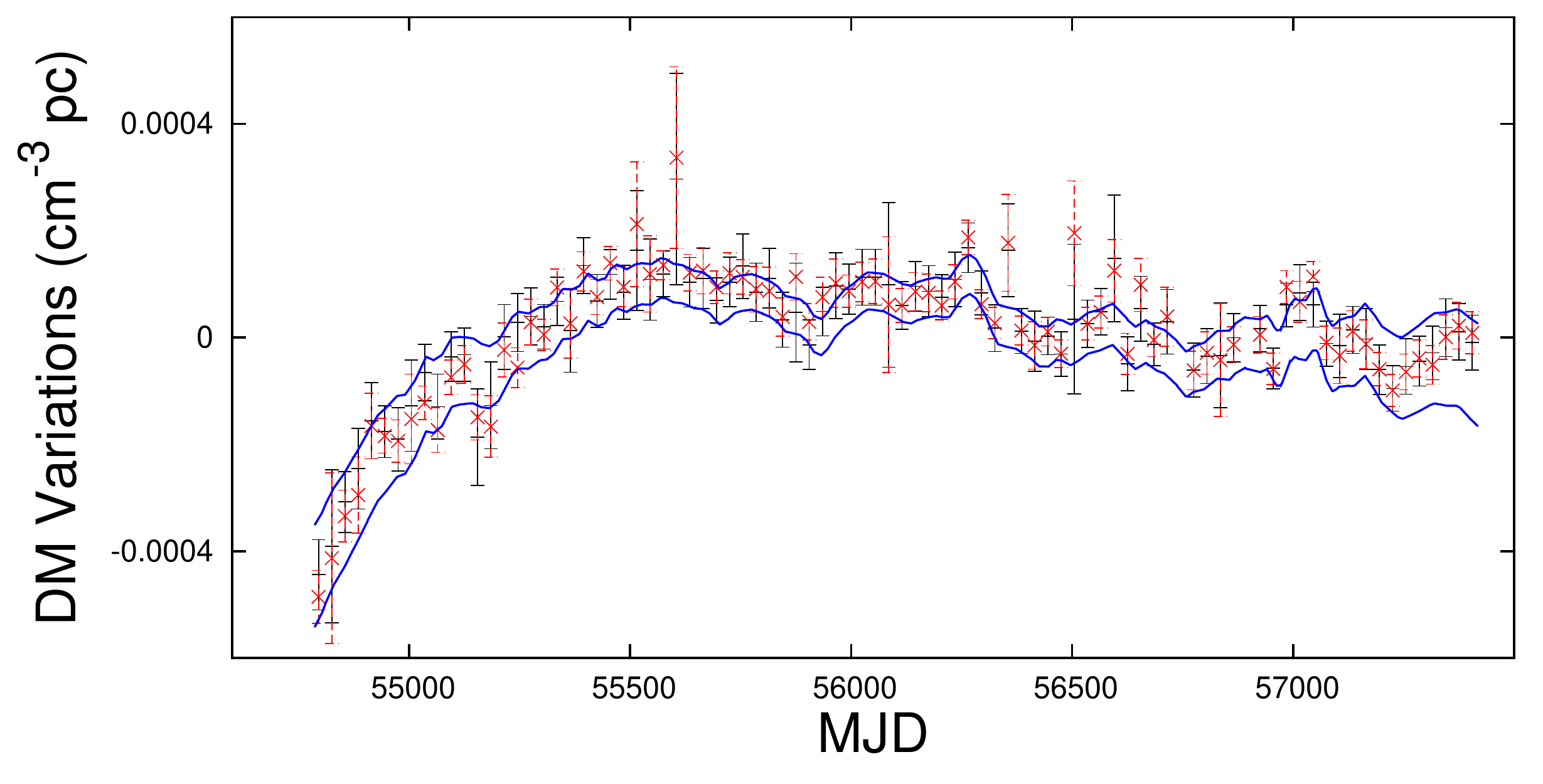}\\
\includegraphics[trim = 0 0 0 0, clip, width=150mm]{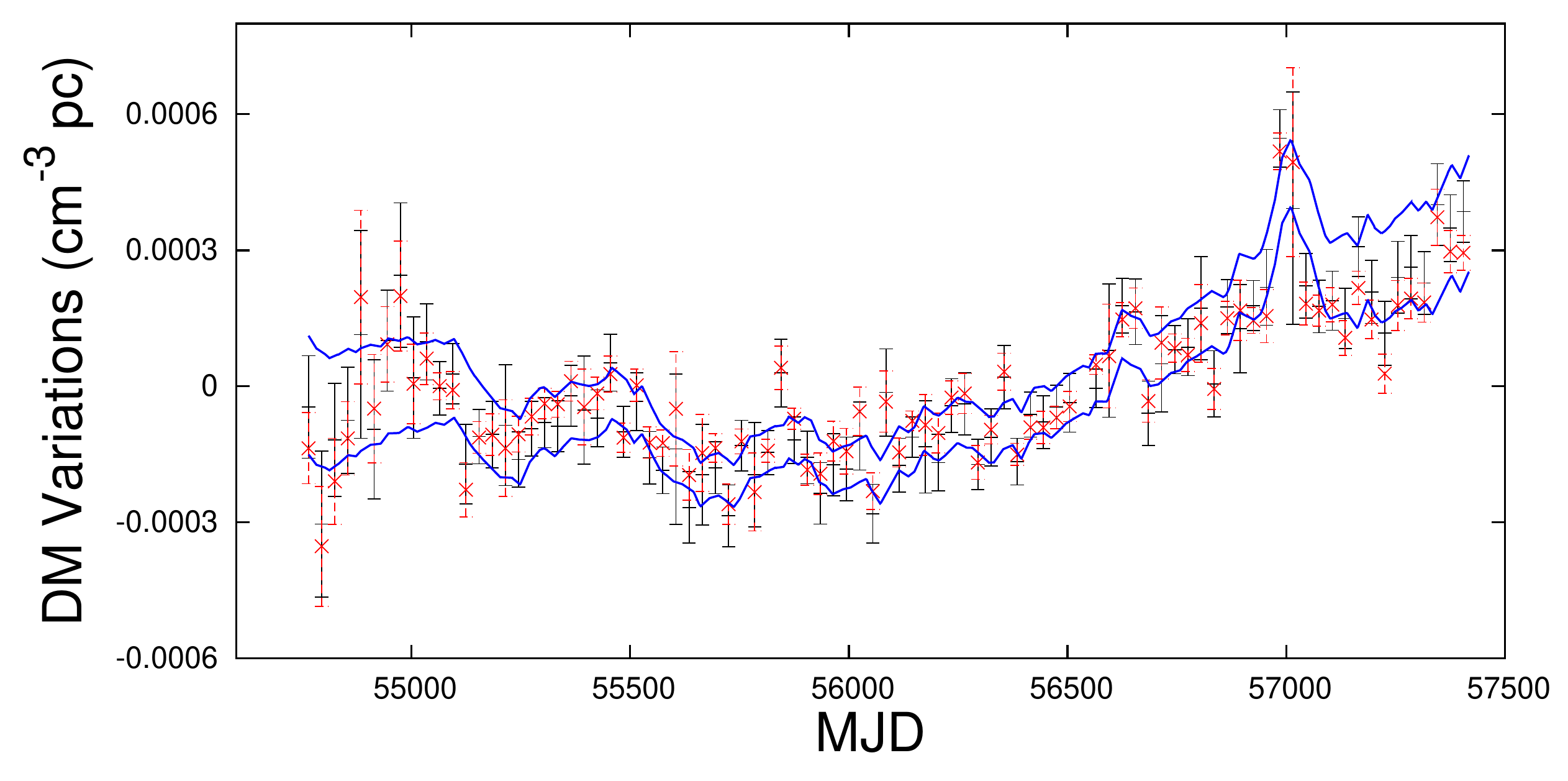}\\
\includegraphics[trim = 0 0 0 0, clip, width=150mm]{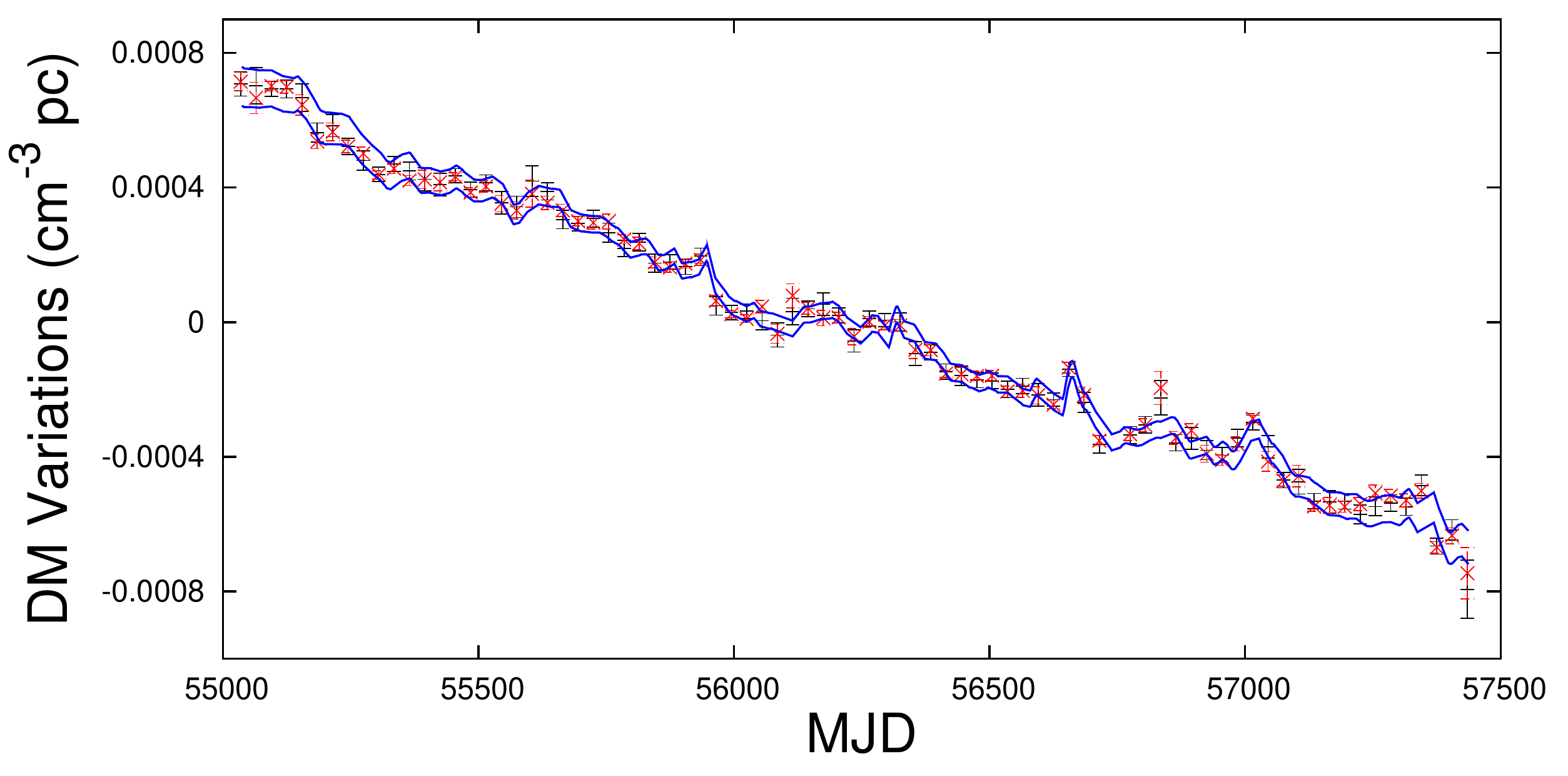}
\end{array}$
\end{center}
\vspace{-0.6cm}
\caption{A comparison of different models for the DM variations in PSRs J1713+0747 (top panel), J1744$-$1134 (middle panel), and J1909$-$3744 (bottom panel) data sets.  We compare the DMX parameterisation with a 30 day window from a ToA analysis (black points with 1$\sigma$ uncertainties), and from a profile domain analysis (red points with 1$\sigma$ uncertainties), and find the profile domain analysis consistently results in improved constraints on DM by an average of 40\%. We also compare the result of using a time-stationary smooth power-law model (blue lines representing the 1$\sigma$ confidence interval) and find it is consistent with the DMX model.  }
\label{figure:J1909WidthDM}
\end{figure*}

We obtain consistent results using the full data set and performing the sampling with the GHS.  In Fig~\ref{figure:HighSNProf} (top-left panel) we show log-intensity for the profile data over the MJD range included in the data set for the PSR J1713+0747 20-cm data in units of S/N. In the middle-left and bottom-left panels we then show the absolute value of the residuals in units of S/N after subtracting out the mean profile model, and additionally after subtracting out our model for stochasticity in the pulse profile. We find that the model for profile stochasticity has successfully modelled the shape variation, with the residuals that remain being noise-like for the duration of the data set.  For comparison, in Fig.~\ref{figure:HighSNProf} we also show the log-intensity (top-right panel) and absolute value of the residuals (middle-right panel) for the 10~cm PSR J1909$-$3744 data set for which no significant profile stochasticity was detected.

We stress that, because we are using the same basis to determine the profile stochasticity as for the mean profile model, and because it is being evaluated simultaneously with the mean profile model, this variation is not simply the result of mismodelling the mean profile.  Instead it reflects genuine variation in the shape of the profile between epochs.  To demonstrate this we show in the bottom-right panel of Fig.~\ref{figure:HighSNProf} the residuals from the fully time-averaged 20~cm data (red points) with the residuals from the epoch for which the most significant profile stochasticity was detected (black points).  The average residuals are an order of magnitude smaller than the shape variation in the individual epoch.

\subsection{Jitter}
\label{Section:PJitter}

\begin{table}
\caption{Parameter estimates for pulse jitter models.  Upper limits are quoted at the 95\% level.}
\begin{tabular}{cccc}
\hline\hline
Jitter Model          &    J1713+0747      &	J1744$-$1134		&    J1909$-$3744   \\
                            & ns                &            ns               &          ns  \\
\hline
PJ$_\mathrm{10cm}$	        &	$<$140 	&	$<$290	    & 	  $<$80      \\
WJ$_\mathrm{10cm}$       &     $<$90 	&	  $<$260      &    $<$75	    \\
IPJ$_\mathrm{10cm}$	   &    --		&	$<$300	&		--	\\
\hline
\hline
PJ$_\mathrm{20cm}$              &	$<$87 	&	240 $\pm$ 35	    & 	  100 $\pm$ 10      \\
PJ$_\mathrm{50cm}$  	        &	$<$390 	&	$<$300	    & 	  $<$100      \\
\hline
\end{tabular}
\label{Table:JitterLimits}
\end{table}

In the 10-cm data we find no evidence for jitter in any of the three data sets, with changes in the log evidence of less than two when including  jitter parameters compared to model M0, along with our model for evolution~(E1).  In Table~\ref{Table:JitterLimits} we list the 95\% upper limits on the different jitter models considered for each data set.

We find that the limit on the~(IPJ)  jitter model in PSR J1744$-$1134 is completely consistent with the~(PJ) model.  This can be understood by considering the S/N ratio of the interpulse at any epoch.  Even in the brightest observations the interpulse is only just visible.  As such the data is unable to discriminate between a model where both components are shifted by the same amount, as in the~(PJ) model, or where it is the separation between the two components that is allowed to vary, as in the~(IPJ) model.

When performing the analysis with the GHS, we include a separate parameter $\mathcal{J}$ for each of the three bands.  We find the upper limit on the pulse jitter in the 10-cm band is consistent with the analysis performed using \textsc{PolyChord} for all three pulsars.   As we do not compute the evidence for different models, when using the GHS we consider the posterior for a particular  $\mathcal{J}$ to represent an upper limit when there is greater than 5\% probability that the parameter takes a value of less than 1~ns.  Given this definition we only detect significant pulse jitter in the 20-cm bands for PSRs J1744$-$1134 and J1909$-$3744.

\subsection{DM Variations}
\label{Section:DMVar}

All three pulsars in our data set are known to exhibit detectable variations in DM (see, e.g., \cite{2016MNRAS.458.2161L, 2015ApJ...813...65T}).  In Fig.~\ref{figure:J1909WidthDM} we show the estimates we obtain on the DM variations for three different approaches for each pulsar.  We use the DMX parameterisation with a 30-day window both with the ToA data (black points with 1$\sigma$ uncertainties), and in the profile domain (red points with 1$\sigma$ uncertainties).  We compare this with a smooth power-law model in the profile domain (blue lines representing the 1$\sigma$ confidence interval on the signal) where we include frequencies in our model from $1/T$ with $T$ the length of the data set up to and including 30-days, and use a quadratic in DM to act as a proxy to the lower frequency fluctuations.

When using the DMX model, in order to minimise the covariance between the mean DM and the variations in DM, we take the epoch which has the best individual constraints on DM and keep that fixed as a reference.  We then include the mean DM in our fit as a free parameter.

We find that all three models are consistent within Gaussian statistics, with no significant ($>$ 3$\sigma$) discrepancies between the DMX model and the power-law model.  We note that the non-stationary DM event observed in previous analysis of PSR J1713+0747 occurs before the start of the data set at MJD 54780 (e.g., \citealt{2016MNRAS.458.2161L}).

\begin{figure}
\begin{center}$
\begin{array}{c}
\hspace{-1cm}
\includegraphics[trim = 10 10 10 10, clip,width=100mm]{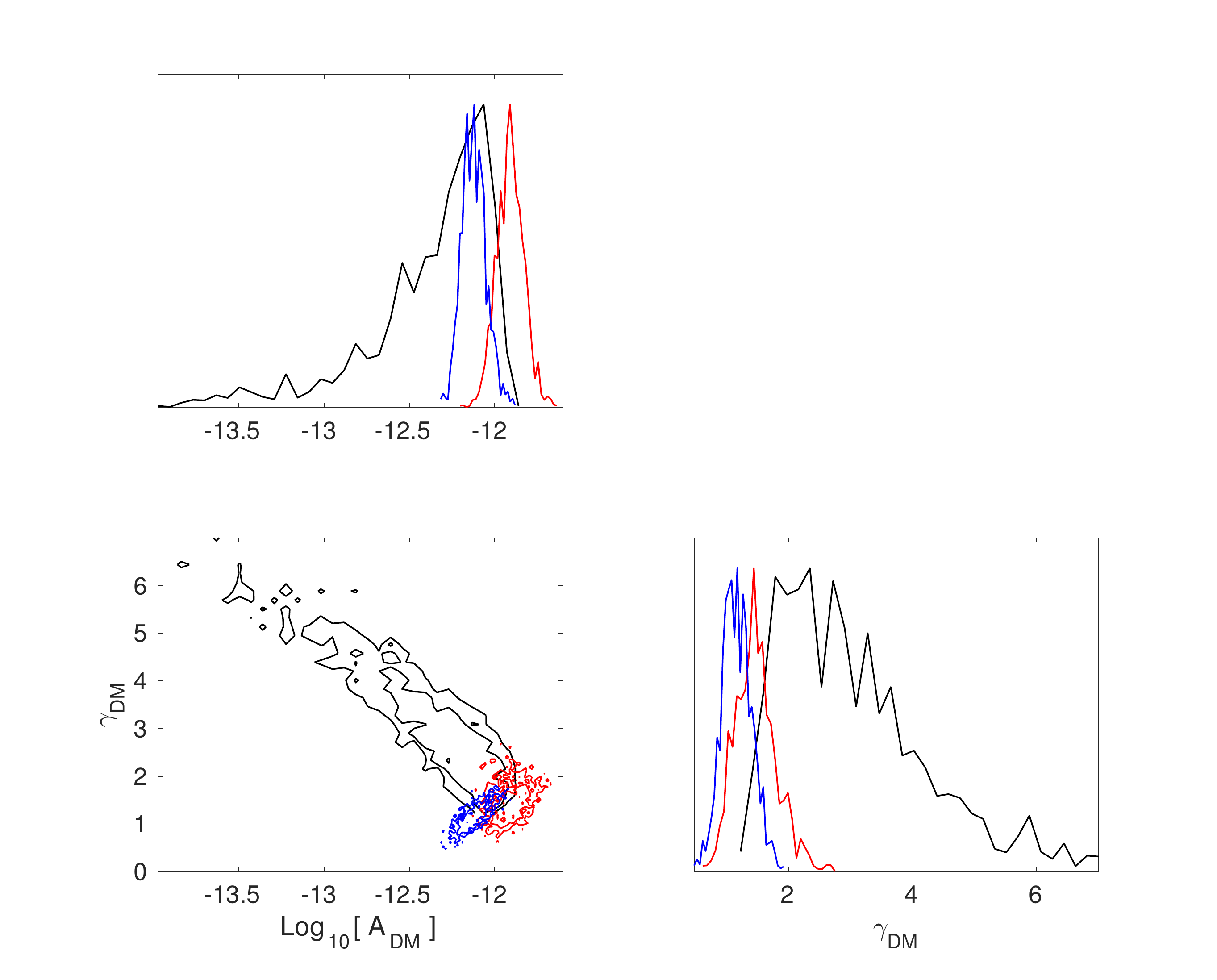} \\
\end{array}$
\end{center}
\caption{One- and two-dimensional marginalised posterior probability distributions for log amplitude and spectral index of the model for power-law DM variations in PSRs J1713+0747 (black lines), J1744$-$1134 (red lines), and J1909$-$3744 (blue lines)}
\label{figure:DMSpecs}
\end{figure}

In Fig.~\ref{figure:DMSpecs} we plot the one- and two-dimensional marginalised posterior probability distributions for the spectral index and log amplitude of the DM power-law model for all three pulsars.  We find the results for PSR J1713+0747 and J1909$-$3744 are consistent with those from an analysis of the first International Pulsar Timing Array data release \citep{2016MNRAS.458.2161L}, however, no significant detection of DM variations in PSR J1744$-$1134  was made in that analysis for a comparison to be made with the results obtained here.


\subsection{Timing Precision}
\label{Section:TimingPrecision}

We find that the profile domain analysis consistently results in higher precision compared to the standard ToA analysis.  Additionally, we find that the measurement precision obtained for the timing model parameters when using the power-law model for DM variations is also superior for all three data sets.  We list the mean and 1$\sigma$ confidence intervals for the timing model parameters obtained from our profile domain analysis of PSRs J1713+0747, J1744$-$1134, and J1909$-$3744 using the power-law model for DM in Table ~\ref{Table:PulsarParams}.

Naturally, the level of improvement when using the power-law model depends both on the pulsar in question and the time scale over which the timing model parameter occurs.  For PSR J1909$-$3744, the binary period is $\sim$ 1.5 days, and so is significantly shorter than the 30-day window used in the DMX model, or the shortest period in the power-law model.  As such, the binary parameters are not highly correlated with either model for DM variations.  We therefore find only a $\sim$3\% improvement in precision when using the power-law model compared to DMX for the binary parameters.
In contrast, the binary period of PSR J1713+0747 is approximately 67~days, and so in this case the binary parameters will be more correlated with the unconstrained DMX model.  For this pulsar we therefore find an average improvement in the measured precision of the binary parameters of 20\% for the power-law model compared to DMX.

\begin{figure}
\begin{center}$
\begin{array}{c}
\hspace{-1cm}
\includegraphics[trim = 10 10 10 10, clip,width=90mm]{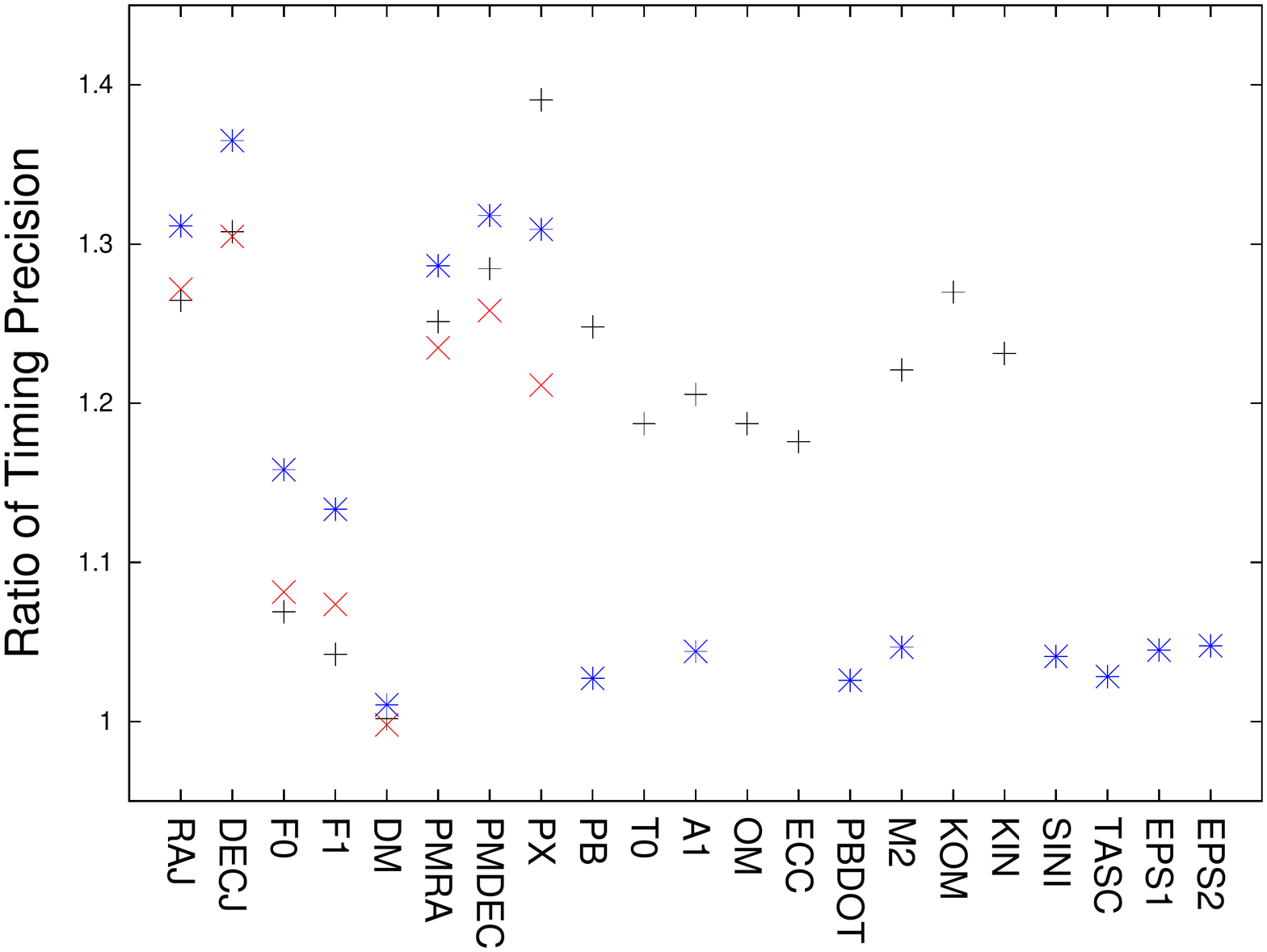} \\
\hspace{-1cm}
\includegraphics[trim = 10 10 10 10, clip,width=90mm]{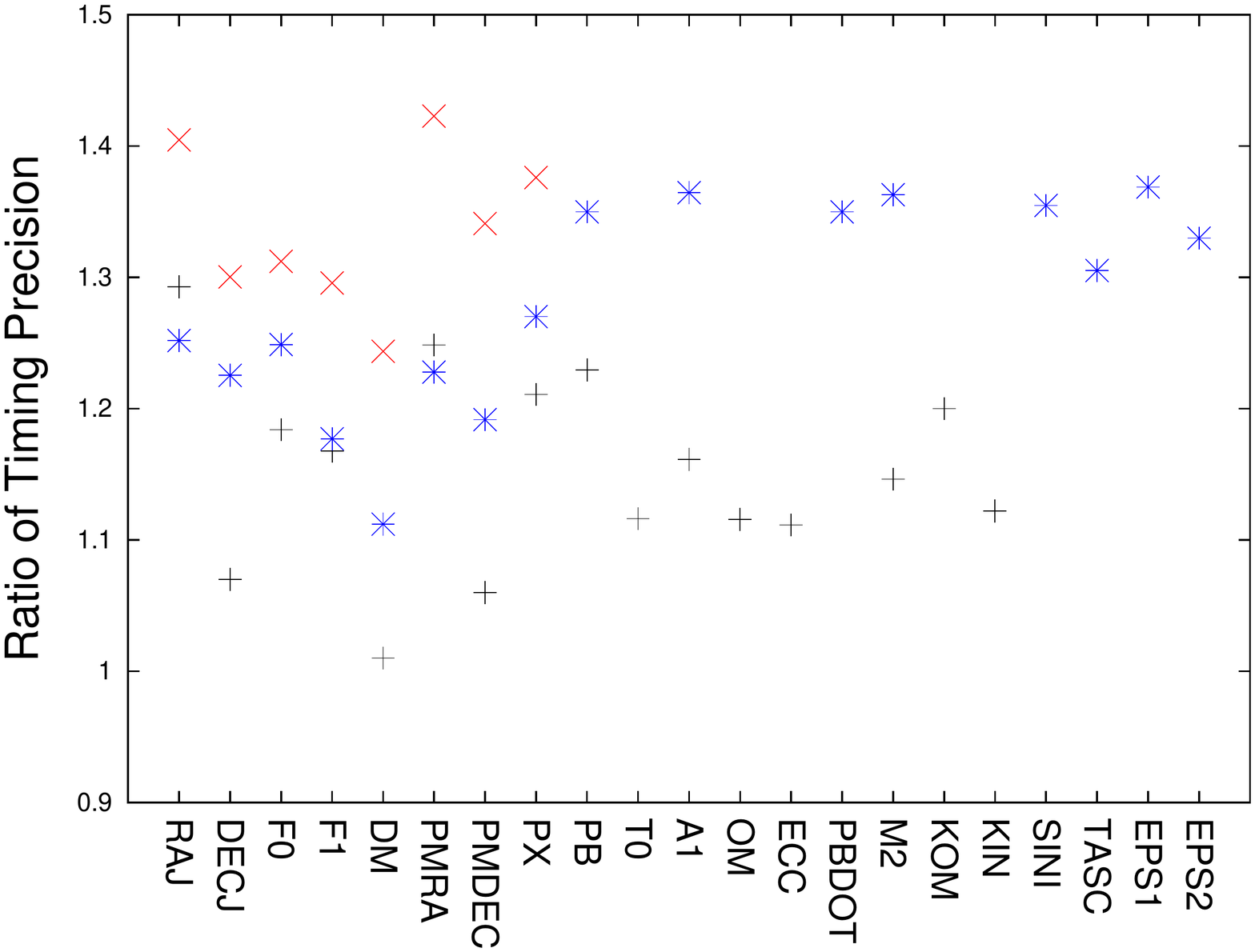} \\
\hspace{-1cm}
\includegraphics[trim = 10 10 10 10, clip,width=90mm]{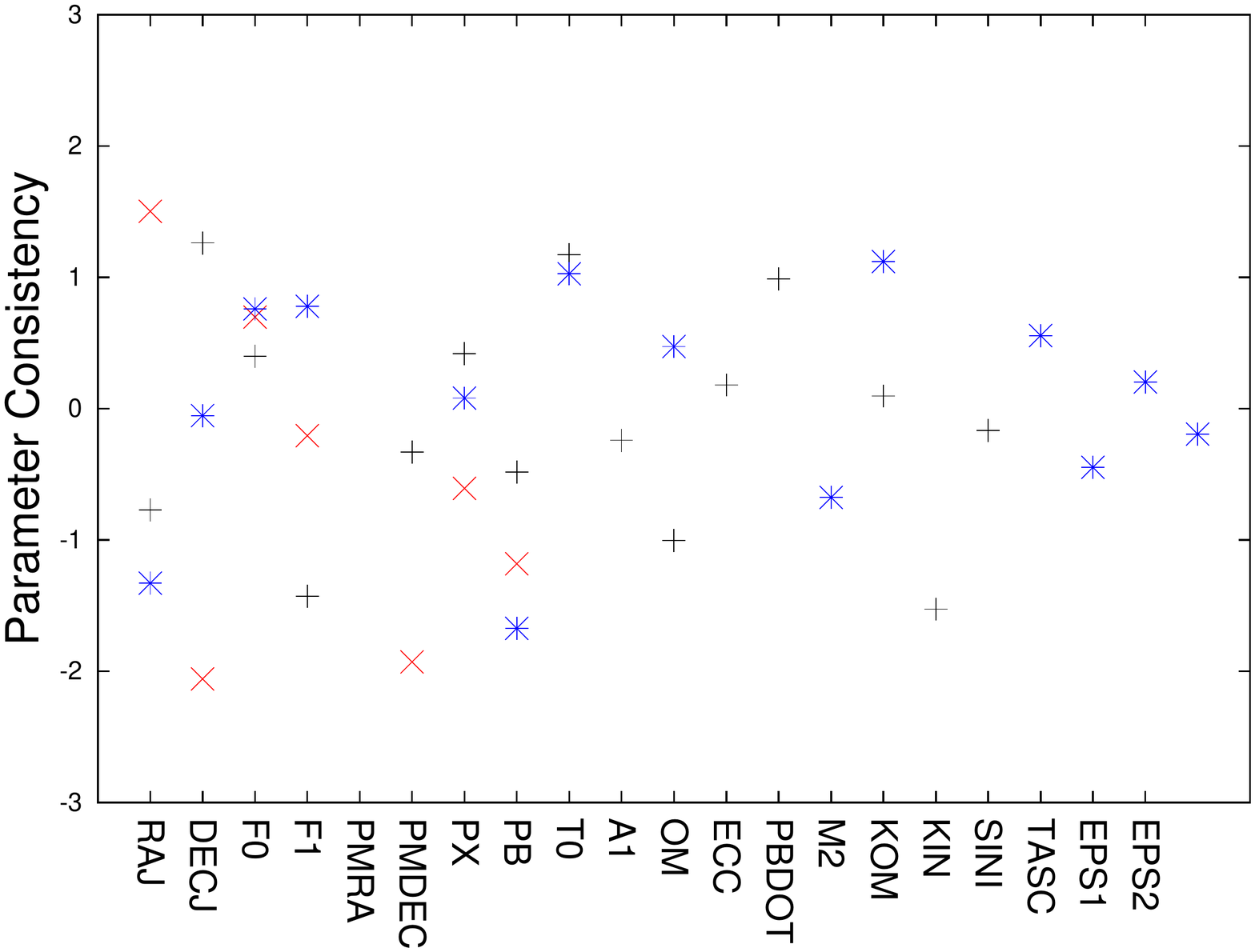}
\end{array}$
\end{center}
\caption{Ratio of the measured precision for the timing model parameters for PSRs J1713+0747 (black +), J1744$-$1134 (red $\times$), and J1909$-$3744 (blue $\ast$) for the DMX and power-law models for DM variations (top panel) and for the ToA and profile domain analysis (middle panel).  Larger numbers correspond to smaller uncertainties in the power-law model (top panel) and in the profile domain analysis (middle panel). (Bottom panel) Difference between the mean timing model parameter estimates for the ToA and profile domain analyses in terms of their uncertainties.}
\label{figure:TimingPrec}
\end{figure}

For all three data sets the astrometric parameters, which have time scales of six months or more, benefit more significantly, with an increase in the measured precision of up to 40\%.  We show the ratios for all timing model parameters for the DMX model compared to the power-law model in the top panel of Fig.~\ref{figure:TimingPrec}.

In the center panel of Fig.~\ref{figure:TimingPrec} we show an equivalent plot comparing the measured timing precision obtained from the standard ToA analysis to the profile domain analysis.  In both cases we use the DMX model for DM variations.  The improvement in the timing precision is more uniform across the different timing parameters in this case.  We find the average improvement is 25\% for the three pulsars analysed.  Given each data set is at least seven years long, for parameters such as parallax for which the measurement precision scales as the square-root of the total observing time, this corresponds to having an additional four years of data of equivalent quality.

As with the simulations, we also perform an analysis using the ToAs formed from the fully frequency-averaged profile data.  In this case we use a separate template for each band which does not evolve across the band.  For PSR J1909$-$3744 and J1744$-$1134 we find a decrease in the measured precision of the timing parameters of up to 20$\%$, with a mean decrease of 15$\%$,  compared to using  the sub-band ToAs.

For PSR J1713+0747, we find a much greater loss of precision, of approximately a factor two,  when using the fully frequency-averaged ToAs.  In this case there is significantly more scatter present in the residuals, and we find the evidence supports an additional EQUAD term with an amplitude of approximately 200~ns, consistent with the properties of the data set used in \cite{2015Sci...349.1522S}.  This is consistent with the effect of combining profile evolution and scintillation seen in Simulation~2 in Section~\ref{Section:Simulations}.

Finally, we compare the parameter estimates obtained from our profile domain analysis both with an analysis of the corresponding ToAs, and with other independent published results.

In the bottom panel of Fig.~\ref{figure:TimingPrec} we show the difference between the mean timing model parameter estimates for the ToA and profile domain analyses in terms of their uncertainties.  For clarity we exclude DM from this plot, as its value is highly dependent on the model used for the DM variations, and whether profile evolution has been incorporated.  We find that the remaining parameters are all consistent within their uncertainties.  In the following two subsections we also compare our profile domain results with parameters published in \cite{2016MNRAS.455.1751R} (R16), \cite{2016MNRAS.458.3341D} (D16), \cite{2016arXiv160300545F} (F16) and \cite{2016ApJ...818...92M} (M16).

\subsubsection{Comparison of Astrometric parameters}

We first consider the astrometric parameters, parallax and proper motion.  Comparing parallax measurements we find we are consistent with previously published values within 2$\sigma$ confidence intervals for both PSRs J1744$-$1134 and J1909$-$3744\footnote{The uncertainty on the parallax in R16 has a typographical error. The corrected value is 0.81(3), which is consistent with our measurement at the 2$\sigma$ level (Reardon, private communication).}, however small discrepancies between the measurements can be seen for PSR J1713+0747.  

For PSR J1713+0747, we find our parallax measurement lies 1$\sigma$, 1.6$\sigma$, and 2.6$\sigma$ below the values reported in R16 ($\pi=0.86(9)$~mas), M16 ($\pi=0.85(3)$~mas), and D16 ($\pi=0.90(3)$~mas) respectively.  In this case a further independent check is provided by  VLBA measurements \citep{2009ApJ...698..250C}, which result in a value of  $\pi=0.95(6)$~mas,  which lies 2.5$\sigma$ above our value.  While none of these offsets are extremely significant, in M16, inconsistency with previously reported values for parallax was attributed to insufficient modeling of DM variations in those earlier publications. In order to check the impact of possible ISM effects, we split our data set up into a 10~cm only, and a combined 20~cm and 40~cm data set, and measure the parallax in each of these separately.  We obtain values of $\pi=0.80(8)$~mas for the combined 20~cm and 40~cm data set, which is consistent with both the results from pulsar timing and the VLBA measurement at the 1.5$\sigma$ level, and a value of $\pi=0.71(5)$~mas for the 10~cm data, which is consistent at the 3$\sigma$ level with VLBA.  Given the 10~cm data is unlikely to suffer significantly from mismodelling of the ISM, and has little RFI or other known sources of systemic noise, we conclude that this marginally lower value is simply the result of random variation, rather than any systematic offset.  

We next compare measurements of proper motion for these three pulsars.  Overall, proper motions are  consistent between M16, D16, R16 and the values we report from our profile domain analysis.  We are only inconsistent at greater than 2$\sigma$ for measurements of the proper motion in declination for PSR J1744$-$1134 obtained in M16 ($\mu_{\delta} = -9.20(8)$~mas\,yr$^{-1}$, 2.6$\sigma$) and proper motion in right ascension for PSR J1713+0747 compared to the value in D16 ($\mu_{\alpha} = -3.888(14)$~mas\,yr$^{-1}$, $2.4\sigma$).  As for parallax, VLBA provides measurements of proper motion for  PSR J1713+0747 of  $\mu_{\delta} = -3.67(18)$~mas\,yr$^{-1}$, and $\mu_{\alpha} = 4.75^{+17}_{-7}$~mas\,yr$^{-1}$  which we are consistent with at the 1.0, and 1.6$\sigma$ level.

\subsubsection{Comparison of Binary parameters}

We find the majority of binary parameters for PSRs J1909$-$3744 and J1713+0747 are consistent with those published in R16, D16 and F16.  For example, measurements of the companion mass for PSR J1909$-$3744 were given as $M_c = 0.2067(19)$~$M_\odot$, 0.213(3)~$M_\odot$, and 0.214(3)~$M_\odot$ in R16, D16 and F16 respectively, which are consistent with the value of $M_c = 0.2074(20)$~$M_\odot$ given in Table~\ref{Table:PulsarParams} to within 2$\sigma$.  Small differences are seen only in the inclination angle for PSR J1909$-$3744 compared to the value reported in D16 ($\sin{i} = 0.99771(13)$, 3$\sigma$).

In \cite{2016MNRAS.458.2161L}, it was shown that unmodeled systematic, or frequency dependent effects can significantly bias the parameter estimates for the stochastic signals in pulsar timing data sets.  Neither in this paper, nor in M16, D16 or R16  have such models been incorporated into the analysis, and so it is possible that inconsistencies at the $\sim 3\sigma$ level could be due to systematic, or frequency dependent effects that are currently unmodeled in these different analyses.  One of the key advantages of the profile domain framework, however, is that eventually we will be able to integrate more physical models for these effects into the analysis.

\begin{table*}
\caption{Timing Model  Parameter Estimates}
\begin{tabular}{llll}
\hline
\multicolumn{4}{c}{Measured Quantities} \\
\hline
                                                & J1713+0747 &                      J1744$-$1134                    &               J1909$-$3744 \\
 \hline
Right ascension, $\alpha$ (hh:mm:ss)\dotfill &  17:13:49.5327239(11)  &  17:44:29.405787(3) &  19:09:47.4346730(12)\\
Declination, $\delta$ (dd:mm:ss)\dotfill & +07:47:37.49794(4) & $-$11:34:54.68136(19)& $-$37:44:14.46674(5)\\
Pulse frequency, $\nu$ (s$^{-1}$)\dotfill & 218.81184037834490(13) & 245.4261197130575(11) & 339.3156872882423(5)\\
First derivative of pulse frequency, $\dot{\nu}$ (s$^{-2}$)\dotfill & $-$4.08391(3)$\times 10^{-16}$   & $-$5.38157(7)$\times 10^{-16}$ & $-$1.614806(3)$\times 10^{-15}$\\
Proper motion in right ascension, $\mu_{\alpha} \cos \delta$ (mas\,yr$^{-1}$)\dotfill & 4.919(4) & 18.801(8)  & $-$9.519(3)\\
Proper motion in declination, $\mu_{\delta}$ (mas\,yr$^{-1}$)\dotfill & $-$3.926(7) & $-$9.43(4) & $-$35.770(9) \\
Parallax, $\pi$ (mas)\dotfill & 0.77(4) & 2.39(7)& 0.870(18) \\
Orbital period, $P_b$ (d)\dotfill & 67.825130973(3)  & - & 1.53344947428(3)\\
Epoch of periastron, $T_0$ (MJD)\dotfill & 51997.5797(4) & - & -  \\
Longitude of periastron, $\omega_0$ (deg)\dotfill & 176.1975(17) & - & - \\
Orbital eccentricity, $e$\dotfill & 7.49409(7)$\times 10^{-5}$ & - & - \\
First derivative of orbital period, $\dot{P_b}$\dotfill & - & - & 5.18(7)$\times 10^{-13}$ \\
Companion mass, $M_c$ ($M_\odot$)\dotfill & 0.277(14) & - & 0.2074(20) \\
Longitude of ascending node, $\Omega$ (degrees)\dotfill & 87(3) \\
Orbital inclination angle, $i$ (degrees)\dotfill & 72.4(8) \\
Sine of inclination angle,$\sin{i}$\dotfill & - & - & 0.99818(9) \\
Projected semi-major axis of orbit, $x$ (lt-s)\dotfill & 32.34242145(16) & - & 1.89799109(4) \\
TASC (MJD)\dotfill & - & - & 53113.95074205(3) \\
EPS1\dotfill & - & - & 2.3(16)$\times 10^{-8}$ \\
EPS2\dotfill & - & - & $-$1.04(9)$\times 10^{-7}$ \\
\hline
\multicolumn{4}{c}{Set Quantities} \\
\hline
Epoch of position determination (MJD)\dotfill & 54500 & 54500 & 54500 \\
Epoch of frequency determination (MJD)\dotfill & 56100 &  54500 & 54500 \\
\hline
\multicolumn{4}{c}{Derived Quantities} \\
\hline
Distance from parallax (kpc) \dotfill & 1.30(6) & 0.419(11) & 1.16(2)\\
Distance from $\dot{P_b}$ (kpc) \dotfill & - & - & 1.174 (16) \\
\hline
\end{tabular}\label{Table:PulsarParams}
\end{table*}

\section{Conclusions}
\label{Section:Conclusions}

We have extended the profile domain analysis framework `Generative Pulsar Timing Analysis' to incorporate wide-band effects such as profile evolution and broadband shape variation in the pulse profile.  We also incorporate models for epoch-to-epoch variation in both pulse width and in the separation in phase of the main pulse and interpulse.    This framework allows for a full timing analysis to be performed using the profile data, rather than forming estimates of the pulse times of arrivals as in the standard pulsar timing paradigm, thereby enabling simultaneous estimation of, for example,  DM variations, pulse shape variation, profile evolution and the pulsar timing model.

In order to handle the large number of profiles ($\sim20000$), and the large number of parameters ($>1000$)  that must be included in an analysis of  wide-band data sets we used Hamiltonian Monte Carlo sampling methods and introduced a new parameterisation which is significantly more efficient in the low signal-to-noise (S/N) regime for many of the stochastic parameters included in the model.  While we only used this likelihood here in the context of pulsar timing analysis, its potential applications are extremely wide ranging.

We applied this analysis framework first to a series of simulations, and then to over seven years of observations for PSRs~J1713$+$0747, J1744$-$1134, and J1909$-$3744  made between 700 and 3600~MHz.  We found that in all but the most trivial simulation the  profile domain analysis resulted in higher precision compared to a standard timing analysis, with an average improvement of 25\%.  For timing parameters that scale as the square-root of observing time, such as parallax, one would need an additional  four years of data with equivalent timing precision for a seven year data set to achieve the same improvement.

We compared a non-time-stationary piecewise model for the variations in the DM with a time-stationary power-law model for each data set and found that they were consistent in each case.  By incorporating the inherent smoothness in the DM variations into the model, however, the measurement precision obtained for the astrometric timing parameters was found to improve by as much as 40\%, and the precision of the binary parameters for PSR J1713+0747 was found to improve by 20\%.  

We also detected  significant wide-band pulse shape variation in the PSR J1713+0747 data set at the level expected for hour-long integrations given previously published estimates of the intrinsic variation in the pulsar (e.g. \citealt{2015ApJ...813...65T}).   Not accounting for this shape variation changes the measured arrival times at the level of $\sim$30~ns, the same order of magnitude as the expected shift due to gravitational waves in the pulsar timing band.

As new instrumentation that continues to increase the simultaneous bandwidth of observations comes online, the ability to deal with effects that combine coherently over the band, such as intrinsic pulse shape variation and profile evolution, will continue to become more important.  The profile domain framework developed in this work and preceding papers provides a statistically robust approach to simultaneously incorporating these effects into pulsar timing analysis.  By directly modeling the effect of shape change, one can improve the sensitivity of the analysis to any process that results in a shift of the arrival time.  Such processes include post-Keplerian effects in highly relativistic systems  \citep{2006Sci...314...97K},  changes in the binary period, which provide tests of the potential evolution in the gravitational constant $G$  \citep{2015ApJ...809...41Z},  and the passage of nHz gravitational waves.  As both models and sampling efficiency improve, and the analysis of progressively more complex data sets becomes tractable,  profile domain analysis will ultimately provide the optimal approach to performing high precision pulsar timing analysis.

\section{Acknowledgements}

The Parkes radio telescope is part of the Australia Telescope National Facility which is funded by the Commonwealth of Australia for operation as a National Facility managed by Commonwealth Science and Industrial Research Organization (CSIRO).  We thank all of the observers, engineers, and Parkes observatory staff members who have assisted with the observations reported in this paper.

LL was supported by a Junior Research Fellowship at Trinity Hall College, Cambridge University.

\bibliographystyle{mn2e}
\bibliography{references}

\begin{thebibliography}{3}
\expandafter\ifx\csname natexlab\endcsname\relax\def\natexlab#1{#1}\fi

\bibitem[{{Manchester} {et~al}\mbox{.}(2013){Manchester}, {Hobbs}, {Bailes},
  {Coles}, {van Straten}, {Keith}, {Shannon}, {Bhat}, {Brown}, {Burke-Spolaor},
  {Champion}, {Chaudhary}, {Edwards}, {Hampson}, {Hotan}, {Jameson}, {Jenet},
  {Kesteven}, {Khoo}, {Kocz}, {Maciesiak}, {Oslowski}, {Ravi}, {Reynolds},
  {Sarkissian}, {Verbiest}, {Wen}, {Wilson}, {Yardley}, {Yan}, \&
  {You}}]{2013PASA...30...17M}
{Manchester} R.~N. {et~al.}, 2013, \pasa, 30, 17

\bibitem[{{Refregier}(2003)}]{2003MNRAS.338...35R}
{Refregier} A., 2003, \mnras, 338, 35

\bibitem[{{Shannon} {et~al}\mbox{.}(2014){Shannon}
  {et~al.}}]{2014MNRAS.443.1463S}
{Shannon} R.~M., {et~al.}, 2014, Mon. Not. R. Astron. Soc., 443, 1463

\end{thebibliography}


\begin{thebibliography}{5}
\expandafter\ifx\csname natexlab\endcsname\relax\def\natexlab#1{#1}\fi

\bibitem[{{Handley}, {Hobson} \& {Lasenby}(2015){Handley}, {Hobson}, \&
  {Lasenby}}]{2015arXiv150201856H}
{Handley} W.~J., {Hobson} M.~P., {Lasenby} A.~N., 2015, ArXiv e-prints

\bibitem[{{Kass} \& {Raftery}(1995)}]{bayesRef}
{Kass} R.~E., {Raftery} A.~E., 1995, Journal of the American Statistical
  Association, 90, 791

\bibitem[{{Neal}(2000)}]{2000physics...9028N}
{Neal} R.~M., 2000, ArXiv Physics e-prints

\bibitem[{{Refregier}(2003)}]{2003MNRAS.338...35R}
{Refregier} A., 2003, \mnras, 338, 35

\bibitem[{{Skilling}(2004)}]{2004AIPC..735..395S}
{Skilling} J., 2004, in American Institute of Physics Conference Series, Vol.
  735, American Institute of Physics Conference Series, {Fischer} R., {Preuss}
  R., {Toussaint} U.~V., eds., pp. 395--405

\end{thebibliography}


\begin{thebibliography}{41}
\expandafter\ifx\csname natexlab\endcsname\relax\def\natexlab#1{#1}\fi

\bibitem[{{Arzoumanian} {et~al}\mbox{.}(2015){Arzoumanian}, {Brazier},
  {Burke-Spolaor}, {Chamberlin}, {Chatterjee}, {Christy}, {Cordes}, {Cornish},
  {Crowter}, {Demorest}, {Dolch}, {Ellis}, {Ferdman}, {Fonseca},
  {Garver-Daniels}, {Gonzalez}, {Jenet}, {Jones}, {Jones}, {Kaspi}, {Koop},
  {Lam}, {Lazio}, {Levin}, {Lommen}, {Lorimer}, {Luo}, {Lynch}, {Madison},
  {McLaughlin}, {McWilliams}, {Nice}, {Palliyaguru}, {Pennucci}, {Ransom},
  {Siemens}, {Stairs}, {Stinebring}, {Stovall}, {Swiggum}, {Vallisneri}, {van
  Haasteren}, {Wang}, \& {Zhu}}]{2015ApJ...813...65T}
{Arzoumanian} Z. {et~al.}, 2015, \apj, 813, 65

\bibitem[{{Chatterjee} {et~al}\mbox{.}(2009){Chatterjee}, {Brisken},
  {Vlemmings}, {Goss}, {Lazio}, {Cordes}, {Thorsett}, {Fomalont}, {Lyne}, \&
  {Kramer}}]{2009ApJ...698..250C}
{Chatterjee} S. {et~al.}, 2009, \apj, 698, 250

\bibitem[{{Cognard} {et~al}\mbox{.}(2013){Cognard}, {Theureau}, {Guillemot},
  {Liu}, {Lassus}, \& {Desvignes}}]{2013sf2a.conf..327C}
{Cognard} I., {Theureau} G., {Guillemot} L., {Liu} K., {Lassus} A., {Desvignes}
  G., 2013, in SF2A-2013: Proceedings of the Annual meeting of the French
  Society of Astronomy and Astrophysics, {Cambresy} L., {Martins} F., {Nuss}
  E., {Palacios} A., eds., pp. 327--330

\bibitem[{{Coles} {et~al}\mbox{.}(2015){Coles}, {Kerr}, {Shannon}, {Hobbs},
  {Manchester}, {You}, {Bailes}, {Bhat}, {Burke-Spolaor}, {Dai}, {Keith},
  {Levin}, {Os{\l}owski}, {Ravi}, {Reardon}, {Toomey}, {van Straten}, {Wang},
  {Wen}, \& {Zhu}}]{2015ApJ...808..113C}
{Coles} W.~A. {et~al.}, 2015, \apj, 808, 113

\bibitem[{{Cordes}(1978)}]{1978ApJ...222.1006C}
{Cordes} J.~M., 1978, \apj, 222, 1006

\bibitem[{{Cordes}, {Shannon} \& {Stinebring}(2016){Cordes}, {Shannon}, \&
  {Stinebring}}]{2016ApJ...817...16C}
{Cordes} J.~M., {Shannon} R.~M., {Stinebring} D.~R., 2016, \apj, 817, 16

\bibitem[{{Dai} {et~al}\mbox{.}(2016){Dai}, {Johnston}, {Bell}, {Coles},
  {Hobbs}, {Ekers}, \& {Lenc}}]{2016MNRAS.462.3115D}
{Dai} S., {Johnston} S., {Bell} M.~E., {Coles} W.~A., {Hobbs} G., {Ekers}
  R.~D., {Lenc} E., 2016, \mnras, 462, 3115

\bibitem[{{Dai} {et~al}\mbox{.}(2015){Dai} {et~al.}}]{2015MNRAS.449.3223D}
{Dai} S., {et~al.}, 2015, \mnras, 449, 3223

\bibitem[{{Demorest} {et~al}\mbox{.}(2013){Demorest}
  {et~al.}}]{2013ApJ...762...94D}
{Demorest} P.~B., {et~al.}, 2013, The Astrophys. J., 762, 94

\bibitem[{{Desvignes} {et~al}\mbox{.}(2016){Desvignes}, {Caballero}, {Lentati},
  {Verbiest}, {Champion}, {Stappers}, {Janssen}, {Lazarus}, {Os{\l}owski},
  {Babak}, {Bassa}, {Brem}, {Burgay}, {Cognard}, {Gair}, {Graikou},
  {Guillemot}, {Hessels}, {Jessner}, {Jordan}, {Karuppusamy}, {Kramer},
  {Lassus}, {Lazaridis}, {Lee}, {Liu}, {Lyne}, {McKee}, {Mingarelli},
  {Perrodin}, {Petiteau}, {Possenti}, {Purver}, {Rosado}, {Sanidas}, {Sesana},
  {Shaifullah}, {Smits}, {Taylor}, {Theureau}, {Tiburzi}, {van Haasteren}, \&
  {Vecchio}}]{2016MNRAS.458.3341D}
{Desvignes} G. {et~al.}, 2016, \mnras, 458, 3341

\bibitem[{{Feroz} \& {Hobson}(2008)}]{2008MNRAS.384..449F}
{Feroz} F., {Hobson} M.~P., 2008, Mon. Not. R. Astron. Soc., 384, 449

\bibitem[{{Feroz}, {Hobson} \& {Bridges}(2009){Feroz}, {Hobson}, \&
  {Bridges}}]{2009MNRAS.398.1601F}
{Feroz} F., {Hobson} M.~P., {Bridges} M., 2009, Mon. Not. R. Astron. Soc., 398,
  1601

\bibitem[{{Fonseca} {et~al}\mbox{.}(2016){Fonseca}, {Pennucci}, {Ellis},
  {Stairs}, {Nice}, {Ransom}, {Demorest}, {Arzoumanian}, {Crowter}, {Dolch},
  {Ferdman}, {Gonzalez}, {Jones}, {Jones}, {Lam}, {Levin}, {McLaughlin},
  {Stovall}, {Swiggum}, \& {Zhu}}]{2016arXiv160300545F}
{Fonseca} E. {et~al.}, 2016, ArXiv e-prints

\bibitem[{{Handley}, {Hobson} \& {Lasenby}(2015){Handley}, {Hobson}, \&
  {Lasenby}}]{2015MNRAS.453.4384H}
{Handley} W.~J., {Hobson} M.~P., {Lasenby} A.~N., 2015, \mnras, 453, 4384

\bibitem[{{Hankins} \& {Cordes}(1981)}]{1981ApJ...249..241H}
{Hankins} T.~H., {Cordes} J.~M., 1981, \apj, 249, 241

\bibitem[{{Hewish} {et~al}\mbox{.}(1968){Hewish}, {Bell}, {Pilkington},
  {Scott}, \& {Collins}}]{1968Natur.217..709H}
{Hewish} A., {Bell} S.~J., {Pilkington} J.~D.~H., {Scott} P.~F., {Collins}
  R.~A., 1968, \nat, 217, 709

\bibitem[{{Keith} {et~al}\mbox{.}(2013){Keith} {et~al.}}]{2013MNRAS.429.2161K}
{Keith} M.~J., {et~al.}, 2013, Mon. Not. R. Astron. Soc., 429, 2161

\bibitem[{{Kramer} {et~al}\mbox{.}(2006){Kramer}, {Stairs}, {Manchester},
  {McLaughlin}, {Lyne}, {Ferdman}, {Burgay}, {Lorimer}, {Possenti}, {D'Amico},
  {Sarkissian}, {Hobbs}, {Reynolds}, {Freire}, \&
  {Camilo}}]{2006Sci...314...97K}
{Kramer} M. {et~al.}, 2006, Science, 314, 97

\bibitem[{{Lee} {et~al}\mbox{.}(2014){Lee}, {Bassa}, {Janssen}, {Karuppusamy},
  {Kramer}, {Liu}, {Perrodin}, {Smits}, {Stappers}, {van Haasteren}, \&
  {Lentati}}]{2014MNRAS.441.2831L}
{Lee} K.~J. {et~al.}, 2014, \mnras, 441, 2831

\bibitem[{{Lentati}, {Alexander} \& {Hobson}(2015){Lentati}, {Alexander}, \&
  {Hobson}}]{2015MNRAS.447.2159L}
{Lentati} L., {Alexander} P., {Hobson} M.~P., 2015, \mnras, 447, 2159

\bibitem[{{Lentati} {et~al}\mbox{.}(2013){Lentati}, {Alexander}, {Hobson},
  {Taylor}, {Gair}, {Balan}, \& {van Haasteren}}]{2013PhRvD..87j4021L}
{Lentati} L., {Alexander} P., {Hobson} M.~P., {Taylor} S., {Gair} J., {Balan}
  S.~T., {van Haasteren} R., 2013, Phys. Rev. D, 87, 104021

\bibitem[{{Lentati}, {Hobson} \& {Alexander}(2014){Lentati}, {Hobson}, \&
  {Alexander}}]{2014MNRAS.444.3863L}
{Lentati} L., {Hobson} M.~P., {Alexander} P., 2014, \mnras, 444, 3863

\bibitem[{{Lentati} \& {Shannon}(2015)}]{2015MNRAS.454.1058L}
{Lentati} L., {Shannon} R.~M., 2015, \mnras, 454, 1058

\bibitem[{{Lentati} {et~al}\mbox{.}(2016){Lentati}, {Shannon}, {Coles},
  {Verbiest}, {van Haasteren}, {Ellis}, {Caballero}, {Manchester},
  {Arzoumanian}, {Babak}, {Bassa}, {Bhat}, {Brem}, {Burgay}, {Burke-Spolaor},
  {Champion}, {Chatterjee}, {Cognard}, {Cordes}, {Dai}, {Demorest},
  {Desvignes}, {Dolch}, {Ferdman}, {Fonseca}, {Gair}, {Gonzalez}, {Graikou},
  {Guillemot}, {Hessels}, {Hobbs}, {Janssen}, {Jones}, {Karuppusamy}, {Keith},
  {Kerr}, {Kramer}, {Lam}, {Lasky}, {Lassus}, {Lazarus}, {Lazio}, {Lee},
  {Levin}, {Liu}, {Lynch}, {Madison}, {McKee}, {McLaughlin}, {McWilliams},
  {Mingarelli}, {Nice}, {Os{\l}owski}, {Pennucci}, {Perera}, {Perrodin},
  {Petiteau}, {Possenti}, {Ransom}, {Reardon}, {Rosado}, {Sanidas}, {Sesana},
  {Shaifullah}, {Siemens}, {Smits}, {Stairs}, {Stappers}, {Stinebring},
  {Stovall}, {Swiggum}, {Taylor}, {Theureau}, {Tiburzi}, {Toomey},
  {Vallisneri}, {van Straten}, {Vecchio}, {Wang}, {Wang}, {You}, {Zhu}, \&
  {Zhu}}]{2016MNRAS.458.2161L}
{Lentati} L. {et~al.}, 2016, \mnras, 458, 2161

\bibitem[{{Liu} {et~al}\mbox{.}(2014){Liu}, {Desvignes}, {Cognard}, {Stappers},
  {Verbiest}, {Lee}, {Champion}, {Kramer}, {Freire}, \&
  {Karuppusamy}}]{2014MNRAS.443.3752L}
{Liu} K. {et~al.}, 2014, \mnras, 443, 3752

\bibitem[{{Manchester}(2015)}]{2015IAUGA..2256190M}
{Manchester} R.~N., 2015, IAU General Assembly, 22, 2256190

\bibitem[{{Manchester} {et~al}\mbox{.}(2013){Manchester}, {Hobbs}, {Bailes},
  {Coles}, {van Straten}, {Keith}, {Shannon}, {Bhat}, {Brown}, {Burke-Spolaor},
  {Champion}, {Chaudhary}, {Edwards}, {Hampson}, {Hotan}, {Jameson}, {Jenet},
  {Kesteven}, {Khoo}, {Kocz}, {Maciesiak}, {Oslowski}, {Ravi}, {Reynolds},
  {Sarkissian}, {Verbiest}, {Wen}, {Wilson}, {Yardley}, {Yan}, \&
  {You}}]{2013PASA...30...17M}
{Manchester} R.~N. {et~al.}, 2013, \pasa, 30, 17

\bibitem[{{Matthews} {et~al}\mbox{.}(2016){Matthews}, {Nice}, {Fonseca},
  {Arzoumanian}, {Crowter}, {Demorest}, {Dolch}, {Ellis}, {Ferdman},
  {Gonzalez}, {Jones}, {Jones}, {Lam}, {Levin}, {McLaughlin}, {Pennucci},
  {Ransom}, {Stairs}, {Stovall}, {Swiggum}, \& {Zhu}}]{2016ApJ...818...92M}
{Matthews} A.~M. {et~al.}, 2016, \apj, 818, 92

\bibitem[{{Narayan}(1992)}]{1992RSPTA.341..151N}
{Narayan} R., 1992, Philosophical Transactions of the Royal Society of London
  Series A, 341, 151

\bibitem[{Nelder \& Mead(1965)}]{NelderMead65}
Nelder J.~A., Mead R., 1965, Computer Journal, 7, 308

\bibitem[{{Os{\l}owski} {et~al}\mbox{.}(2011){Os{\l}owski}, {van Straten},
  {Hobbs}, {Bailes}, \& {Demorest}}]{2011MNRAS.418.1258O}
{Os{\l}owski} S., {van Straten} W., {Hobbs} G.~B., {Bailes} M., {Demorest} P.,
  2011, \mnras, 418, 1258

\bibitem[{{Pennucci}, {Demorest} \& {Ransom}(2014){Pennucci}, {Demorest}, \&
  {Ransom}}]{2014ApJ...790...93P}
{Pennucci} T.~T., {Demorest} P.~B., {Ransom} S.~M., 2014, \apj, 790, 93

\bibitem[{{Ransom} {et~al}\mbox{.}(2009){Ransom}, {Demorest}, {Ford},
  {McCullough}, {Ray}, {DuPlain}, \& {Brandt}}]{2009AAS...21460508R}
{Ransom} S.~M., {Demorest} P., {Ford} J., {McCullough} R., {Ray} J., {DuPlain}
  R., {Brandt} P., 2009, in American Astronomical Society Meeting Abstracts,
  Vol. 214, American Astronomical Society Meeting Abstracts 214, p. 605.08

\bibitem[{{Reardon} {et~al}\mbox{.}(2016){Reardon}, {Hobbs}, {Coles}, {Levin},
  {Keith}, {Bailes}, {Bhat}, {Burke-Spolaor}, {Dai}, {Kerr}, {Lasky},
  {Manchester}, {Os{\l}owski}, {Ravi}, {Shannon}, {van Straten}, {Toomey},
  {Wang}, {Wen}, {You}, \& {Zhu}}]{2016MNRAS.455.1751R}
{Reardon} D.~J. {et~al.}, 2016, \mnras, 455, 1751

\bibitem[{{Refregier}(2003)}]{2003MNRAS.338...35R}
{Refregier} A., 2003, \mnras, 338, 35

\bibitem[{{Shannon} {et~al}\mbox{.}(2015){Shannon}, {Ravi}, {Lentati}, {Lasky},
  {Hobbs}, {Kerr}, {Manchester}, {Coles}, {Levin}, {Bailes}, {Bhat},
  {Burke-Spolaor}, {Dai}, {Keith}, {Os{\l}owski}, {Reardon}, {van Straten},
  {Toomey}, {Wang}, {Wen}, {Wyithe}, \& {Zhu}}]{2015Sci...349.1522S}
{Shannon} R.~M. {et~al.}, 2015, Science, 349, 1522

\bibitem[{{Shannon} {et~al}\mbox{.}(2014){Shannon}
  {et~al.}}]{2014MNRAS.443.1463S}
{Shannon} R.~M., {et~al.}, 2014, Mon. Not. R. Astron. Soc., 443, 1463

\bibitem[{{Stappers} {et~al}\mbox{.}(2011){Stappers}, {Hessels}, {Alexov},
  {Anderson}, {Coenen}, {Hassall}, {Karastergiou}, {Kondratiev}, {Kramer}, {van
  Leeuwen}, {Mol}, {Noutsos}, {Romein}, {Weltevrede}, {Fender}, {Wijers},
  {B{\"a}hren}, {Bell}, {Broderick}, {Daw}, {Dhillon}, {Eisl{\"o}ffel},
  {Falcke}, {Griessmeier}, {Law}, {Markoff}, {Miller-Jones}, {Scheers},
  {Spreeuw}, {Swinbank}, {Ter Veen}, {Wise}, {Wucknitz}, {Zarka}, {Anderson},
  {Asgekar}, {Avruch}, {Beck}, {Bennema}, {Bentum}, {Best}, {Bregman},
  {Brentjens}, {van de Brink}, {Broekema}, {Brouw}, {Br{\"u}ggen}, {de Bruyn},
  {Butcher}, {Ciardi}, {Conway}, {Dettmar}, {van Duin}, {van Enst}, {Garrett},
  {Gerbers}, {Grit}, {Gunst}, {van Haarlem}, {Hamaker}, {Heald}, {Hoeft},
  {Holties}, {Horneffer}, {Koopmans}, {Kuper}, {Loose}, {Maat},
  {McKay-Bukowski}, {McKean}, {Miley}, {Morganti}, {Nijboer}, {Noordam},
  {Norden}, {Olofsson}, {Pandey-Pommier}, {Polatidis}, {Reich},
  {R{\"o}ttgering}, {Schoenmakers}, {Sluman}, {Smirnov}, {Steinmetz}, {Sterks},
  {Tagger}, {Tang}, {Vermeulen}, {Vermaas}, {Vogt}, {de Vos}, {Wijnholds},
  {Yatawatta}, \& {Zensus}}]{2011A&A...530A..80S}
{Stappers} B.~W. {et~al.}, 2011, Astronomy \& Astrophysics, 530, A80

\bibitem[{{Taylor}, {Ashdown} \& {Hobson}(2008){Taylor}, {Ashdown}, \&
  {Hobson}}]{2008MNRAS.389.1284T}
{Taylor} J.~F., {Ashdown} M.~A.~J., {Hobson} M.~P., 2008, \mnras, 389, 1284

\bibitem[{{Taylor}(1992)}]{1992RSPTA.341..117T}
{Taylor} J.~H., 1992, Royal Society of London Philosophical Transactions Series
  A, 341, 117

\bibitem[{{Zhu} {et~al}\mbox{.}(2015){Zhu}, {Stairs}, {Demorest}, {Nice},
  {Ellis}, {Ransom}, {Arzoumanian}, {Crowter}, {Dolch}, {Ferdman}, {Fonseca},
  {Gonzalez}, {Jones}, {Jones}, {Lam}, {Levin}, {McLaughlin}, {Pennucci},
  {Stovall}, \& {Swiggum}}]{2015ApJ...809...41Z}
{Zhu} W.~W. {et~al.}, 2015, \apj, 809, 41

\end{thebibliography}

\end{document}